\documentclass[a4paper,aps,prd,twocolumn,showpacs,longbibliography]{revtex4-1}
\pdfoutput=1

\usepackage{enumerate}
\usepackage{graphicx}
\usepackage[caption=false]{subfig}
\usepackage{hyperref}

\usepackage{bm}
\usepackage{amsmath}
\usepackage{amssymb}

\DeclareMathOperator{\SNR}{SNR}
\DeclareMathOperator{\E}{E}
\DeclareMathOperator{\Var}{Var}
\DeclareMathOperator{\median}{median}

\newcommand{\calA}{\mathcal{A}}
\newcommand{\calF}{\mathcal{F}}
\newcommand{\calJ}{\mathcal{J}}

\newcommand{\calN}{\mathcal{N}}
\newcommand{\calP}{\mathcal{P}}
\newcommand{\calT}{\mathcal{T}}

\newcommand{\Ans}[1][n]{A_{#1}^{*}}
\newcommand{\Fstat}{^{0}}

\newcommand{\diff}{\Delta}
\newcommand{\gct}{^{\textrm{gc}}}
\newcommand{\mat}[1]{\boldsymbol{\mathrm{#1}}}
\newcommand{\ndot}[2][s]{#2^{(#1)}}
\newcommand{\phase}{^{\phi}}

\newcommand{\relerr}[2]{\varepsilon(#1,#2)}
\newcommand{\sums}[1][N]{\sum_{\ell=1}^{#1}}
\newcommand{\ua}{_{a}}
\newcommand{\ub}{_{b}}
\newcommand{\umax}{_{\textrm{max}}}
\newcommand{\umin}{_{\textrm{min}}}
\newcommand{\unc}{^\textrm{u}}
\newcommand{\unoint}{_\textrm{ni}}
\newcommand{\upureint}{_\textrm{pi}}
\newcommand{\useg}{_{\ell}}
\renewcommand{\vec}[1]{\boldsymbol{#1}}

\newcommand{\sig}[1]{#1^\mathrm{s}}
\newcommand{\sigvec}[1]{\sig{\vec{#1}}}
\newcommand{\coh}[1]{\widetilde{#1}}
\newcommand{\cohvec}[1]{\coh{\vec{#1}}}
\newcommand{\cohs}[1]{\coh{#1}\useg}
\newcommand{\cohsmat}[1]{\cohs{\mat{#1}}}
\newcommand{\cohsvec}[1]{\cohs{\vec{#1}}}
\newcommand{\semi}[1]{\widehat{#1}}
\newcommand{\semimat}[1]{\semi{\mat{#1}}}
\newcommand{\semivec}[1]{\semi{\vec{#1}}}

\newcommand{\commitDATE}{2015-10-28 17:36:55 +0100}
\newcommand{\commitID}{commitID: 2ded9fa}
\newcommand{\commitSTATUS}{CLEAN}

\defcitealias{Wette.Prix.2013a}{Paper~I}
\newcommand{\PaperI}{\citetalias{Wette.Prix.2013a}}
\defcitealias{Wette.2014a}{Paper~II}
\newcommand{\PaperII}{\citetalias{Wette.2014a}}

\begin{document}

\title{Parameter-space metric for all-sky semicoherent searches for \\ gravitational-wave pulsars}
\author{Karl Wette}
\email{karl.wette@aei.mpg.de}
\affiliation{Max-Planck-Institut f\"ur Gravitationsphysik (Albert-Einstein-Institut), Callinstra\ss{}e 38, 30167 Hannover, Germany}

\date{\commitDATE; \commitID-\commitSTATUS}

\begin{abstract}
The sensitivity of all-sky searches for gravitational-wave pulsars is primarily limited by the finite availability of computing resources.
Semicoherent searches are a widely-used method of maximizing sensitivity to gravitational-wave pulsars at fixed computing cost: the data from a gravitational-wave detector are partitioned into a number of segments, each segment is coherently analyzed, and the analysis results from each segment are summed together.
The generation of template banks for the coherent analysis of each segment, and for the summation, requires knowledge of the metrics associated with the coherent and semicoherent parameter spaces respectively.
We present a useful approximation to the semicoherent parameter-space metric, analogous to that presented in Wette and Prix [Phys.~Rev.~D \textbf{88}, 123005 (2013)] for the coherent metric.
The new semicoherent metric is compared to previous work in Pletsch [Phys.~Rev.~D \textbf{82}, 042002 (2010)], and Brady and Creighton [Phys.~Rev.~D \textbf{61}, 082001 (2000)].
We find that semicoherent all-sky searches require orders of magnitude more templates than previously predicted.
\end{abstract}


\maketitle

\section{Introduction}\label{sec:introduction}

Gravitational-wave pulsars~\footnote{
As noted in~\cite{Cutler.2012a}, the term ``gravitational-wave pulsar'' is a slight misnomer, since the signals they emit are more accurately described as ``sinusoidal'' than ``pulsating''.
} are rotating neutron stars which are hypothesized, though not yet observed, to emit gravitational radiation in the form of long-lived, continuous, quasisinusoidal signals.
The maximum amplitude of a gravitational-wave pulsar signal has been most recently studied in~\cite[][]{JohnsonMcDaniel.Owen.2013a}, and the evolution of the population of Galactic neutron stars emitting at an assumed amplitude has been modeled in~\cite{Knispel.Allen.2008a,Wade.etal.2012a}.
The sensitivity of the interferometric gravitational-wave observatories LIGO~\cite{Abbott.etal.2009f,Harry.etal.2010a} and Virgo~\cite{Accadia.etal.2012a,Acernese.etal.2015a} are expected to improve significantly within this decade, while a third large-scale observatory, KAGRA~\cite{Somiya.2012a}, is under construction.
Nevertheless, it remains likely that gravitational-wave pulsars will remain at best marginally detectable for the foreseeable future.

The detection of a gravitational-wave pulsar, in the absence of an observed electromagnetic counterpart~\cite[cf.][]{Aasi.etal.2014b}, is a significant data-analysis challenge, requiring both large-scale computing resources and highly-optimized data analysis algorithms.
The former is provided by the distributed computing project Einstein@Home~\footnote{
\url{http://www.einsteinathome.org/}.
}, which is dedicated to searching for both gravitational-wave and electromagnetic pulsars; to date, the project has discovered many new radio and gamma-ray pulsars~\cite[e.g.][]{Knispel.etal.2013a,Pletsch.etal.2013a}.
Research into the latter has led to advances in the understanding of optimal and/or robust detection statistics~\cite[e.g.][]{Jaranowski.etal.1998a,Prix.Krishnan.2009a,Dergachev.2010a,Cutler.2012a,Keitel.etal.2014a}, the construction of matched-filtering template banks~\cite[e.g.][]{Brady.etal.1998a,Prix.2007a,Prix.2007b,Messenger.etal.2009a,Wette.Prix.2013a,Wette.2014a,Pisarski.Jaranowski.2015a},
hierarchical search methods~\cite[e.g.][]{Brady.Creighton.2000a,Krishnan.etal.2004a,Pletsch.2010a} and their optimization given limited computing resources~\cite[e.g.][]{Cutler.etal.2005a,Prix.Shaltev.2012a}, the estimation of search sensitivity~\cite[e.g.][]{Wette.2012a}, and pipelines for post-processing promising detection candidates~\cite[e.g.][]{Shaltev.Prix.2013a,Behnke.etal.2015a}.

This paper builds on recent work on all-sky, broadband, \emph{coherent} searches for \emph{isolated} gravitational-wave pulsars.
(Similar studies of the parameter-space metric for gravitational-wave pulsars in \emph{binary} systems can be found in e.g.~\cite{Messenger.2011a,Leaci.Prix.2015a}.)
In~\cite{Wette.Prix.2013a} (hereafter~\PaperI), a useful approximation to the \emph{parameter-space metric} is developed, which is important for the correct construction of banks of matched-filtering templates.
Previous work~\cite[e.g.][]{Astone.etal.2002b,Pletsch.Allen.2009a} had assumed restrictions on the time-span of data which could be coherently analyzed; these restrictions were relaxed in~\PaperI.
In~\cite{Wette.2014a} (hereafter~\PaperII), a practical implementation of an optimal template bank using lattice template placement is presented.
The number of templates in the bank was found to be in good agreement with the prediction of~\cite{Brady.etal.1998a}.

This paper presents a useful approximation to the parameter-space metric for all-sky \emph{semicoherent} searches for gravitational-wave pulsars, building on the work of~\PaperI.
In a semicoherent search, the data are divided into a number of \emph{segments}, each of which are coherently searched using a per-segment \emph{coherent template bank}; the search results from each segment are then summed together using a \emph{semicoherent template bank}~\footnote{
The terms ``coarse template bank'' and ``fine template bank'' are also used to refer to the coherent and semicoherent template banks respectively, e.g.\ in~\cite{Prix.Shaltev.2012a}.
}.
Each semicoherent template preserves a consistent (instantaneous) signal frequency $f(t)$ between each segment, but not necessarily a consistent signal amplitude.
While semicoherent searches are less sensitive than a single-segment fully-coherent search of the same data, they are also computationally cheaper, and are therefore able to analyze data spanning years or more, considerably longer than would be possible using coherent methods alone.

Section~\ref{sec:background} of this paper reviews pertinent background information.
Section~\ref{sec:semic-supersky-param} derives the new semicoherent parameter-space metric, and validates it using numerical simulations.
Section~\ref{sec:comp-prev-work} compares the new metric to previous work~\cite{Pletsch.2010a,Brady.Creighton.2000a}, and Section~\ref{sec:summary} summarizes the conclusions of the paper.

\section{Background}\label{sec:background}

\begin{table}
\begin{tabular*}{\linewidth}{@{\extracolsep{\fill}}ll}
\hline\hline
Quantity & Symbol \\
\hline
Number of segments & N \\
Segment index & $\ell$ \\
\hline
Phase evolution parameters of signal & $\sigvec\lambda$ \\
Difference with respect to signal parameters & $\sig\diff$ \\
\hline
Time-span of coherently-analyzed data & $\coh T$ \\
Parameters of coherent template & $\cohsvec\lambda$ \\
Parameters of nearest coherent template & $\cohsvec\lambda(\cdots)$ \\
Coherent $\calF$-statistic & $2\cohs\calF$ \\
Coherent signal-to-noise ratio & $\cohs\SNR^2$ \\
Coherent noncentrality parameter & $\cohs\rho^2$ \\
Coherent $\calF$-statistic metric & $\cohsmat g\Fstat$ \\
Coherent $\calF$-statistic mismatch & $\cohs\mu\Fstat$ \\
Coherent phase metric & $\cohsmat g\phase$ \\
Coherent reduced supersky metric & $\cohsmat g$ \\
Coherent reduced supersky metric mismatch & $\cohs\mu$ \\
\hline
Time-span of semicoherently-analyzed data & $\semi T$ \\
Parameters of semicoherent template & $\semivec\lambda$ \\
Difference with respect to semicoherent template & $\semi\diff$ \\
Semicoherent $\calF$-statistic & $2\semi\calF$ \\
Semicoherent signal-to-noise ratio & $\semi\SNR^2$ \\
Semicoherent noncentrality parameter & $\semi\rho^2$ \\
Semicoherent $\calF$-statistic metric & $\semimat g\Fstat$ \\
Semicoherent $\calF$-statistic mismatch & $\semi\mu\Fstat$ \\
Semicoherent phase metric & $\semimat g\phase$ \\
Semicoherent reduced supersky metric & $\semimat g$ \\
Reduced supersky mismatch without interpolation & $\semi\mu\unoint$ \\
Reduced supersky mismatch with interpolation & $\semi\mu$ \\
Semicoherent global correlation metric & $\semimat g\gct$ \\
Global correlation mismatch without interpolation & $\semi\mu\gct\unoint$ \\
Global correlation mismatch with interpolation & $\semi\mu\gct$ \\
\hline\hline
\end{tabular*}
\caption{\label{tab:notation}
Summary of notation used in this paper.
The decoration of signal, coherent, and semicoherent quantities with $\sig\cdot$, $\coh\cdot$, and $\semi\cdot$ respectively follows the convention of~\cite{Prix.Shaltev.2012a}.
}
\end{table}

This section briefly reviews basics of gravitational-wave pulsar searches (Section~\ref{sec:coher-grav-wave}), and introduces the coherent parameter-space metric of~\PaperI\ (Section~\ref{sec:coher-supersky-param}).
It then gives a summary of the theory and notation of semicoherent searches (Section~\ref{sec:semic-grav-wave}); for a similar review see~\cite{Prix.Shaltev.2012a}.
The notation introduced in this and subsequent sections is summarized in Table~\ref{tab:notation}.

\subsection{Coherent gravitational-wave pulsar searches}\label{sec:coher-grav-wave}

The signal from a gravitational-wave pulsar observed in a detector at time $t$ is modeled~\cite{Jaranowski.etal.1998a} by the function $h(t, \vec\calA, \vec\lambda)$.
The four parameters $\vec\calA$ control the amplitude modulation of the signal, and the vector of parameters $\vec\lambda$ control its phase evolution.
For all-sky searches for isolated pulsars, $\vec\lambda$ includes the sky position of the pulsar, its initial frequency $f \equiv \ndot[0] f \equiv f(t_0)$ at some reference time $t_0$, and frequency derivatives (or \emph{spindowns}) $\ndot f \equiv d^sf / dt^s |_{t=t_0}$.

A coherent search using the \emph{$\calF$-statistic}~\cite{Jaranowski.etal.1998a,Cutler.Schutz.2005a} matched-filters data from a gravitational-wave detector against $h(t, \vec\calA, \vec\lambda)$, and analytically maximizes the log-likelihood ratio over the $\vec\calA$.
The resultant detection statistic $2\cohs\calF(\vec\lambda)$ is thus a function only of $\vec\lambda$.
(The subscript $\cdot\useg$ here is the \emph{segment index}, which will become relevant for semicoherent searches in Section~\ref{sec:semic-grav-wave}.)
To further maximize $2\cohs\calF(\vec\lambda)$ over the parameter space $\calP$ of possible $\vec\lambda \in \calP$, a finite set of parameters $\{\cohsvec\lambda\} \subset \calP$ -- the coherent template bank -- is generated, and $2\cohs\calF(\cohsvec\lambda)$ is computed for all $\cohsvec\lambda \in \{\cohsvec\lambda\}$.
The most promising pulsar candidate, assuming Gaussian detector noise~\cite[cf.][]{Keitel.etal.2014a}, will be the maximum of $2\cohs\calF(\cohsvec\lambda)$ over $\{\cohsvec\lambda\}$.

We adopt the following definition for the \emph{signal-to-noise ratio}:
\begin{equation}
\label{eq:SNR-def}
\SNR^2 \equiv \frac{ \E[2\calF\text{ near signal}] - \E[2\calF\text{ in noise}] }{ \sqrt{ \Var[2\calF\text{ in noise}] / 8 } } \,.
\end{equation}
where $\E[\cdot]$ denotes the expectation and $\Var[\cdot]$ the variance.
(The rationale for the normalization of the variance by $8$ can be seen in Eqs.~\eqref{eq:coh-SNR-def} and~\eqref{eq:semi-SNR-def} below.)
In noise, $2\cohs\calF(\cohsvec\lambda)$ is a central $\chi^2$ statistic with $4$ degrees of freedom; near a signal, it is a noncentral $\chi^2$ statistic with $4$ degrees of freedom and noncentrality parameter $\cohs\rho^2(\vec\calA, \sigvec\lambda; \cohsvec\lambda)$, where $\cohsvec\lambda$ denotes the phase parameters of the matched-filter template and $\sigvec\lambda$ the true phase parameters of the signal.
It follows that the signal-to-noise ratio of the coherent $\calF$-statistic is
\begin{equation}
\label{eq:coh-SNR-def}
\begin{split}
\cohs\SNR^2(\vec\calA, \sigvec\lambda; \cohsvec\lambda) &= \frac{ \big[ 4 + \cohs\rho^2(\vec\calA, \sigvec\lambda; \cohsvec\lambda) \big] - [ 4 ] }{ \sqrt{ [ 8 ] / 8 } } \\
&= \cohs\rho^2(\vec\calA, \sigvec\lambda; \cohsvec\lambda) \,.
\end{split}
\end{equation}
In the gravitational-wave pulsar literature, $\cohs\rho^2$ is often also referred to as the signal-to-noise ratio; however, as noted in~\cite{Leaci.Prix.2015a}, this is true \emph{only} for a fully-coherent search, as will be seen in Section~\ref{sec:semic-grav-wave}.

Since the true phase parameters of a pulsar, denoted $\sigvec\lambda$, will never exactly match one of the $\{\cohsvec\lambda\}$, some loss of signal-to-noise ratio is inevitable.
The \emph{mismatch} quantifies the loss in signal-to-noise ratio $\cohs\SNR^2(\vec\calA, \sigvec\lambda; \cohsvec\lambda)$ at a template $\cohsvec\lambda \ne \sigvec\lambda$, relative to that of a perfect match $\cohs\SNR^2(\vec\calA, \sigvec\lambda; \sigvec\lambda)$; it is defined to be
\begin{equation}
\label{eq:coh-Fstat-mismatch-SNR-def}
\cohs\mu\Fstat(\vec\calA, \sigvec\lambda; \cohsvec\lambda) \equiv 1 - \frac{ \cohs\SNR^2(\vec\calA, \sigvec\lambda; \cohsvec\lambda) }{ \cohs\SNR^2(\vec\calA, \sigvec\lambda; \sigvec\lambda) } \,.
\end{equation}
Substitution of Eq.~\eqref{eq:coh-SNR-def} gives the mismatch alternatively in terms of the noncentrality parameters:
\begin{equation}
\label{eq:coh-Fstat-mismatch-def}
\cohs\mu\Fstat(\vec\calA, \sigvec\lambda; \cohsvec\lambda) = 1 - \frac{ \cohs\rho^2(\vec\calA, \sigvec\lambda; \cohsvec\lambda) }{ \cohs\rho^2(\vec\calA, \sigvec\lambda; \sigvec\lambda) } \,.
\end{equation}
The definition of mismatch in terms of loss of signal-to-noise ratio is consistent with the data analysis of searches for signals from compact binary coalescence~\cite[e.g.][]{Owen.1996a,Owen.Sathyaprakash.1999a,Abbott.etal.2008e}; definition in terms of loss of noncentrality parameter is more common in gravitational-wave pulsar data analysis~\cite[e.g.][]{Prix.Shaltev.2012a}.
In this paper, by taking Eq.~\eqref{eq:coh-Fstat-mismatch-SNR-def} as the primary definition of mismatch, we aim to highlight the difference between signal-to-noise ratio and noncentrality parameter for semicoherent searches (see Section~\ref{sec:semic-grav-wave}).
Ultimately, it will be seen that the two definitions of mismatch given by Eqs.~\eqref{eq:coh-Fstat-mismatch-SNR-def} and~\eqref{eq:coh-Fstat-mismatch-def} are equivalent.

Efficient template bank construction seeks to minimize the average $\cohs\mu\Fstat$, over all possible $\sigvec\lambda$, at fixed number of templates $\cohs\calN \equiv |\{\cohsvec\lambda\}|$; this can be achieved using lattice template placement (see~\PaperII).
To aid in constructing the template bank, the parameter-space metric is introduced~\cite{Owen.1996a,Prix.2007a}.
A second-order Taylor expansion of Eq.~\eqref{eq:coh-Fstat-mismatch-def} with respect to small parameter differences $\sig\diff\cohsvec\lambda \equiv \cohsvec\lambda - \sigvec\lambda$ gives
\begin{equation}
\label{eq:coh-Fstat-mismatch-metric}
\cohs\mu\Fstat(\vec\calA, \sigvec\lambda; \cohsvec\lambda) \approx \sig\diff\cohsvec\lambda \cdot \cohsmat g\Fstat(\vec\calA, \sigvec\lambda) \cdot \sig\diff\cohsvec\lambda \,, \\
\end{equation}
where $\cdot$ denotes the vector/matrix inner product.
The parameter-space metric is represented by the matrix of components
\begin{equation}
\label{eq:coh-Fstat-metric-def}
\cohsmat g\Fstat(\vec\calA, \sigvec\lambda) \equiv \frac{-1}{2 \cohs\rho^2(\vec\calA, \sigvec\lambda; \sigvec\lambda)} \left. \frac{\partial \cohs\rho^2(\vec\calA, \sigvec\lambda; \vec\lambda)}{\partial \vec\lambda} \right|_{\vec\lambda = \sigvec\lambda} \,,
\end{equation}
and is a function of $\vec\calA$ and $\sigvec\lambda$.

Lattice template placement, however, requires Eq.~\eqref{eq:coh-Fstat-mismatch-metric} to be constant with respect to $\vec\calA$ and $\sigvec\lambda$.
A useful approximation which partially satisfies this requirement is
\begin{equation}
\label{eq:coh-Fstat-mismatch-phase-metric}
\cohs\mu\Fstat(\vec\calA, \sigvec\lambda; \cohsvec\lambda) \approx \sig\diff\cohsvec\lambda \cdot \cohsmat g\phase(\sigvec\lambda) \cdot \sig\diff\cohsvec\lambda \,; \\
\end{equation}
the \emph{phase metric} $\cohsmat g\phase(\sigvec\lambda)$ approximates $\cohsmat g\Fstat(\vec\calA, \sigvec\lambda)$, and is given by
\begin{equation}
\label{eq:coh-phase-metric-def}
\cohsmat g\phase(\sigvec\lambda) \equiv \big\langle \partial_{\vec\lambda}\phi \otimes \partial_{\vec\lambda}\phi \big\rangle_t - \big\langle \partial_{\vec\lambda}\phi \big\rangle_t \otimes \big\langle \partial_{\vec\lambda}\phi \big\rangle_t |_{\vec\lambda = \sigvec\lambda} \,,
\end{equation}
where $\partial_{\vec\lambda}$ is the vector of derivatives with respect to $\vec\lambda$, $\otimes$ denotes the vector outer product, and the $\langle\cdot\rangle_t$ operator time-averages over the segment time-span $\coh T$.
The pulsar phase $\phi(t,\vec\lambda)$ is given by (e.g.~\PaperI)
\begin{equation}
\label{eq:phase-def}
\frac{ \phi(t, \vec\lambda) }{2\pi} \approx \sum_{s=0}^{s\umax} \ndot f \frac{ (t-t_0)^{s+1} }{ (s+1)! }
+ \frac{ \vec r(t) \cdot \vec n }{c} f\umax \,,
\end{equation}
where $\vec r(t)$ is the detector position relative to the Solar System barycenter, $\vec n$ is the pulsar sky position, and $f\umax$ is a constant chosen conservatively to be the maximum of $f(t) \equiv d\phi(t, \vec\lambda) / dt$ over $\coh T$.
It follows from Eq.~\eqref{eq:coh-phase-metric-def} that $\cohsmat g\phase(\sigvec\lambda)$ will also be constant with respect to any phase parameter $\lambda^i$ in which $\phi(t,\vec\lambda)$ is linear.
As seen from Eq.~\eqref{eq:phase-def}, such parameters include the frequency $f$ and spindowns $\ndot f$, but not the sky position
$\vec n = (\cos\alpha \cos\delta, \sin\alpha \cos\delta, \sin\delta)$ when parameterized in terms of right ascension $\alpha$ and declination $\delta$.

\subsection{Coherent supersky parameter-space metric}\label{sec:coher-supersky-param}

\PaperI\ presented a close approximation to $\cohsmat g\phase(\sigvec\lambda)$ which is constant with respect to all phase parameters $\vec\lambda$, including sky position: the \emph{reduced supersky metric} $\cohsmat g$ (denoted $\mathbf{g}_{\mathrm{rss}}$ in~\PaperI).
It is derived from $\cohsmat g\phase(\sigvec\lambda)$ by the following procedure, simplified from~\PaperI:
\begin{enumerate}[(i)]

\item \label{item:coh-metric-supersky}
$\cohsmat g\phase(\sigvec\lambda)$ is computed using the three components of $\vec n$ in equatorial coordinates as sky coordinates; this metric is the \emph{unconstrained supersky metric} $\cohsmat g\unc$ (denoted $\mathbf{g}_{\mathrm{ss}}$ in~\PaperI).
Since $\phi(t,\vec\lambda)$ is linear in $\vec n$, this yields a metric which is constant with respect to all phase parameters; however, the two-dimensional sky is now embedded in three dimensions, which is undesirable for template placement.

\item \label{item:coh-metric-lin-transf}
A linear coordinate transformation is performed on $\cohsmat g\unc$ which exploits the near-degeneracy between the signal phase's dependence on the Earth's orbital motion, and its dependence on the pulsar's frequency and spindowns.
The transformation also removes from $\cohsmat g\unc$ any correlation between sky and frequency/spindown coordinates.

\item \label{item:coh-metric-eigen-vec}
The eigenvectors and eigenvalues of the $3 \times 3$ block of the transformed $\cohsmat g\unc$ pertaining to the sky parameters $\vec n$ are computed.
The eigenvector corresponding to the smallest eigenvalue is identified; by definition, offsets $\sig\diff\cohsvec\lambda$ parallel to this eigenvector will change the mismatch $\sig\diff\cohsvec\lambda \cdot \cohsmat g\unc \cdot \sig\diff\cohsvec\lambda$ by the smallest amount possible.

\item \label{item:coh-metric-reduced}
The transformed metric $\cohsmat g\unc$ is projected onto the hyperplane in the coordinate space of $\vec n$ perpendicular to the eigenvector identified in step~\eqref{item:coh-metric-eigen-vec}.
This reduces the number of sky parameters back to two, and yields the metric $\cohsmat g$.
The phase parameters associated with $\cohsmat g$ are the sky parameters $(\cohs{n\ua}, \cohs{n\ub})$, and the frequency/spindowns $\ndot{\cohs\nu} \equiv \ndot f + \cohsvec{\Delta^s} \cdot \vec n$, where the $\cohsvec{\Delta^s}$ are vectors computed during step~\eqref{item:coh-metric-lin-transf}.

\label{item:coh-metric-end}
\end{enumerate}
We denote the linear transformation of the components of $\cohsmat g\unc$ described in step~\eqref{item:coh-metric-lin-transf} by the operator $\calT$, and the projection described in step~\eqref{item:coh-metric-reduced} by the operator $\calJ$, i.e.
\begin{equation}
\cohsmat g \equiv \calJ \calT \big[ \cohsmat g\unc \big] \,.
\end{equation}

\PaperI\ demonstrated, using numerical simulations, that $\cohsmat g$ is a good approximation to $\cohsmat g\phase(\sigvec\lambda)$ for segment time-spans $\coh T \gtrsim 1$~day, maximum mismatches $\cohs\mu\phase \lesssim 0.6$, and spindowns up to $\ndot[2] f$.

\subsection{Semicoherent gravitational-wave pulsar searches}\label{sec:semic-grav-wave}

\begin{figure}
\includegraphics[width=\linewidth]{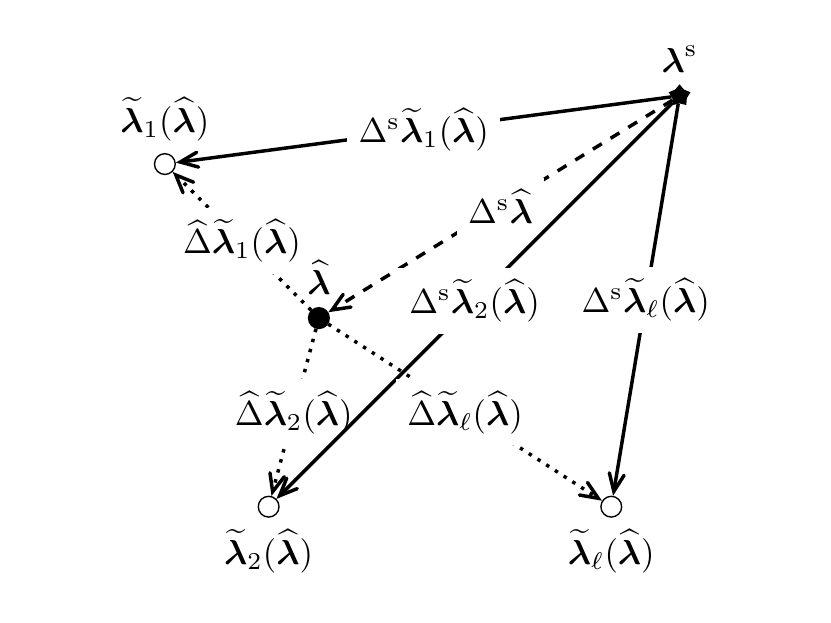}
\caption{\label{fig:mismatch_diagram}
Relationship between the signal parameters $\sigvec\lambda$, its nearest semicoherent template $\semivec\lambda$, and its $N$ nearest coherent templates $\cohvec\lambda_1, \cohvec\lambda_2, \dots, \cohsvec\lambda, \dots$, following Fig.~1 of~\cite{Prix.Shaltev.2012a}.
}
\end{figure}

This section describes a common semicoherent search method for gravitational-wave pulsars, referred to as ``interpolating StackSlide'' in~\cite{Prix.Shaltev.2012a}.

The gravitational-wave detector data are partitioned in time into $N$ segments, labeled with the index $\ell = 1, 2, \dots, N$.
For simplicity we assume that the time-spans of all segments are identical and equal to $\coh T$; the mid-time of each segment is denoted $t_\ell$.
In each segment $\ell$, the coherent $\calF$-statistic $2\cohs\calF(\cohsvec\lambda)$ is calculated on a coherent template bank $\{\cohsvec\lambda\} \in \calP$ which is constructed using the coherent metric $\cohsmat g$ at a common reference time $t_0$.
As the metrics $\cohsmat g$ are generally different for each segment, so too will be the template banks $\{\cohsvec\lambda\}$.

The semicoherent $\calF$-statistic is computed on a distinct, semicoherent template bank $\{\semivec\lambda\} \in \calP$.
For each semicoherent template $\semivec\lambda$, the coherent template in each segment which is nearest to $\semivec\lambda$, as measured by the respective coherent metric $\cohsmat g$, is determined; these $N$ templates are denoted $\cohsvec\lambda(\semivec\lambda)$.
The metric mismatch between $\semivec\lambda$ and $\cohsvec\lambda(\semivec\lambda)$ is denoted
\begin{equation}
\label{eq:coh-near-mismatch-def}
\cohs\mu\big( \semivec\lambda; \cohsvec\lambda(\semivec\lambda) \big) \equiv \semi\diff\cohsvec\lambda(\semivec\lambda) \cdot \cohsmat g \cdot \semi\diff\cohsvec\lambda(\semivec\lambda) \,,
\end{equation}
where $\semi\diff\cohsvec\lambda(\semivec\lambda) \equiv \cohsvec\lambda(\semivec\lambda) - \semivec\lambda$; see Fig.~\ref{fig:mismatch_diagram}.
Finally, the $N$ coherent $\calF$-statistics computed at the $\cohsvec\lambda(\semivec\lambda)$, denoted $2\cohs\calF\big(\cohsvec\lambda(\semivec\lambda)\big)$, are retrieved and summed to give
\begin{equation}
\label{eq:semi-Fstat-def}
2\semi\calF\big(\cohsvec\lambda(\semivec\lambda)\big) \equiv \sums 2\cohs\calF\big(\cohsvec\lambda(\semivec\lambda)\big) \,.
\end{equation}

In noise, $2\semi\calF\big(\cohsvec\lambda(\semivec\lambda)\big)$ is a central $\chi^2$ statistic with $4N$ degrees of freedom; near a signal, it is a noncentral $\chi^2$ statistic with $4N$ degrees of freedom and noncentrality parameter
\begin{equation}
\label{eq:semi-rhosqr-def}
\semi\rho^2\big( \sigvec\lambda; \cohsvec\lambda(\semivec\lambda) \big) \equiv \sums \cohs\rho^2\big( \sigvec\lambda; \cohsvec\lambda(\semivec\lambda) \big) \,,
\end{equation}
where $\cohs\rho^2\big( \sigvec\lambda; \cohsvec\lambda(\semivec\lambda) \big)$ are the noncentrality parameters of the $N$ coherent $\calF$-statistics.
(Henceforth we suppress the dependencies on the $\vec\calA$.)
By Eq.~\eqref{eq:SNR-def}, the signal-to-noise ratio of the semicoherent $\calF$-statistic is therefore
\begin{equation}
\label{eq:semi-SNR-def}
\begin{split}
\semi\SNR^2\big( \sigvec\lambda; \cohsvec\lambda(\semivec\lambda) \big) &= \frac{ \big[ 4N + \semi\rho^2\big( \sigvec\lambda; \cohsvec\lambda(\semivec\lambda) \big) \big] - [ 4N ] }{ \sqrt{ [ 8N ] / 8 } } \\
&= \frac{1}{\sqrt{N}} \semi\rho^2\big( \sigvec\lambda; \cohsvec\lambda(\semivec\lambda) \big) \,.
\end{split}
\end{equation}
Note that Eq.~\eqref{eq:semi-SNR-def} is a factor of $\sqrt{N}$ less than the signal-to-noise ratio of a \emph{fully-coherent} search using the same amount of data $N \coh T$.

The definition of the semicoherent mismatch follows from Eq.~\eqref{eq:coh-Fstat-mismatch-SNR-def}:
\begin{align}
\label{eq:semi-Fstat-mismatch-SNR-def}
\semi\mu\Fstat\big( \sigvec\lambda; \cohsvec\lambda(\semivec\lambda) \big) &\equiv 1 - \frac{ \semi\SNR^2\big( \sigvec\lambda; \cohsvec\lambda(\semivec\lambda) \big) }{ \semi\SNR^2(\sigvec\lambda; \sigvec\lambda) } \,,
\end{align}
where $\semi\SNR^2(\sigvec\lambda; \sigvec\lambda)$ is the signal-to-noise ratio at perfect match in both coherent and semicoherent templates.
Substituting Eq.~\eqref{eq:semi-SNR-def} demonstrates that the semicoherent mismatch can equivalently be defined as the loss in noncentrality parameter, c.f.\ Eq.~(18) of~\cite{Prix.Shaltev.2012a}:
\begin{equation}
\label{eq:semi-Fstat-mismatch-rhosqr-equiv}
\semi\mu\Fstat\big( \sigvec\lambda; \cohsvec\lambda(\semivec\lambda) \big) = 1 - \frac{ \semi\rho^2\big( \sigvec\lambda; \cohsvec\lambda(\semivec\lambda) \big) }{ \semi\rho^2(\sigvec\lambda; \sigvec\lambda) } \,.
\end{equation}
where $\semi\rho^2(\sigvec\lambda; \sigvec\lambda)$ is the noncentrality parameter at perfect match in both coherent and semicoherent templates.
It is assumed that the coherent noncentrality parameter at perfect match is constant over all segments, which implies
\begin{equation}
\label{eq:coh-rhosqr-match}
\semi\rho^2(\sigvec\lambda; \sigvec\lambda) \equiv N \cohs\rho^2( \sigvec\lambda; \sigvec\lambda ) \,.
\end{equation}
Substituting Eqs.~\eqref{eq:semi-rhosqr-def}, and~\eqref{eq:coh-rhosqr-match} into Eq.~\eqref{eq:semi-Fstat-mismatch-rhosqr-equiv} yields
\begin{align}
\semi\mu\Fstat\big( \sigvec\lambda; \cohsvec\lambda(\semivec\lambda) \big) &= 1 - \frac{1}{N} \sums \frac{ \cohs\rho^2\big( \sigvec\lambda; \cohsvec\lambda(\semivec\lambda) \big) }{ \cohs\rho^2( \sigvec\lambda; \sigvec\lambda ) } \\
\label{eq:semi-Fstat-mismatch}
&= \frac{1}{N} \sums \cohs\mu\Fstat\big( \sigvec\lambda; \cohsvec\lambda(\semivec\lambda) \big) \,,
\end{align}
where $\cohs\mu\Fstat\big( \sigvec\lambda; \cohsvec\lambda(\semivec\lambda) \big)$ is the coherent mismatch in the $\ell$th segment between the nearest template $\cohsvec\lambda(\semivec\lambda)$ and the signal $\sigvec\lambda$.
Further substitution of Eq.~\eqref{eq:coh-Fstat-mismatch-phase-metric} results in
\begin{equation}
\label{eq:semi-Fstat-mismatch-coh-metric}
\semi\mu\Fstat\big( \sigvec\lambda; \cohsvec\lambda(\semivec\lambda) \big) \approx \frac{1}{N} \sums \sig\diff\cohsvec\lambda(\semivec\lambda) \cdot \cohsmat g\phase(\sigvec\lambda) \cdot \sig\diff\cohsvec\lambda(\semivec\lambda) \,,
\end{equation}
where (see Fig.~\ref{fig:mismatch_diagram})
\begin{align}
\sig\diff\cohsvec\lambda(\semivec\lambda) &\equiv \cohsvec\lambda(\semivec\lambda) - \sigvec\lambda \\
&= \big( \cohsvec\lambda(\semivec\lambda) - \semivec\lambda \big) + \big( \semivec\lambda - \sigvec\lambda \big) \\
\label{eq:sig-diff-cohlambda-semilambda-expand}
&= \semi\diff\cohsvec\lambda(\semivec\lambda) + \sig\diff\semivec\lambda \,.
\end{align}
Finally, substituting Eq.~\eqref{eq:sig-diff-cohlambda-semilambda-expand} into Eq.~\eqref{eq:semi-Fstat-mismatch-coh-metric} gives
\begin{equation}
\label{eq:semi-Fstat-mismatch-expand}
\begin{split}
\semi\mu\Fstat\big( \sigvec\lambda; \cohsvec\lambda(\semivec\lambda) \big) &\approx \frac{1}{N} \sums \sig\diff\semivec\lambda \cdot \cohsmat g\phase(\sigvec\lambda) \cdot \sig\diff\semivec\lambda \\
&+ \frac{1}{N} \sums \semi\diff\cohsvec\lambda(\semivec\lambda) \cdot \cohsmat g\phase(\sigvec\lambda) \cdot \semi\diff\cohsvec\lambda(\semivec\lambda) \\
&+ \frac{2}{N} \sums \semi\diff\cohsvec\lambda(\semivec\lambda) \cdot \cohsmat g\phase(\sigvec\lambda) \cdot \sig\diff\semivec\lambda \,.
\end{split}
\end{equation}

The first term on the right-hand side of Eq.~\eqref{eq:semi-Fstat-mismatch-expand} is the mismatch between the signal $\sigvec\lambda$ and the semicoherent template $\semivec\lambda$ in the ideal case of \emph{no interpolation}, i.e.\ supposing $\semi\diff\cohsvec\lambda(\semivec\lambda) = 0$ for all $\ell$.
It may be written as
\begin{align}
\label{eq:semi-Fstat-mismatch-no-interp}
\semi\mu\Fstat\unoint(\sigvec\lambda; \semivec\lambda) &\equiv \frac{1}{N} \sums \cohs\mu\Fstat( \sigvec\lambda; \semivec\lambda) \\
&\approx \frac{1}{N} \sums \sig\diff\semivec\lambda \cdot \cohsmat g\phase(\sigvec\lambda) \cdot \sig\diff\semivec\lambda \\
\label{eq:semi-Fstat-mismatch-no-interp-factor}
&\approx \sig\diff\semivec\lambda \cdot \semimat g\phase(\sigvec\lambda) \cdot \sig\diff\semivec\lambda \,,
\end{align}
where
\begin{equation}
\label{eq:semi-metric-def}
\semimat g\phase(\sigvec\lambda) \equiv \frac{1}{N} \sums \cohsmat g\phase(\sigvec\lambda)
\end{equation}
is the semicoherent parameter-space metric; cf.\ Eq.~(2.9) of~\cite{Brady.Creighton.2000a}.
The second term on the right-hand side of Eq.~\eqref{eq:semi-Fstat-mismatch-expand} is the mismatch \emph{purely} from interpolation,
\begin{align}
\label{eq:semi-Fstat-mismatch-pure-interp}
\semi\mu\Fstat\upureint\big( \semivec\lambda; \cohsvec\lambda(\semivec\lambda) \big) &\equiv \frac{1}{N} \sums \cohs\mu\Fstat\big( \semivec\lambda; \cohsvec\lambda(\semivec\lambda) \big) \\
&\approx \frac{1}{N} \sums \semi\diff\cohsvec\lambda(\semivec\lambda) \cdot \cohsmat g\phase(\sigvec\lambda) \cdot \semi\diff\cohsvec\lambda(\semivec\lambda) \,.
\end{align}

The average of the third term on the right-hand side of Eq.~\eqref{eq:semi-Fstat-mismatch-expand} will tend to zero when $N$ is large, and when $\semi\mu\Fstat\unoint$ is averaged over many different signals $\sigvec\lambda$~\cite{Prix.Shaltev.2012a}.
This is because the term is linear in the offsets $\semi\diff\cohsvec\lambda(\semivec\lambda)$ and $\sig\diff\semivec\lambda$, which are presumed to be equally distributed about zero.
Likewise, under the same conditions the average of Eq.~\eqref{eq:semi-Fstat-mismatch-pure-interp} will tend to $\langle \cohs\mu\Fstat \rangle$, the average coherent mismatch.
The average of Eq.~\eqref{eq:semi-Fstat-mismatch-SNR-def} will therefore tend to
\begin{equation}
\label{eq:semi-Fstat-mismatch-avg}
\langle \semi\mu\Fstat \rangle \rightarrow \langle \semi\mu\Fstat\unoint \rangle + \langle \cohs\mu\Fstat \rangle \,.
\end{equation}
It follows from Eqs.~\eqref{eq:semi-Fstat-mismatch-avg} and~\eqref{eq:semi-Fstat-mismatch-no-interp} that the average semicoherent mismatch $\langle \semi\mu\Fstat \rangle$ is minimized if $\semivec\lambda \in \{\semivec\lambda\}$ is chosen to be nearest to the signal $\sigvec\lambda$, as measured by the semicoherent metric $\semimat g\phase$, thus minimizing $\langle \semi\mu\Fstat\unoint \rangle$.

Therefore, the semicoherent template bank $\{\semivec\lambda\}$ should be constructed similarly to the coherent template banks $\{\cohsvec\lambda\}$, using $\semimat g\phase$ to minimize $\semi\mu\Fstat\unoint$, and hence $\langle \semi\mu\Fstat \rangle$, at fixed number of semicoherent templates $\semi\calN \equiv |\{\semivec\lambda\}|$.
This may be achieved using lattice template placement (see~\PaperII), but only if $\semimat g\phase$ is constant with respect to $\vec\calA$ and $\sigvec\lambda$.
The construction of such a metric, starting from the semicoherent phase metric $\semimat g\phase(\sigvec\lambda) \equiv \sums \cohsmat g\phase(\sigvec\lambda) / N$, is the subject of the next section.

\section{Semicoherent supersky parameter-space metric}\label{sec:semic-supersky-param}

This section introduces a close approximation to $\semimat g\phase(\sigvec\lambda)$ which is constant with respect to all phase parameters $\lambda$: the semicoherent reduced supersky metric $\semimat g$.
Its derivation, presented in Section~\ref{sec:derivation}, closely follows that of the coherent metric $\cohsmat g$, which is described in Section~\ref{sec:coher-supersky-param} and~\PaperI.
The new metric is validated using numerical simulations presented in Section~\ref{sec:valid-using-numer}.

\subsection{Derivation}\label{sec:derivation}

\begin{figure*}
\centering
\subfloat[]{\includegraphics[width=0.2\linewidth]{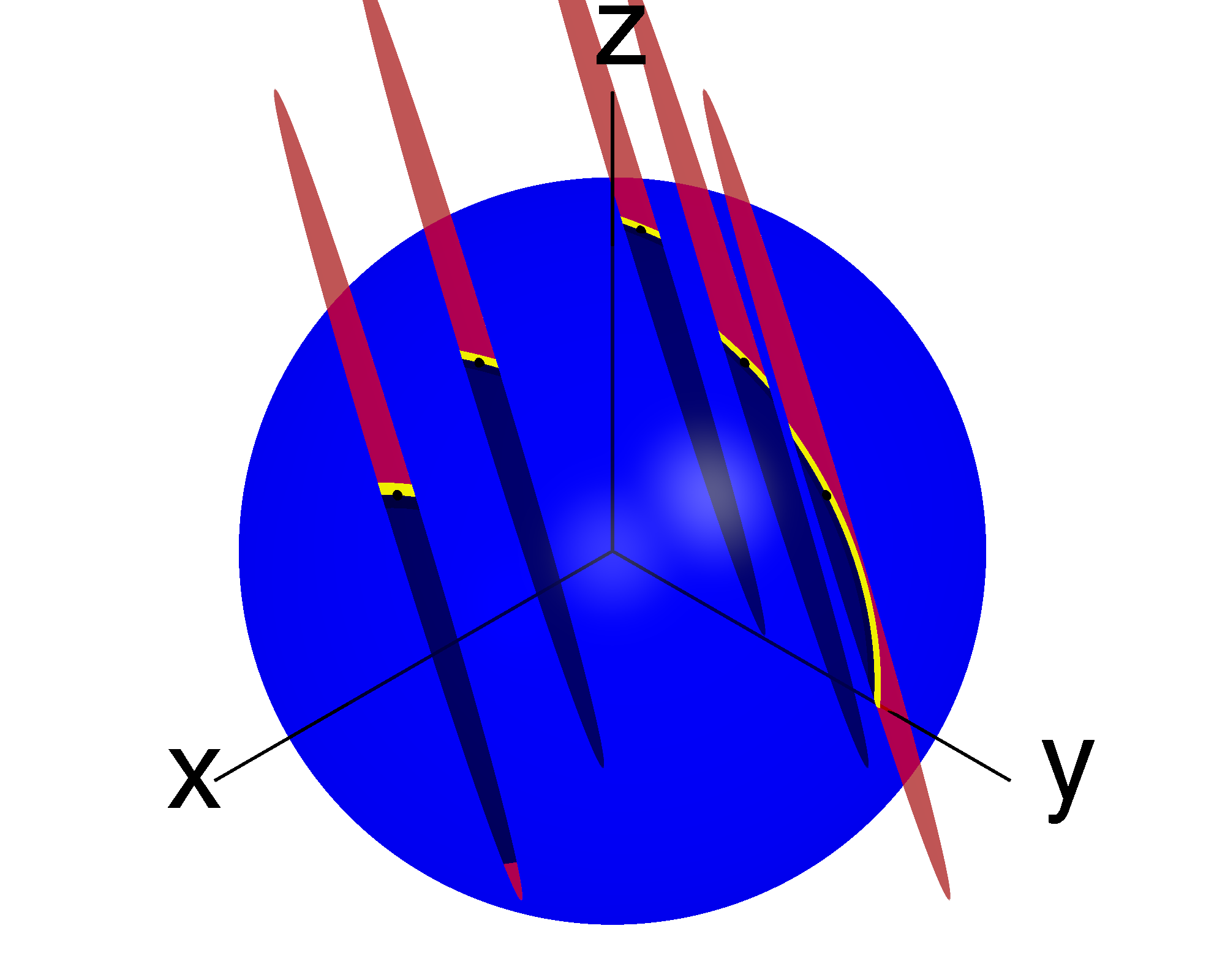}\label{fig:seg1_ssky_ellipses}}
\subfloat[]{\includegraphics[width=0.2\linewidth]{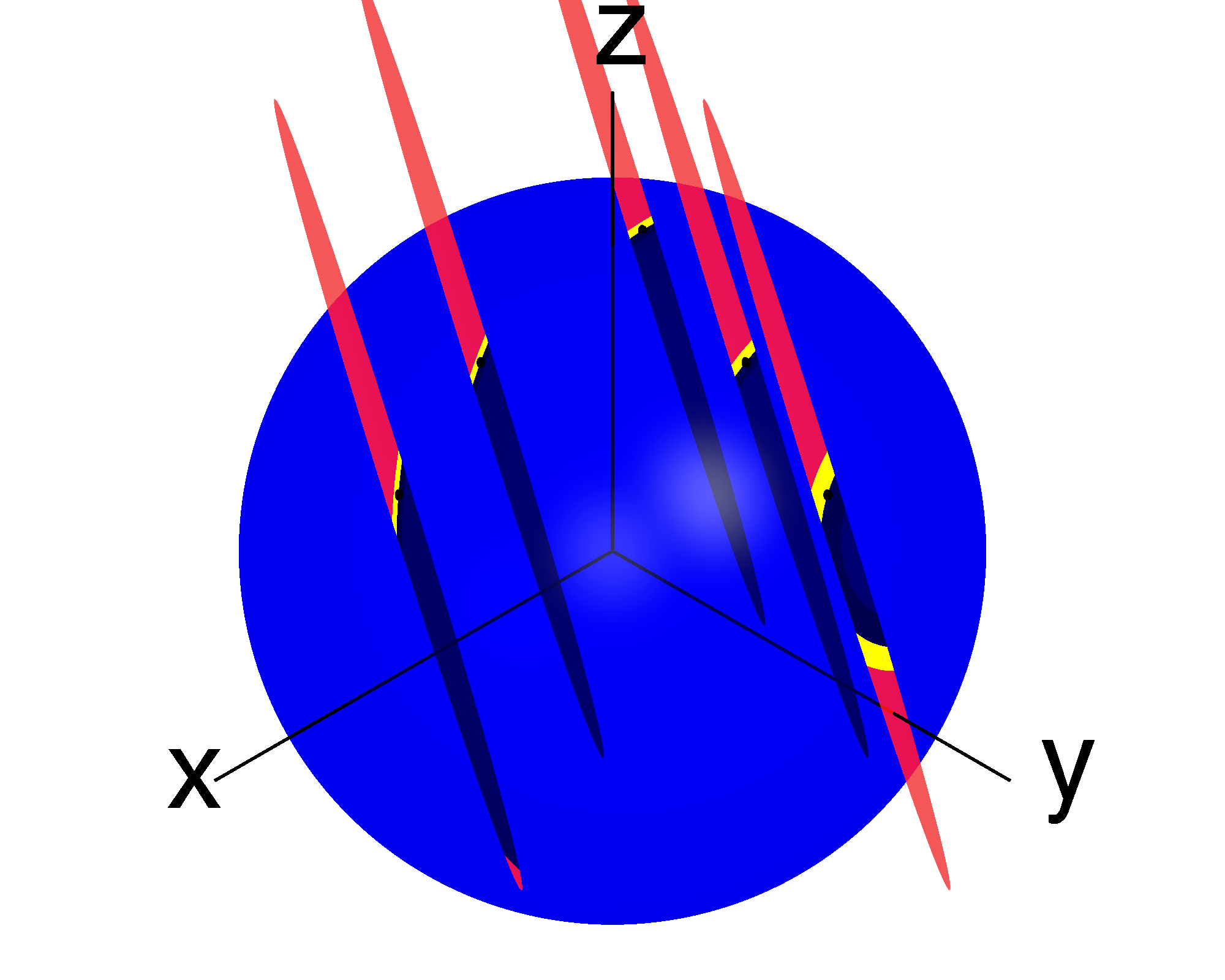}\label{fig:seg2_ssky_ellipses}}
\subfloat[]{\includegraphics[width=0.2\linewidth]{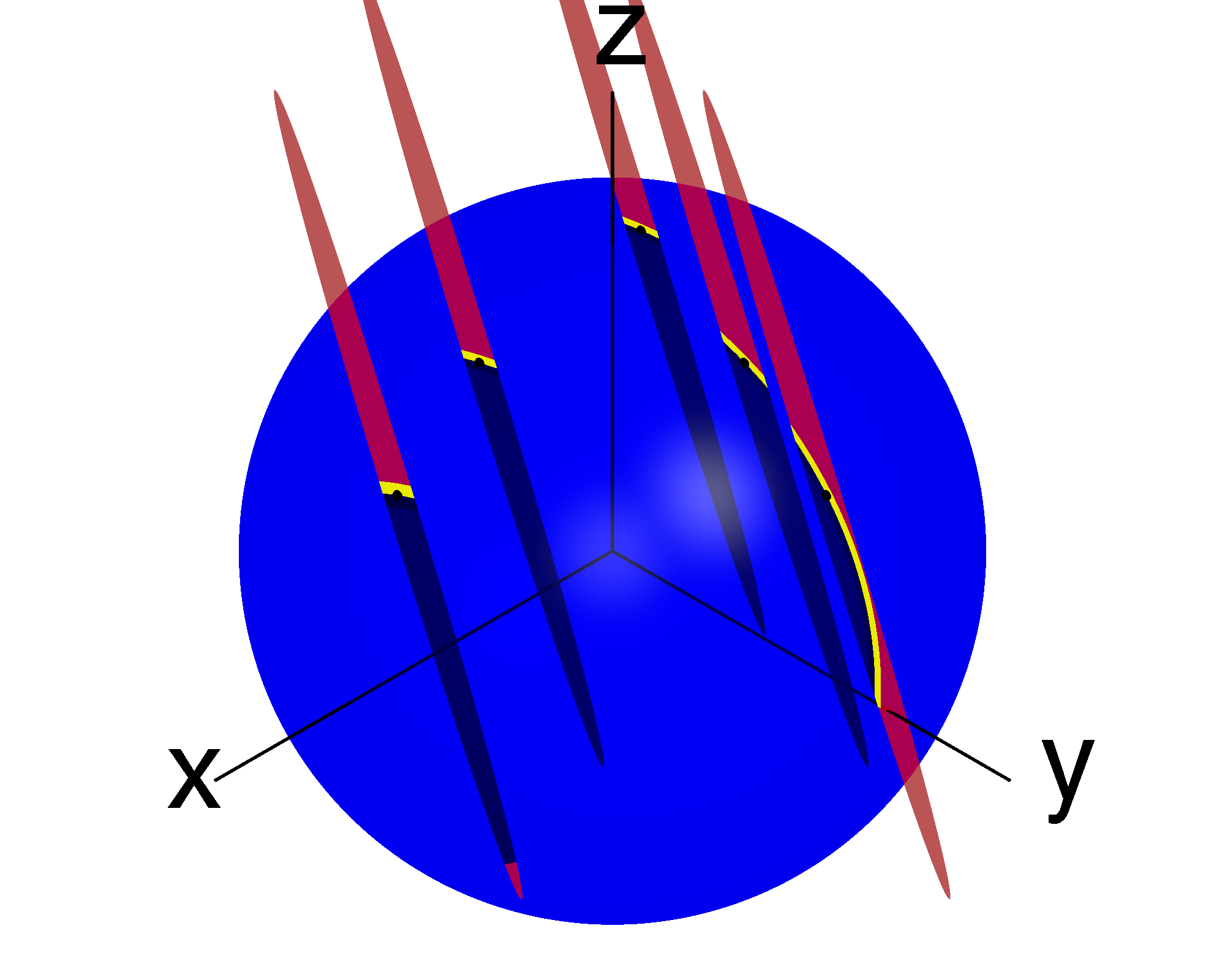}\label{fig:seg3_ssky_ellipses}}
\subfloat[]{\includegraphics[width=0.2\linewidth]{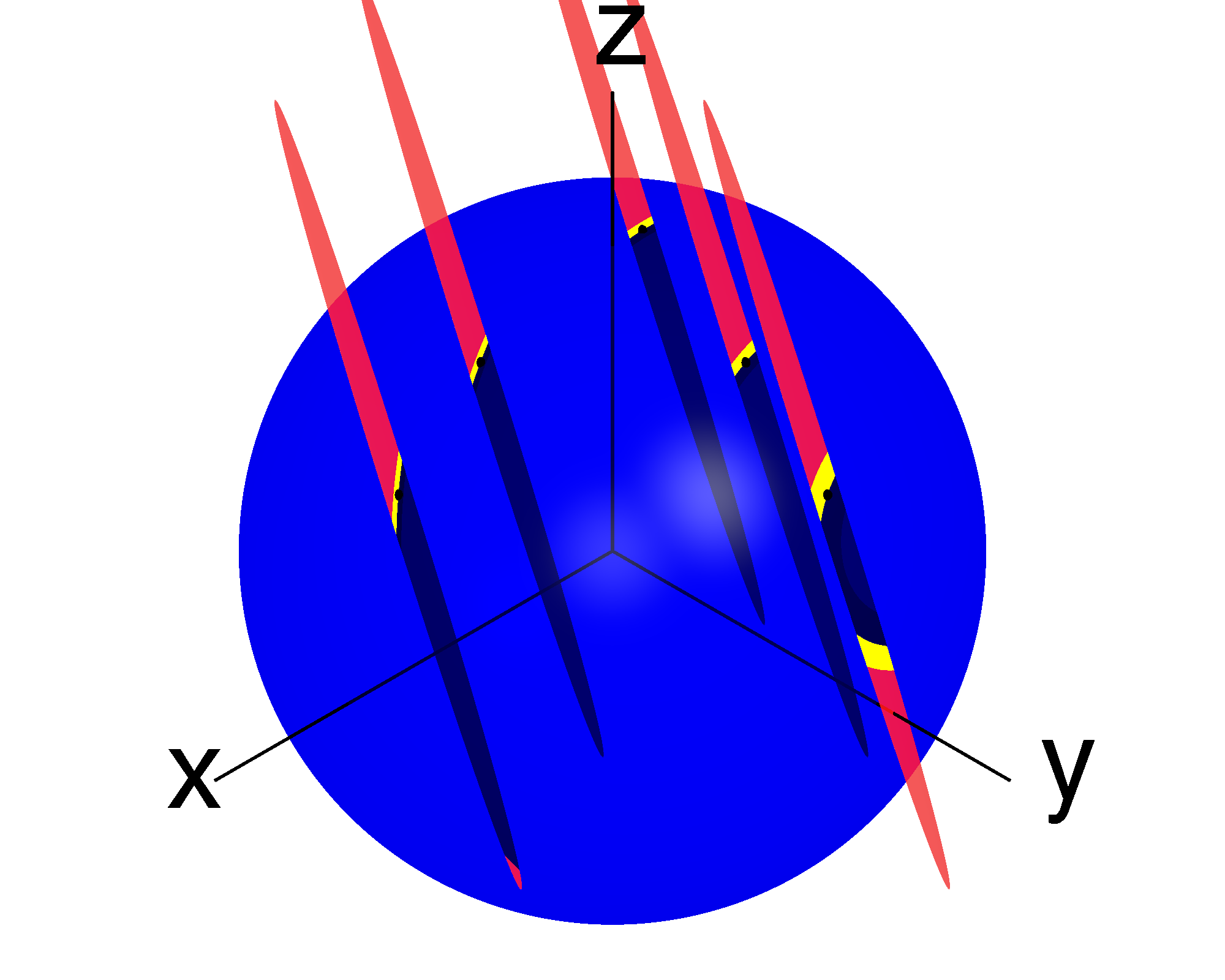}\label{fig:seg4_ssky_ellipses}}
\subfloat[]{\includegraphics[width=0.2\linewidth]{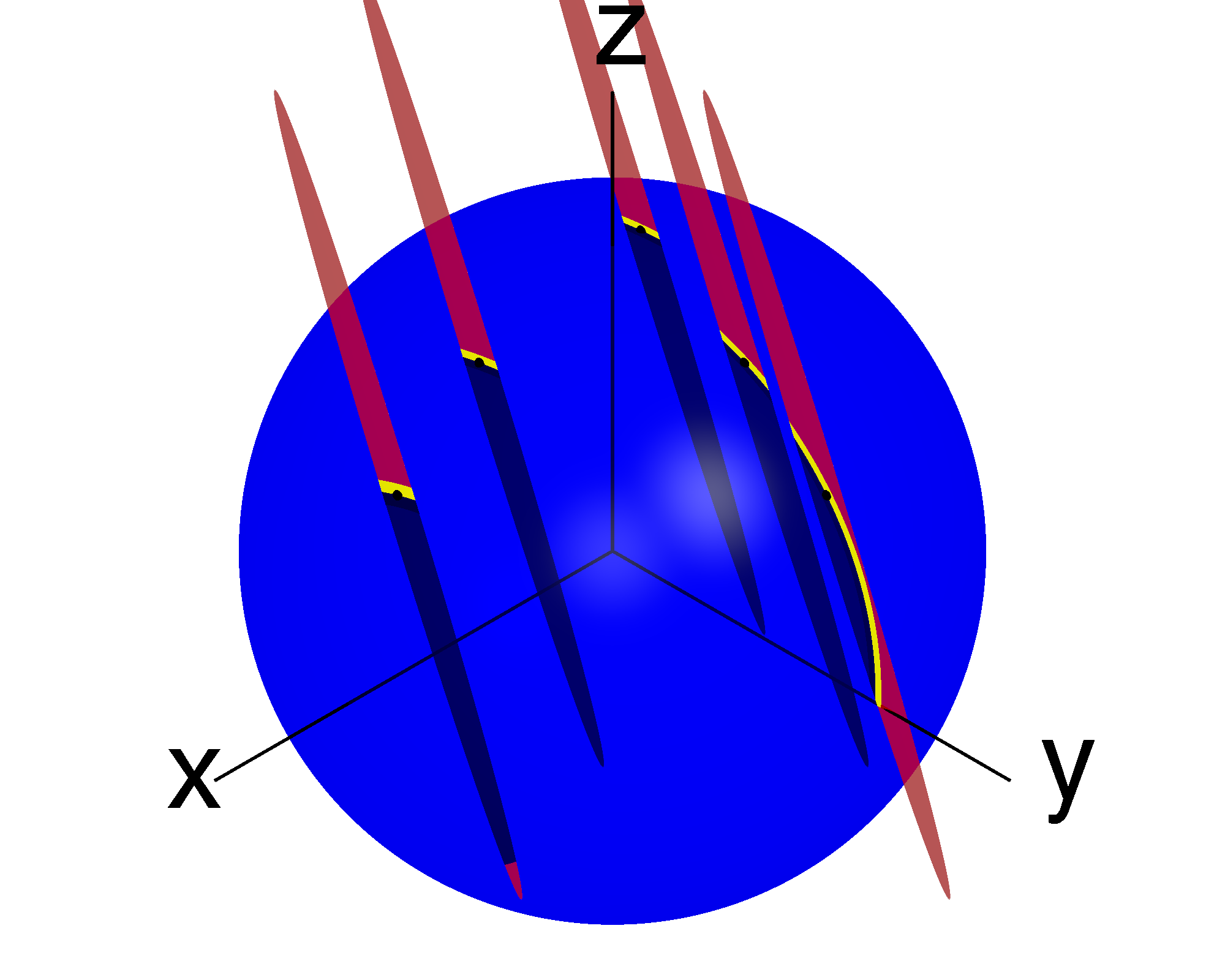}\label{fig:seg5_ssky_ellipses}}\\
\subfloat[]{\includegraphics[width=0.5\linewidth]{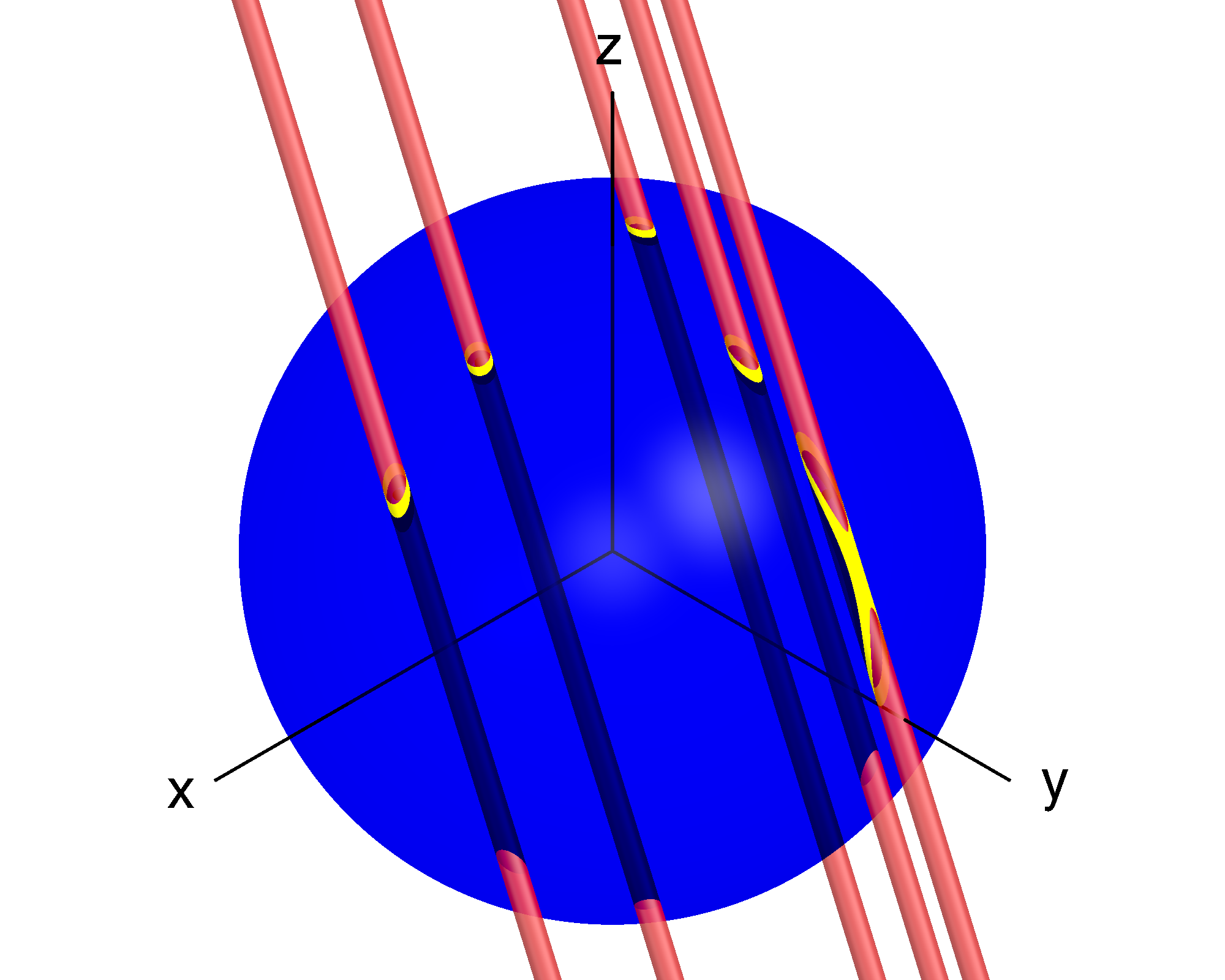}\label{fig:avg_ssky_ellipses}}
\subfloat[]{\includegraphics[width=0.5\linewidth]{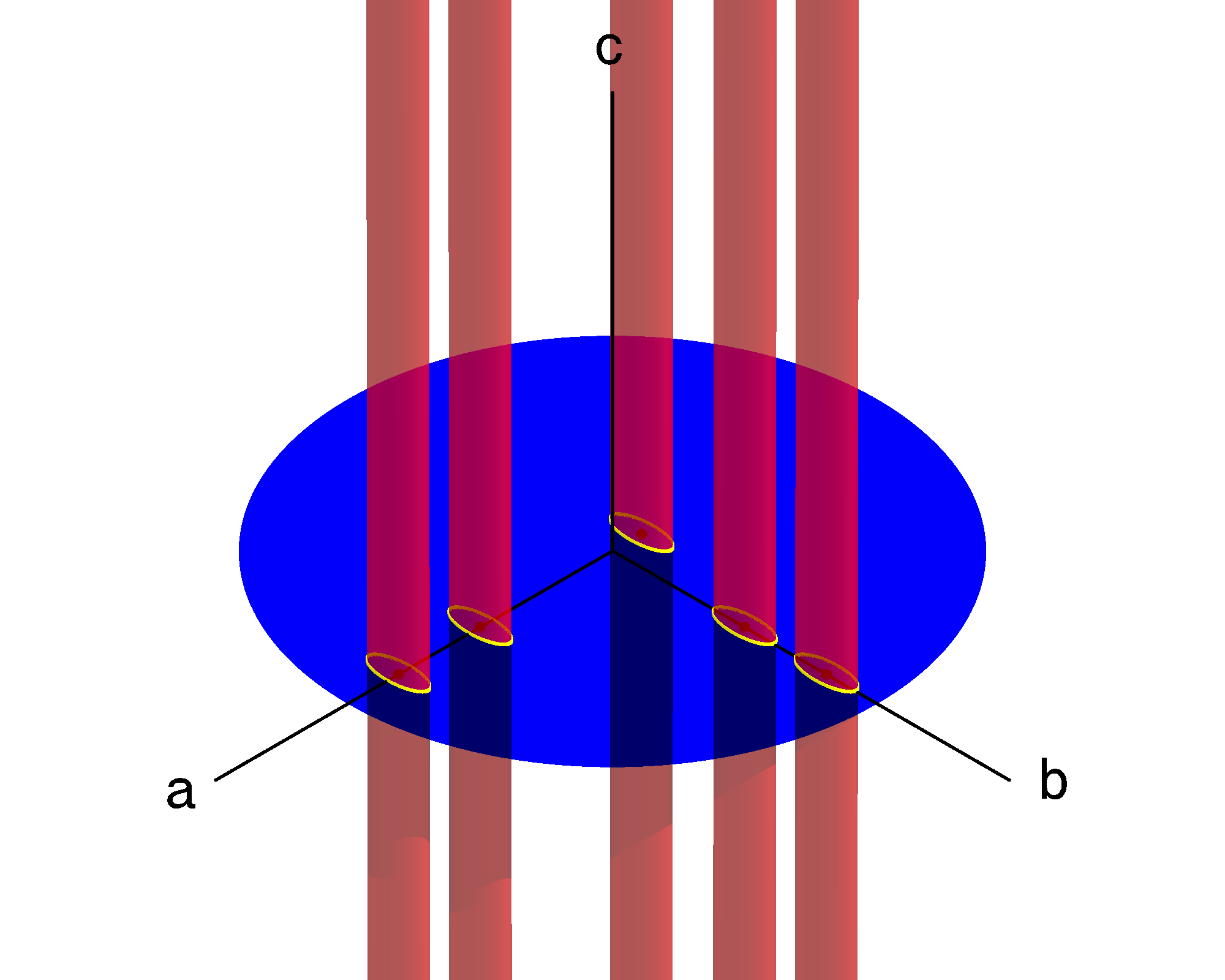}\label{fig:avg_aligned_ssky_ellipses}}
\caption{\label{fig:ssky_ellipses}
Metric ellipsoids of the coherent and semicoherent supersky metrics, at 5 example sky points.
\protect\subref{fig:seg1_ssky_ellipses}--\protect\subref{fig:seg5_ssky_ellipses}:
Metric ellipsoids (red) of 5 coherent unconstrained supersky metrics $\cohsmat g\unc$, in equatorial coordinates $(x,y,z)$, for $\coh\mu\umax = 30$, $\coh T = 4$~days, $f\umax = 1$~kHz, $t_0 = \text{UTC 2015-01-01 00:00:00}$, and $t_\ell - t_0 = -180, -90, 0, +90, +180$~days respectively.
Their intersection with the sky sphere $|\vec n| = 1$ (blue) reproduces the coherent physical sky metric (yellow).
\protect\subref{fig:avg_ssky_ellipses}: Metric ellipsoids (red) of the semicoherent unconstrained supersky metric $\semimat g\unc = \sums[5] \cohsmat g\unc / 5$, for $\semi\mu\umax = 1500$.
Its intersection with the sky sphere (blue) reproduces the semicoherent physical sky metric (yellow).
\protect\subref{fig:avg_aligned_ssky_ellipses}: Metric ellipsoids (red) of the transformed metric $\cal T \semimat g\unc$, in coordinates $(a,b,c)$ given by its sky eigenvectors, for $\semi\mu\umax = 1500$.
Its projection onto the disk $c = 0$ (blue) generates the semicoherent supersky metric $\semimat g = \calJ \calT \semimat g\unc$ in $(\semi n\ua, \semi n\ub)$ coordinates (yellow).
}
\end{figure*}

First, $\cohsmat g\phase(\sigvec\lambda)$ are computed for all segments, at a common reference time $t_0$, using the components of $\vec n$ as sky coordinates; these metrics are the coherent unconstrained supersky metrics $\cohsmat g\unc$.
Second, the $\cohsmat g\unc$ are averaged to give $\semimat g\unc \equiv \sums \cohsmat g\unc / N$, the semicoherent unconstrained supersky metric.
Finally, the procedure described in Section~\ref{sec:coher-supersky-param} and~\PaperI, denoted by the operators $\calJ \calT$, is applied to $\semimat g\unc$ to give the semicoherent supersky metric
\begin{equation}
\semimat g \equiv \frac{\calJ \calT}{N} \left[ \sums \cohsmat g\unc \right] \,.
\end{equation}
The phase parameters associated with $\semimat g$ are the sky parameters $(\semi n\ua, \semi n\ub)$, and the frequency/spindowns $\ndot{\semi\nu} \equiv \ndot f + \semivec{\Delta^s} \cdot \vec n$, where the $\semivec{\Delta^s}$ are vectors computed when applying $\calT$.

Figure~\ref{fig:ssky_ellipses} illustrates this process using an example setup of $N = 5$~segments, each spanning $\coh T = 4$~days, with segment mid-times $t_\ell$ distributed evenly over a total time-span $\semi T = 364$~days.
The effect of the changing segment mid-time on the metrics $\cohsmat g\unc$ is to rotate their \emph{metric ellipsoids} (i.e.\ the volumes that satisfy $\sig\Delta\cohsvec\lambda \cdot \cohsmat g\unc \cdot \sig\Delta\cohsvec\lambda \le \coh\mu\umax$ for some $\coh\mu\umax$) around their longest axis, as may be discerned from Figs.~\ref{fig:seg1_ssky_ellipses}--\ref{fig:seg5_ssky_ellipses}.
The metric ellipsoid of the semicoherent metric $\semimat g\unc$, in this case, retains the same orientation of the longest axis as the coherent metrics, and still results in a non-constant physical metric (Fig.~\ref{fig:avg_ssky_ellipses}).
Applying $\calT$ to $\semimat g\unc$, however, reorients the metric ellipsoids along the $c$ axis (Fig.~\ref{fig:avg_aligned_ssky_ellipses}), corresponding to the eigenvector of the $3 \times 3$ sky block of $\semimat g\unc$ with the smallest eigenvalue.
Applying $\calJ$ then projects $\calT \semimat g\unc$ onto the disk $c = 0$, giving the metric $\semimat g$.

Note that, contrary to Eq.~\eqref{eq:semi-metric-def}, $\semimat g$ is \emph{not} a simple average of the metrics $\cohsmat g$, i.e.
\begin{equation}
\label{eq:semi-not-coh-avg}
\frac{\calJ \calT}{N} \left[ \sums \cohsmat g\unc \right] = \semimat g \ne \frac{1}{N} \sums \cohsmat g = \sums \frac{ \calJ \calT \big[ \cohsmat g\unc \big] }{N} \,.
\end{equation}
There are two reasons why Eq.~\eqref{eq:semi-not-coh-avg} does not hold.
First, Eqs.~\eqref{eq:semi-Fstat-mismatch-no-interp-factor} and~\eqref{eq:semi-metric-def} hold only if the $\sig\diff\semivec\lambda$, and hence the $\cohsmat g$, are expressed in a common coordinate system; each $\cohsmat g$, however, is expressed in a per-segment coordinate system $(\cohs{n\ua}, \cohs{n\ub}, \ndot{\cohs\nu})$.
Second, the operator $\calJ$ in general projects each $\calT \cohsmat g\unc$ onto distinct hyperplanes in the coordinate space of $\vec n$, and is therefore a nonlinear operator which does not commute with the averaging in Eq.~\eqref{eq:semi-metric-def}.

\subsection{Validation using numerical simulations}\label{sec:valid-using-numer}

The ability of the semicoherent supersky metric $\semimat g$ to predict the $\calF$-statistic mismatch $\semi\mu\Fstat$ is characterized using numerical simulations, the results of which are presented in this section.

\subsubsection{Simulation procedure}\label{sec:simulation-procedure}

\begin{table*}
\begin{tabular*}{\linewidth}{l@{\extracolsep{\fill}}r@{\extracolsep{2.5\tabcolsep}}r@{\extracolsep{2.5\tabcolsep}}r@{\extracolsep{2.5\tabcolsep}}r@{\extracolsep{2.5\tabcolsep}}r@{\extracolsep{2.5\tabcolsep}}r@{\extracolsep{\fill}}r@{\extracolsep{2.5\tabcolsep}}r@{\extracolsep{2.5\tabcolsep}}r@{\extracolsep{2.5\tabcolsep}}r@{\extracolsep{2.5\tabcolsep}}r@{\extracolsep{2.5\tabcolsep}}r@{\extracolsep{\fill}}r@{\extracolsep{2.5\tabcolsep}}r@{\extracolsep{2.5\tabcolsep}}r@{\extracolsep{2.5\tabcolsep}}r@{\extracolsep{2.5\tabcolsep}}r@{\extracolsep{2.5\tabcolsep}}r@{\extracolsep{\fill}}r@{\extracolsep{2.5\tabcolsep}}r@{\extracolsep{2.5\tabcolsep}}r@{\extracolsep{2.5\tabcolsep}}r@{\extracolsep{2.5\tabcolsep}}r@{\extracolsep{2.5\tabcolsep}}r@{\extracolsep{2.5\tabcolsep}}}
\hline\hline
 & \multicolumn{24}{c}{ $\coh T$ / days} \\
 & \multicolumn{6}{c}{1} & \multicolumn{6}{c}{3} & \multicolumn{6}{c}{5} & \multicolumn{6}{c}{7} \\
\hline
 $N$ &  $\semi T$/d &  $\eta$ &  $S^{0}\unoint$ &  $S^{0}$ &  $S^{.5}\unoint$ &  $S^{.5}$ &  $\semi T$/d &  $\eta$ &  $S^{0}\unoint$ &  $S^{0}$ &  $S^{.5}\unoint$ &  $S^{.5}$ &  $\semi T$/d &  $\eta$ &  $S^{0}\unoint$ &  $S^{0}$ &  $S^{.5}\unoint$ &  $S^{.5}$ &  $\semi T$/d &  $\eta$ &  $S^{0}\unoint$ &  $S^{0}$ &  $S^{.5}\unoint$ &  $S^{.5}$ \\
\hline
3 & 3 & 1.00 &  $0.9$ &  $1.2$ &  $2.4$ &  $2.8$ & 9 & 1.00 &  $1.1$ &  $1.8$ &  $3.1$ &  $\infty$ & 15 & 1.00 &  $1.0$ &  $2.0$ &  $2.5$ &  $\infty$ & 21 & 1.00 &  $1.0$ &  $1.9$ &  $1.9$ &  $4.9$ \\
5 & 5 & 1.00 &  $0.9$ &  $1.3$ &  $2.4$ &  $3.1$ & 15 & 1.00 &  $1.1$ &  $2.1$ &  $2.7$ &  $\infty$ & 25 & 1.00 &  $1.0$ &  $2.3$ &  $1.8$ &  $\infty$ & 35 & 1.00 &  $0.8$ &  $2.2$ &  $1.3$ &  $6.0$ \\
7 & 7 & 1.00 &  $0.9$ &  $1.3$ &  $2.4$ &  $3.4$ & 21 & 1.00 &  $1.1$ &  $2.3$ &  $2.5$ &  $\infty$ & 35 & 1.00 &  $0.8$ &  $2.6$ &  $1.4$ &  $\infty$ & 49 & 1.00 &  $0.9$ &  $2.4$ &  $1.5$ &  $5.7$ \\
9 & 9 & 1.00 &  $0.9$ &  $1.3$ &  $2.4$ &  $3.5$ & 27 & 1.00 &  $1.0$ &  $2.4$ &  $2.0$ &  $\infty$ & 45 & 1.00 &  $0.8$ &  $2.8$ &  $1.4$ &  $\infty$ & 63 & 1.00 &  $1.2$ &  $2.5$ &  $1.9$ &  $\infty$ \\
11 & 11 & 1.00 &  $0.9$ &  $1.3$ &  $2.4$ &  $3.6$ & 33 & 1.00 &  $0.9$ &  $2.6$ &  $1.7$ &  $\infty$ & 55 & 1.00 &  $1.0$ &  $3.0$ &  $1.6$ &  $\infty$ & 77 & 1.00 &  $1.4$ &  $2.5$ &  $2.2$ &  $\infty$ \\
13 & 13 & 1.00 &  $0.9$ &  $1.4$ &  $2.4$ &  $3.7$ & 39 & 1.00 &  $0.9$ &  $2.7$ &  $1.5$ &  $\infty$ & 65 & 1.00 &  $1.2$ &  $3.1$ &  $1.9$ &  $\infty$ & 91 & 1.00 &  $1.6$ &  $2.5$ &  $2.5$ &  $\infty$ \\
15 & 15 & 1.00 &  $0.9$ &  $1.4$ &  $2.4$ &  $3.9$ & 45 & 1.00 &  $0.8$ &  $2.8$ &  $1.4$ &  $\infty$ & 75 & 1.00 &  $1.3$ &  $3.2$ &  $2.1$ &  $\infty$ & 105 & 1.00 &  $1.7$ &  $2.5$ &  $2.7$ &  $\infty$ \\
17 & 17 & 1.00 &  $0.9$ &  $1.4$ &  $2.3$ &  $3.9$ & 51 & 1.00 &  $0.9$ &  $2.9$ &  $1.4$ &  $\infty$ & 85 & 1.00 &  $1.4$ &  $3.4$ &  $2.3$ &  $\infty$ & 119 & 1.00 &  $1.8$ &  $2.5$ &  $3.0$ &  $\infty$ \\
19 & 19 & 1.00 &  $0.9$ &  $1.4$ &  $2.3$ &  $3.9$ & 57 & 1.00 &  $0.9$ &  $3.0$ &  $1.5$ &  $\infty$ & 95 & 1.00 &  $1.5$ &  $3.4$ &  $2.5$ &  $\infty$ & 133 & 1.00 &  $1.8$ &  $2.4$ &  $3.2$ &  $\infty$ \\
\hline
 $\semi T$/d &  $N$ &  $\eta$ &  $S^{0}\unoint$ &  $S^{0}$ &  $S^{.5}\unoint$ &  $S^{.5}$ &  $N$ &  $\eta$ &  $S^{0}\unoint$ &  $S^{0}$ &  $S^{.5}\unoint$ &  $S^{.5}$ &  $N$ &  $\eta$ &  $S^{0}\unoint$ &  $S^{0}$ &  $S^{.5}\unoint$ &  $S^{.5}$ &  $N$ &  $\eta$ &  $S^{0}\unoint$ &  $S^{0}$ &  $S^{.5}\unoint$ &  $S^{.5}$ \\
\hline
120 & 60 & 0.50 &  $1.5$ &  $2.7$ &  $2.5$ &  $\infty$ & 20 & 0.50 &  $1.9$ &  $3.9$ &  $2.7$ &  $\infty$ & 12 & 0.50 &  $2.0$ &  $3.1$ &  $2.9$ &  $\infty$ & 8 & 0.47 &  $1.9$ &  $2.2$ &  $3.0$ &  $5.3$ \\
120 & 90 & 0.75 &  $1.9$ &  $3.0$ &  $2.4$ &  $\infty$ & 30 & 0.75 &  $1.7$ &  $4.2$ &  $2.7$ &  $\infty$ & 18 & 0.75 &  $2.0$ &  $3.4$ &  $2.9$ &  $\infty$ & 12 & 0.70 &  $2.1$ &  $2.3$ &  $3.0$ &  $\infty$ \\
120 & 120 & 1.00 &  $1.2$ &  $1.9$ &  $2.5$ &  $\infty$ & 40 & 1.00 &  $1.5$ &  $3.8$ &  $2.7$ &  $\infty$ & 24 & 1.00 &  $1.7$ &  $3.6$ &  $2.8$ &  $\infty$ & 17 & 0.99 &  $1.9$ &  $2.4$ &  $3.0$ &  $\infty$ \\
240 & 120 & 0.50 &  $2.1$ &  $4.5$ &  $4.0$ &  $\infty$ & 40 & 0.50 &  $2.8$ &  $4.9$ &  $4.8$ &  $\infty$ & 24 & 0.50 &  $2.9$ &  $3.7$ &  $5.0$ &  $\infty$ & 17 & 0.50 &  $2.8$ &  $2.2$ &  $5.4$ &  $6.0$ \\
240 & 180 & 0.75 &  $3.1$ &  $3.3$ &  $4.1$ &  $\infty$ & 60 & 0.75 &  $2.2$ &  $5.0$ &  $4.7$ &  $\infty$ & 36 & 0.75 &  $2.9$ &  $4.1$ &  $5.1$ &  $\infty$ & 25 & 0.73 &  $2.9$ &  $2.2$ &  $5.0$ &  $\infty$ \\
240 & 240 & 1.00 &  $1.6$ &  $3.0$ &  $4.0$ &  $\infty$ & 80 & 1.00 &  $1.8$ &  $4.9$ &  $4.7$ &  $\infty$ & 48 & 1.00 &  $2.0$ &  $4.2$ &  $5.0$ &  $\infty$ & 34 & 0.99 &  $2.7$ &  $2.3$ &  $5.0$ &  $\infty$ \\
360 & 180 & 0.50 &  $2.3$ &  $3.1$ &  $5.0$ &  $\infty$ & 60 & 0.50 &  $3.2$ &  $5.0$ &  $6.0$ &  $\infty$ & 36 & 0.50 &  $3.3$ &  $4.3$ &  $\infty$ &  $\infty$ & 25 & 0.49 &  $3.1$ &  $2.3$ &  $\infty$ &  $\infty$ \\
360 & 270 & 0.75 &  $3.8$ &  $3.2$ &  $4.8$ &  $\infty$ & 90 & 0.75 &  $2.6$ &  $5.2$ &  $\infty$ &  $\infty$ & 54 & 0.75 &  $3.5$ &  $4.7$ &  $6.0$ &  $\infty$ & 38 & 0.74 &  $3.2$ &  $2.4$ &  $\infty$ &  $\infty$ \\
360 & 360 & 1.00 &  $1.8$ &  $3.2$ &  $4.6$ &  $\infty$ & 120 & 1.00 &  $1.9$ &  $5.2$ &  $5.5$ &  $\infty$ & 72 & 1.00 &  $2.0$ &  $5.0$ &  $\infty$ &  $\infty$ & 51 & 0.99 &  $3.0$ &  $2.4$ &  $\infty$ &  $\infty$ \\
\hline\hline
\end{tabular*}
\caption{\label{tab:segment_list_table}
Properties of the 72 search setups used to test the semicoherent supersky metric, and of the numerical simulations based on those setups, as described in Section~\ref{sec:valid-using-numer}.
Top half: for 36 setups indexed by number of segments $N$ (rows) and segment time-span $\coh T$ (columns), the total time-span $\semi T$ and segment duty cycle $\eta = N \coh T / \semi T$ are given.
Bottom half: for 36 setups indexed by $\semi T$ (rows) and $\coh T$ (columns), $N$ and $\eta$ are given.
For all setups, the log-probabilities $S^{x}\unoint \equiv -\log_{10} p \big(\relerr{\semi\mu\Fstat\unoint}{\semi\mu\unoint} > x \big)$ and $S^{x} \equiv -\log_{10} p \big(\relerr{\semi\mu\Fstat}{\semi\mu} > 0 \big)$, $x \in \{ 0, 0.5 \}$, are also given.
}
\end{table*}

The simulation procedure closely follows that given in Section~III of~\PaperII\ (which expands upon Appendix~A of~\PaperI) and is briefly described below:
\begin{enumerate}[(i)]

\item\label{item:num-sim-input}
The simulations take a search setup as input, comprising $N$ segments with segment time-span $\coh T$ equally distributed over a total time-span $\semi T$.
The mean mid-time $\sums t_\ell / N$ of each setup is set to the reference time $t_0 = \text{UTC 2015-01-01 00:00:00}$.
Other input parameters are the maximum mismatches of the coherent and semicoherent template banks, $\coh\mu\umax$ and $\semi\mu\umax$ respectively, and the frequency of simulated signals $f$.
When computing metrics and simulating gravitational-wave signals, the LIGO Livingston detector~\cite{Abbott.etal.2009f} is assumed.

\item\label{item:num-sim-nearest-semi}
A signal parameter vector $\sigvec\lambda = (\vec n, f, \ndot[1] f)$ is generated at reference time $t_0$, with $f$ fixed and other parameters taking random values uniformly distributed over the sky $|\vec n| = 1$ and in spindown $\ndot[1] f$.
Then, the nearest semicoherent template to $\sigvec\lambda$, $\semivec\lambda$, is determined assuming a semicoherent template bank constructed from an $\Ans[4]$ lattice using $\semimat g$ with maximum mismatch $\semi\mu\umax$ (see Section~III~A of~\PaperII\ for details).

\item\label{item:num-sim-no-interp-mismatch}
The metric mismatch between $\sigvec\lambda$ and $\semivec\lambda$ assuming \emph{no} interpolation is calculated using
\begin{equation}
\label{eq:semi-ssky-mismatch-no-interp}
\semi\mu\unoint \equiv \sig\diff\semivec\lambda \cdot \semimat g \cdot \sig\diff\semivec\lambda \,.
\end{equation}
The corresponding $\calF$-statistic mismatch $\semi\mu\Fstat\unoint$ is computed, using the software package \textsc{LALSuite}~\footnote{
Available at \url{https://www.lsc-group.phys.uwm.edu/daswg/projects/lalsuite.html}.
}, by simulating a gravitational-wave signal with parameters $\sigvec\lambda$, computing the $\calF$-statistic in each segment at templates $\sigvec\lambda$ and $\semivec\lambda$ to determine the signal-to-noise ratios $\cohs\SNR^2(\sigvec\lambda; \sigvec\lambda)$ and $\cohs\SNR^2(\sigvec\lambda; \semivec\lambda)$, and calculating $\semi\mu\Fstat\unoint$ using Eqs.~\eqref{eq:coh-Fstat-mismatch-SNR-def} and~\eqref{eq:semi-Fstat-mismatch-no-interp}.

\item\label{item:num-sim-nearest-coh}
In each segment $\ell$, the nearest coherent template to $\semivec\lambda$, $\cohsvec\lambda(\semivec\lambda)$, is determined assuming a coherent template bank constructed from an $\Ans[4]$ lattice using $\cohsmat g$ with maximum mismatch $\coh\mu\umax$, following step~\eqref{item:num-sim-nearest-semi}.

\item\label{item:num-sim-mismatch}
The metric mismatch between $\sigvec\lambda$ and $\semivec\lambda$ \emph{with} interpolation is calculated using
\begin{equation}
\label{eq:semi-ssky-mismatch}
\semi\mu \equiv \frac{1}{N} \sums \sig\diff\cohsvec\lambda(\semivec\lambda) \cdot \cohsmat g \cdot \sig\diff\cohsvec\lambda(\semivec\lambda) \,.
\end{equation}
The corresponding $\calF$-statistic mismatch $\semi\mu\Fstat$ is computed following step~\eqref{item:num-sim-nearest-coh} by simulating a gravitational-wave signal with parameters $\sigvec\lambda$, computing the $\calF$-statistic in each segment at templates $\sigvec\lambda$ and $\cohsvec\lambda(\semivec\lambda)$ to determine the signal-to-noise ratios $\cohs\SNR^2(\sigvec\lambda; \sigvec\lambda)$ and $\cohs\SNR^2\big(\sigvec\lambda; \cohsvec\lambda(\semivec\lambda)\big)$, and calculating $\semi\mu\Fstat$ using Eqs.~\eqref{eq:coh-Fstat-mismatch-SNR-def} and~\eqref{eq:semi-Fstat-mismatch}.

\end{enumerate}

The input parameters to the simulation procedure are: 72 search setups given in Table~\ref{tab:segment_list_table}; two values of $f / \text{Hz} \in \{ 100, 1000 \}$; and five combinations of maximum template-bank mismatches $(\coh\mu\umax, \semi\mu\umax) \in \{ (0.1,0.1), (0.3,0.3), (0.5,0.5), (0.1,0.5), (0.5,0.1) \}$.
A total of $3.1{\times}10^{9}$ coherent $\calF$-statistic values were computed.

\subsubsection{Simulation results: relative errors}\label{sec:simul-results:-relat}

\begin{figure*}
\subfloat[]{\includegraphics[width=0.49\linewidth]{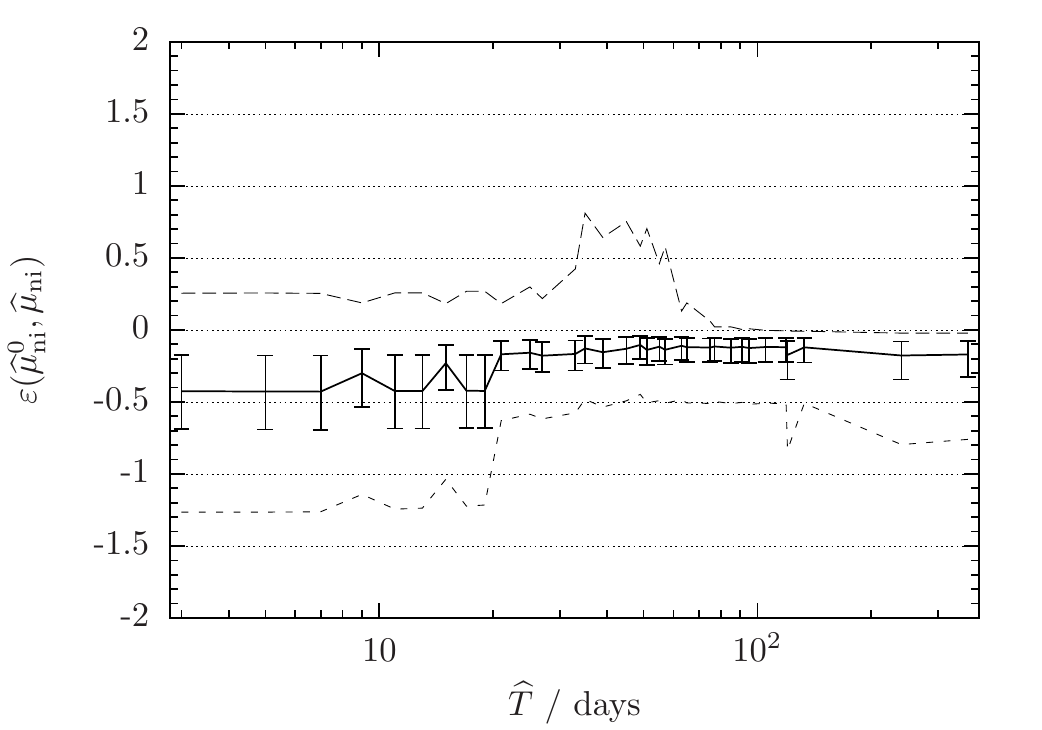}\label{fig:mu_re_twoF_rssky_semiT_noi}}
\subfloat[]{\includegraphics[width=0.49\linewidth]{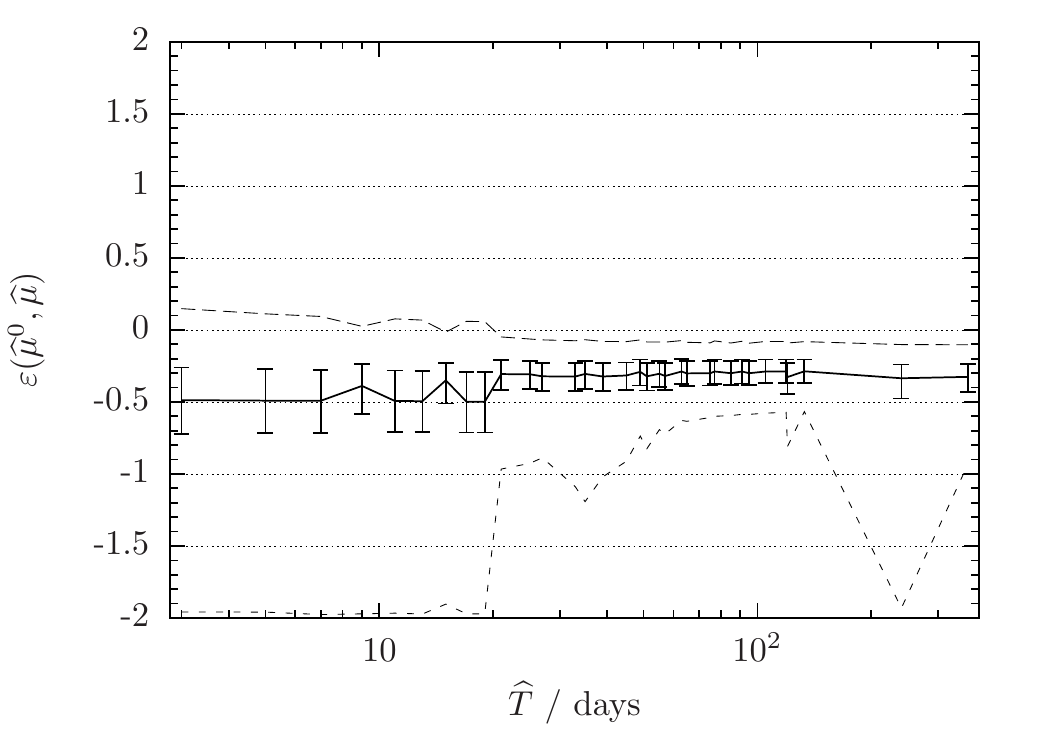}\label{fig:mu_re_twoF_rssky_semiT_tot}}\\
\subfloat[]{\includegraphics[width=0.49\linewidth]{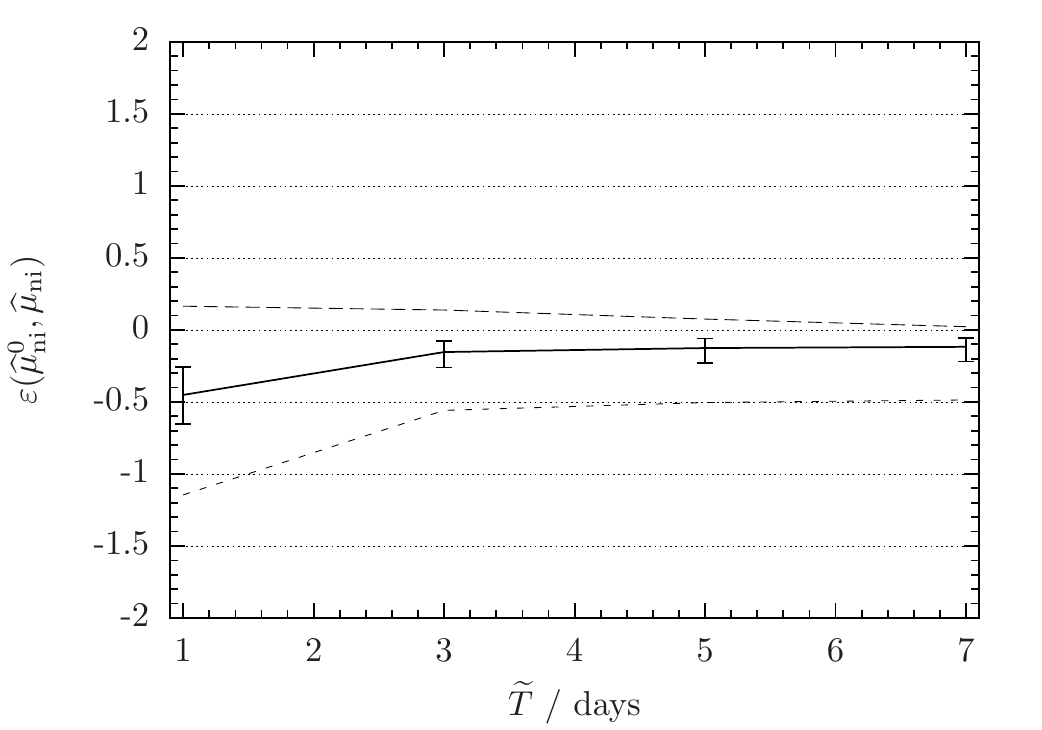}\label{fig:mu_re_twoF_rssky_cohT_noi}}
\subfloat[]{\includegraphics[width=0.49\linewidth]{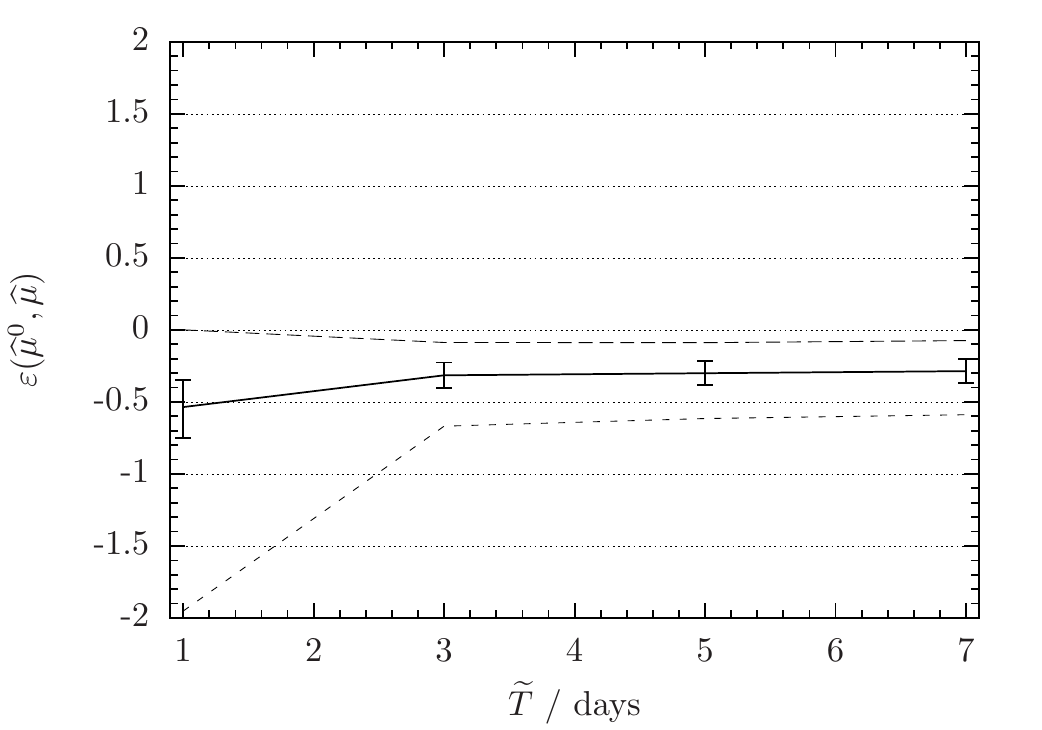}\label{fig:mu_re_twoF_rssky_cohT_tot}}\\
\subfloat[]{\includegraphics[width=0.49\linewidth]{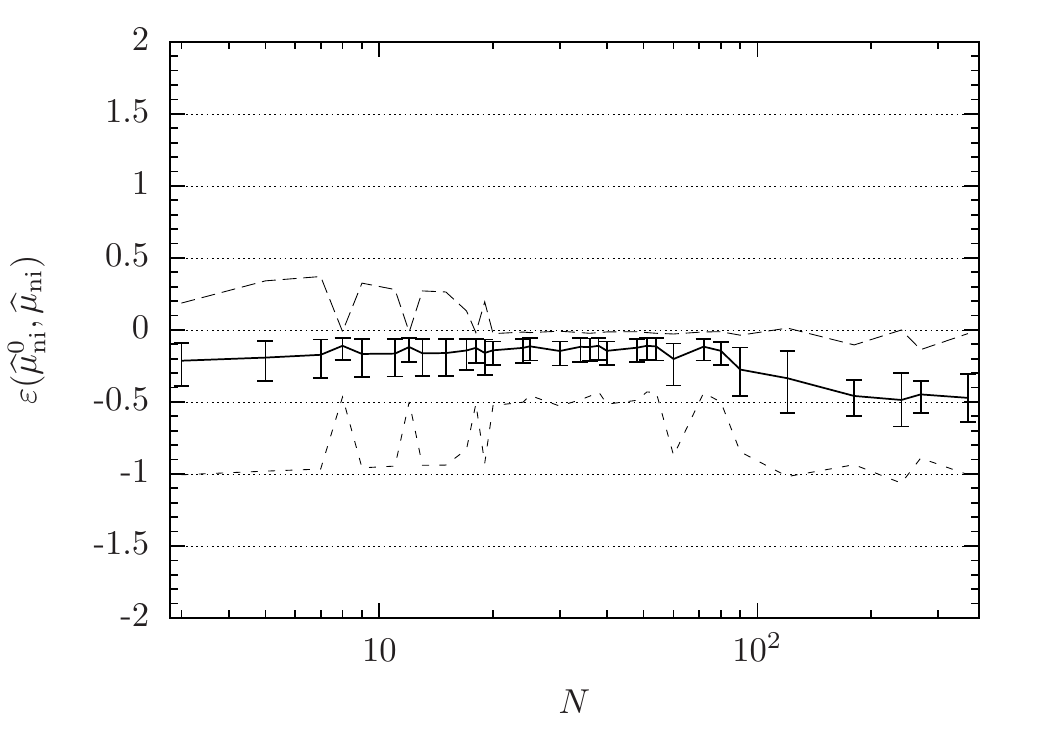}\label{fig:mu_re_twoF_rssky_N_noi}}
\subfloat[]{\includegraphics[width=0.49\linewidth]{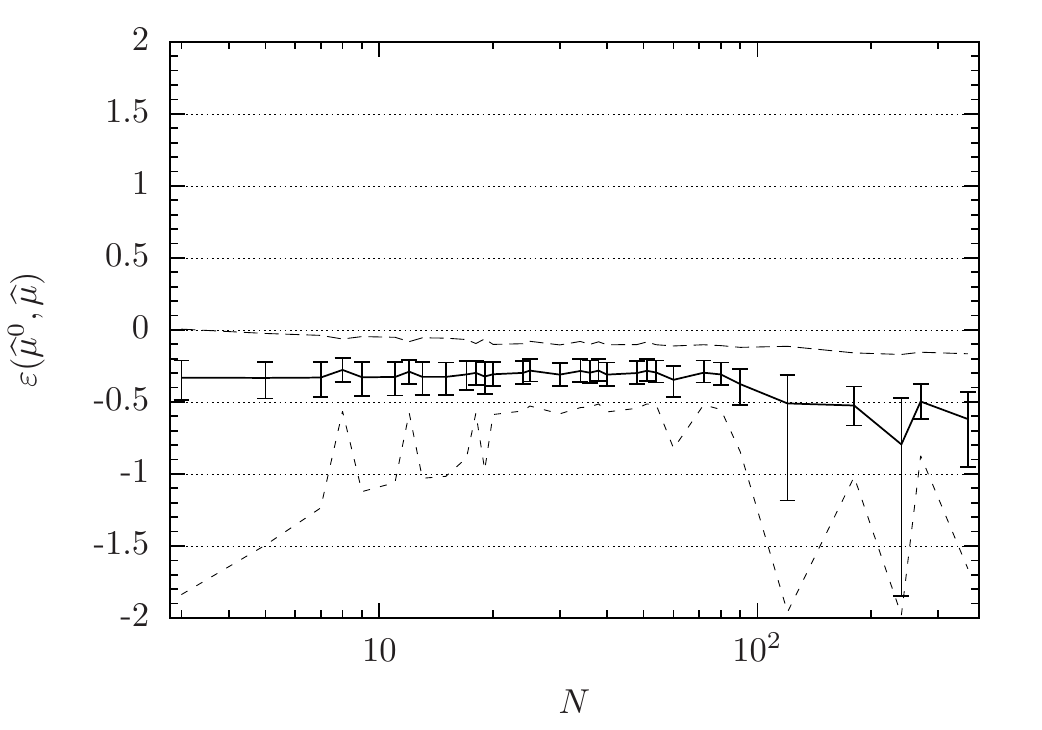}\label{fig:mu_re_twoF_rssky_N_tot}}
\caption{\label{fig:mu_re_twoF_rssky_TN}
Relative errors $\relerr{\semi\mu\Fstat\unoint}{\semi\mu\unoint}$ (left column) and $\relerr{\semi\mu\Fstat}{\semi\mu}$ (right column), at constant \protect\subref{fig:mu_re_twoF_rssky_semiT_noi}\protect\subref{fig:mu_re_twoF_rssky_semiT_tot}~total time-span $\semi T$, \protect\subref{fig:mu_re_twoF_rssky_cohT_noi}\protect\subref{fig:mu_re_twoF_rssky_cohT_tot}~segment time-span $\coh T$, and \protect\subref{fig:mu_re_twoF_rssky_N_noi}\protect\subref{fig:mu_re_twoF_rssky_N_tot}~number of segments $N$.
All other simulation parameters are averaged over.
Plotted are the median (solid line), the 25th--75th percentile range (error bars), and the 2.5th (lower, short-dashed line) and 97.5th (upper, long-dashed line) percentiles.
}
\end{figure*}

\begin{figure*}
\subfloat[]{\includegraphics[width=0.49\linewidth]{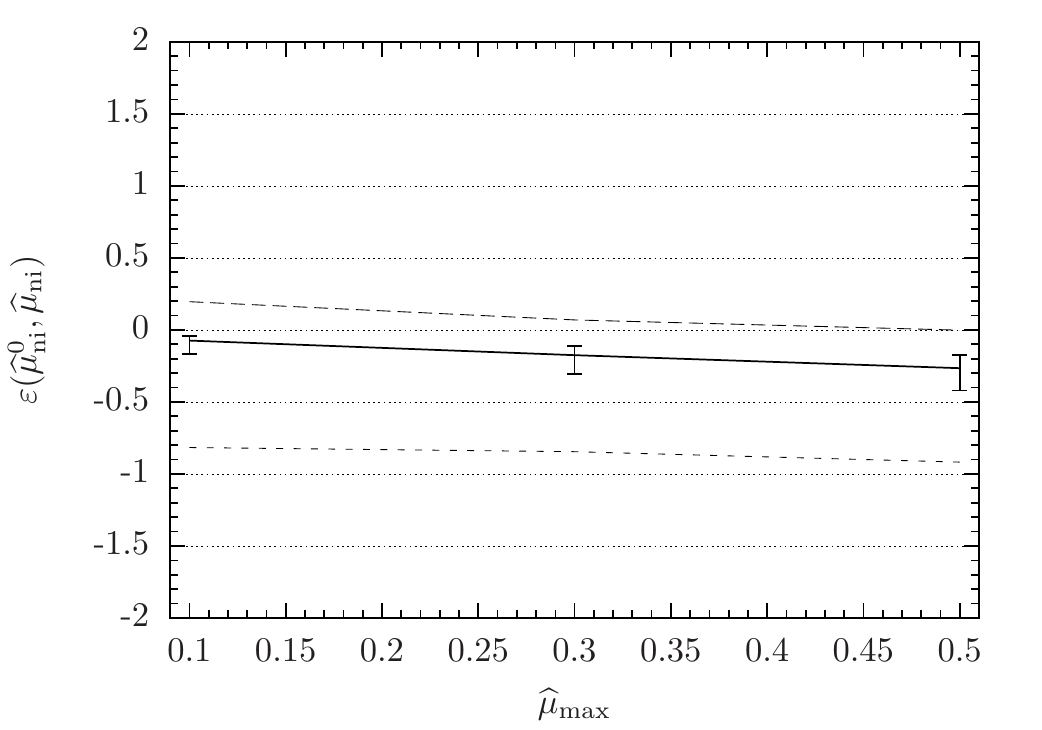}\label{fig:mu_re_twoF_rssky_semimumax_noi}}
\subfloat[]{\includegraphics[width=0.49\linewidth]{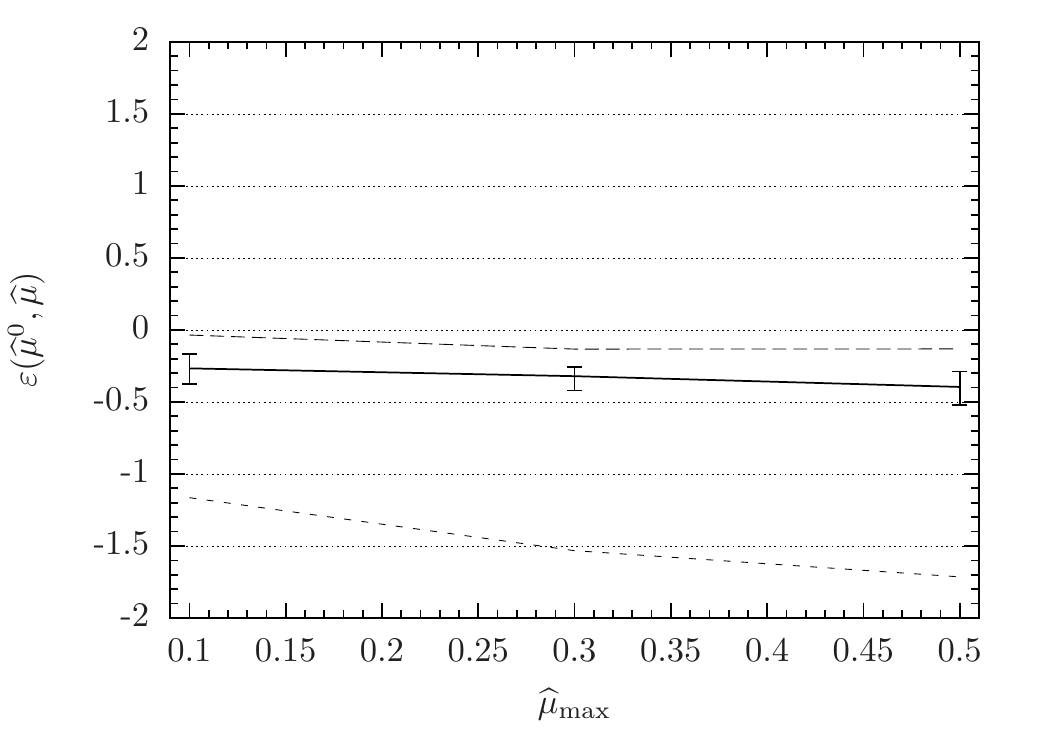}\label{fig:mu_re_twoF_rssky_semimumax_tot}}\\
\subfloat[]{\includegraphics[width=0.49\linewidth]{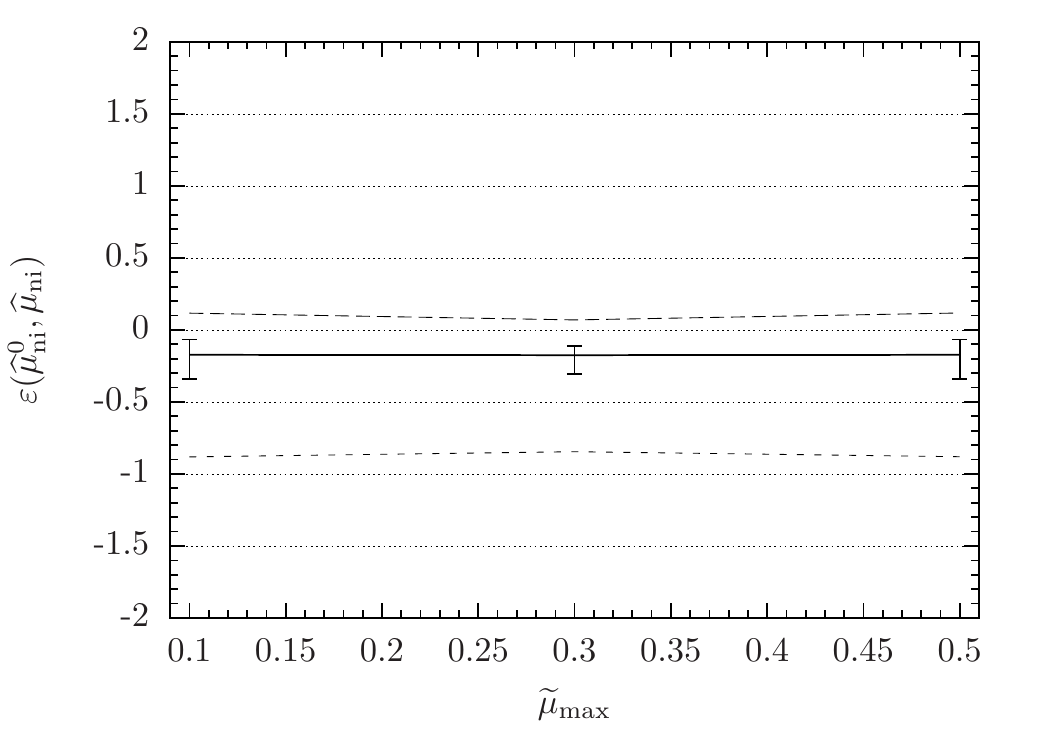}\label{fig:mu_re_twoF_rssky_cohmumax_noi}}
\subfloat[]{\includegraphics[width=0.49\linewidth]{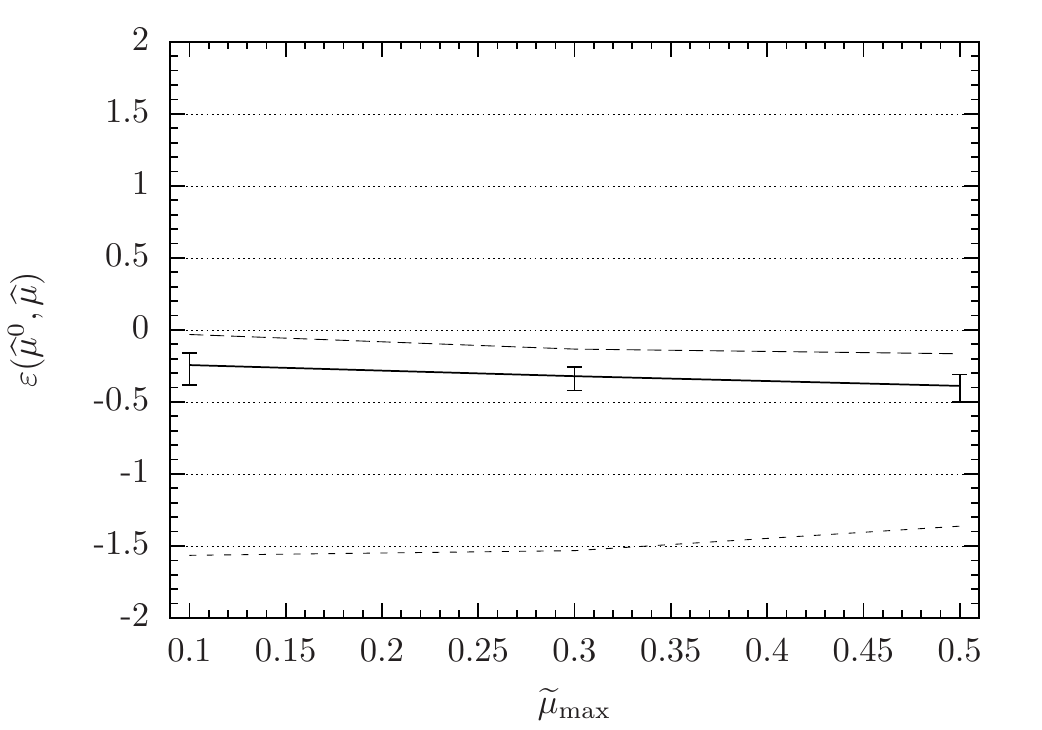}\label{fig:mu_re_twoF_rssky_cohmumax_tot}}\\
\subfloat[]{\includegraphics[width=0.49\linewidth]{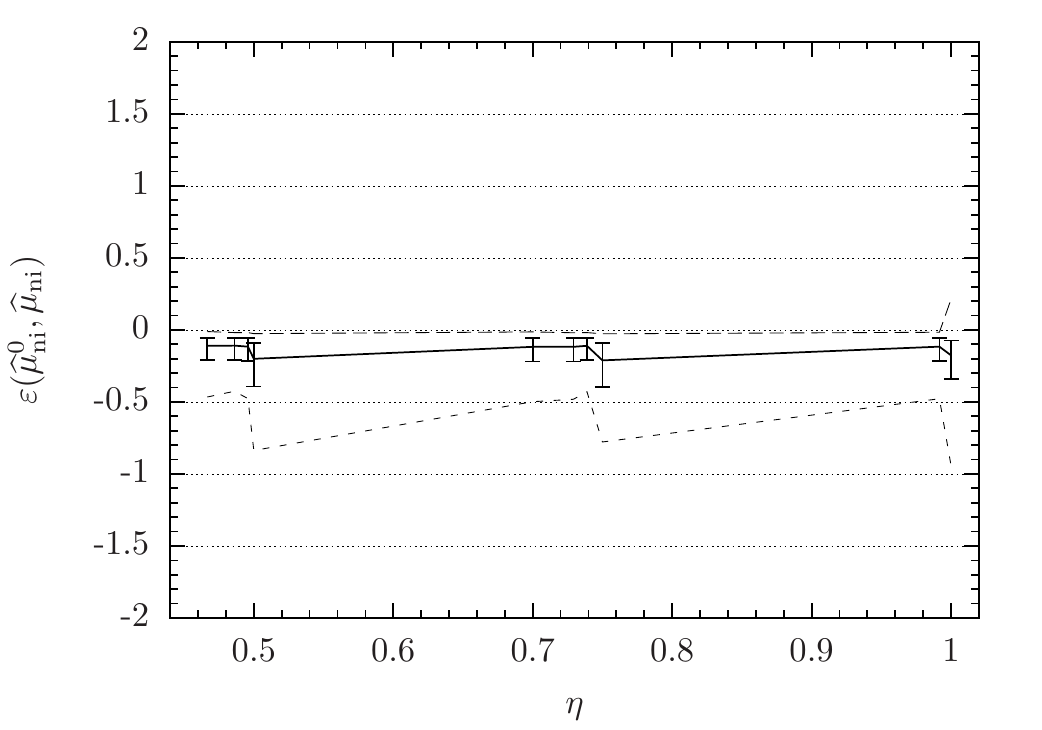}\label{fig:mu_re_twoF_rssky_eta_noi}}
\subfloat[]{\includegraphics[width=0.49\linewidth]{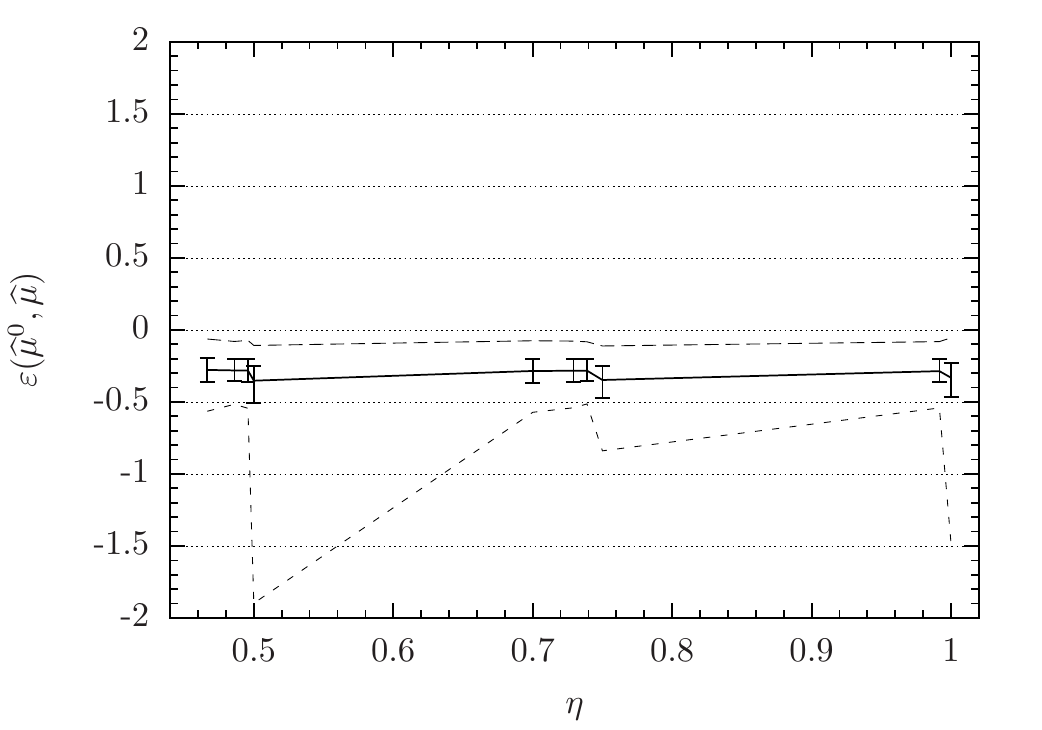}\label{fig:mu_re_twoF_rssky_eta_tot}}
\caption{\label{fig:mu_re_twoF_rssky_mueta}
Relative errors $\relerr{\semi\mu\Fstat\unoint}{\semi\mu\unoint}$ (left column), and $\relerr{\semi\mu\Fstat}{\semi\mu}$ (right column), at constant \protect\subref{fig:mu_re_twoF_rssky_semimumax_noi}\protect\subref{fig:mu_re_twoF_rssky_semimumax_tot}~maximum semicoherent template bank mismatch $\semi\mu\umax$, \protect\subref{fig:mu_re_twoF_rssky_cohmumax_noi}\protect\subref{fig:mu_re_twoF_rssky_cohmumax_tot}~maximum coherent template bank mismatch $\coh\mu\umax$, and \protect\subref{fig:mu_re_twoF_rssky_eta_noi}\protect\subref{fig:mu_re_twoF_rssky_eta_tot}~segment duty cycle $\eta = N \coh T / \semi T$.
All other simulation parameters are averaged over.
Plotted are the median (solid line), the 25th--75th percentile range (error bars), and the 2.5th (lower, short-dashed line) and 97.5th (upper, long-dashed line) percentiles.
}
\end{figure*}

\begin{figure*}
\subfloat[]{\includegraphics[width=0.49\linewidth]{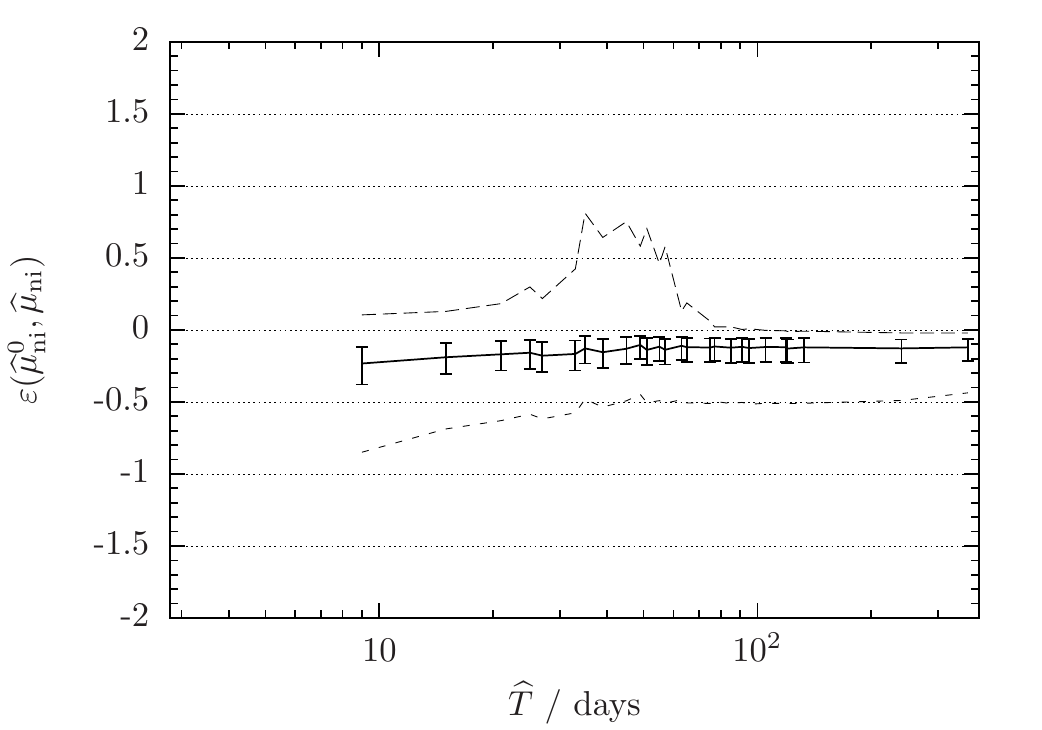}\label{fig:mu_re_twoF_rssky_semiT_cohTgt1_noi}}
\subfloat[]{\includegraphics[width=0.49\linewidth]{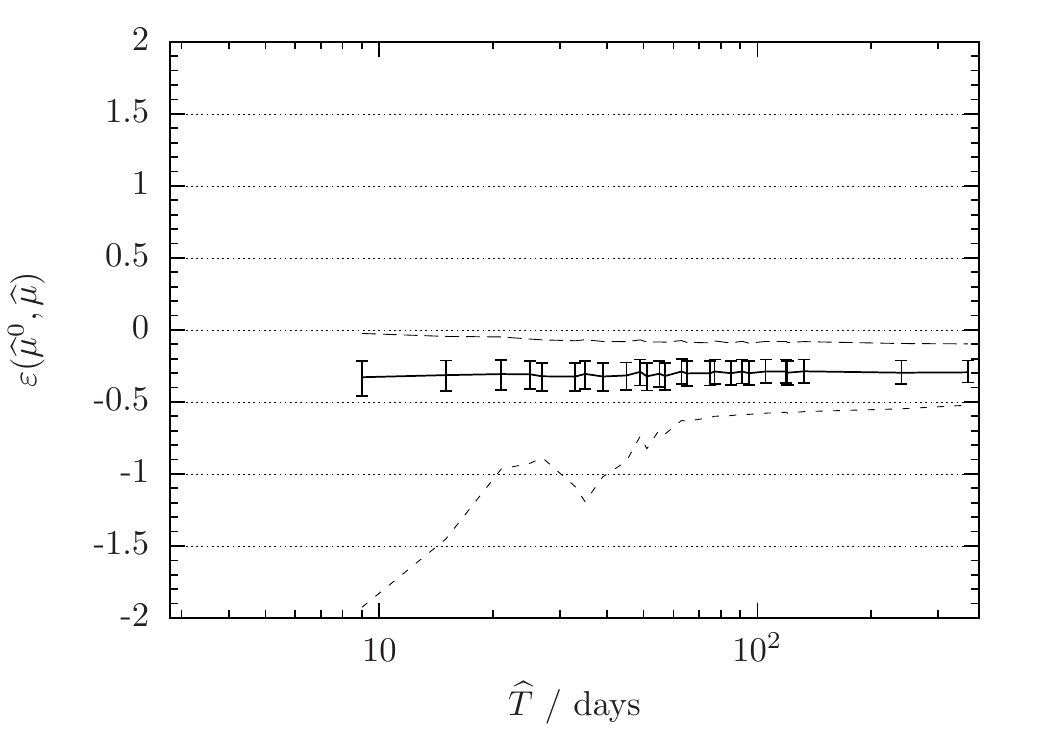}\label{fig:mu_re_twoF_rssky_semiT_cohTgt1_tot}}\\
\subfloat[]{\includegraphics[width=0.49\linewidth]{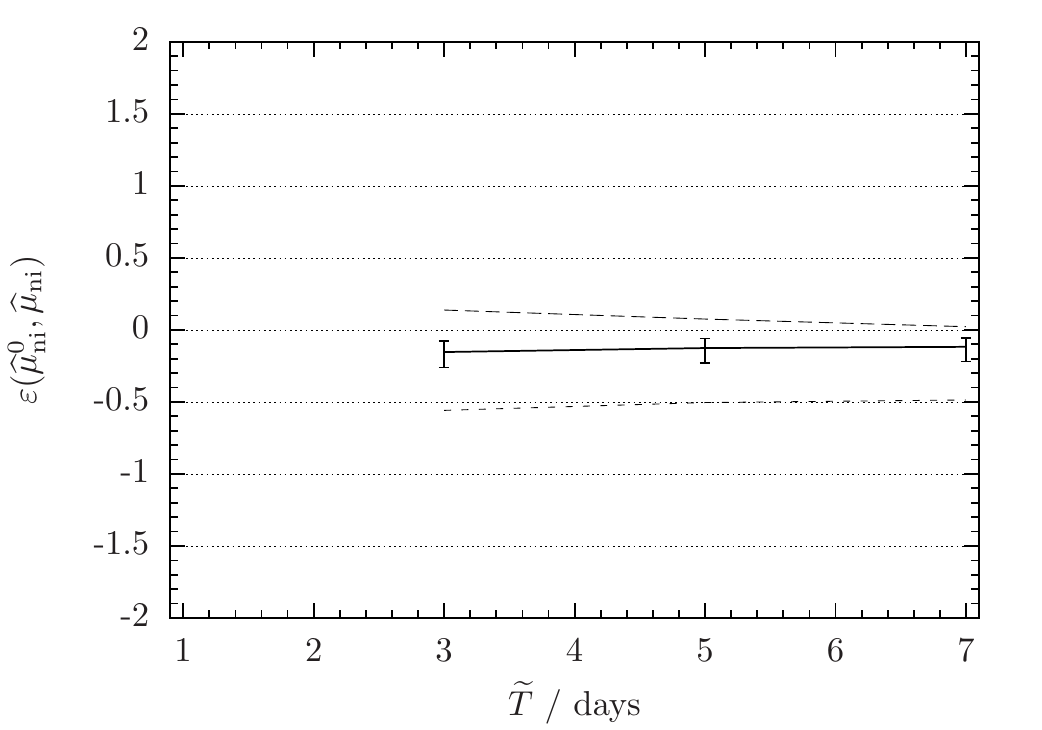}\label{fig:mu_re_twoF_rssky_cohT_cohTgt1_noi}}
\subfloat[]{\includegraphics[width=0.49\linewidth]{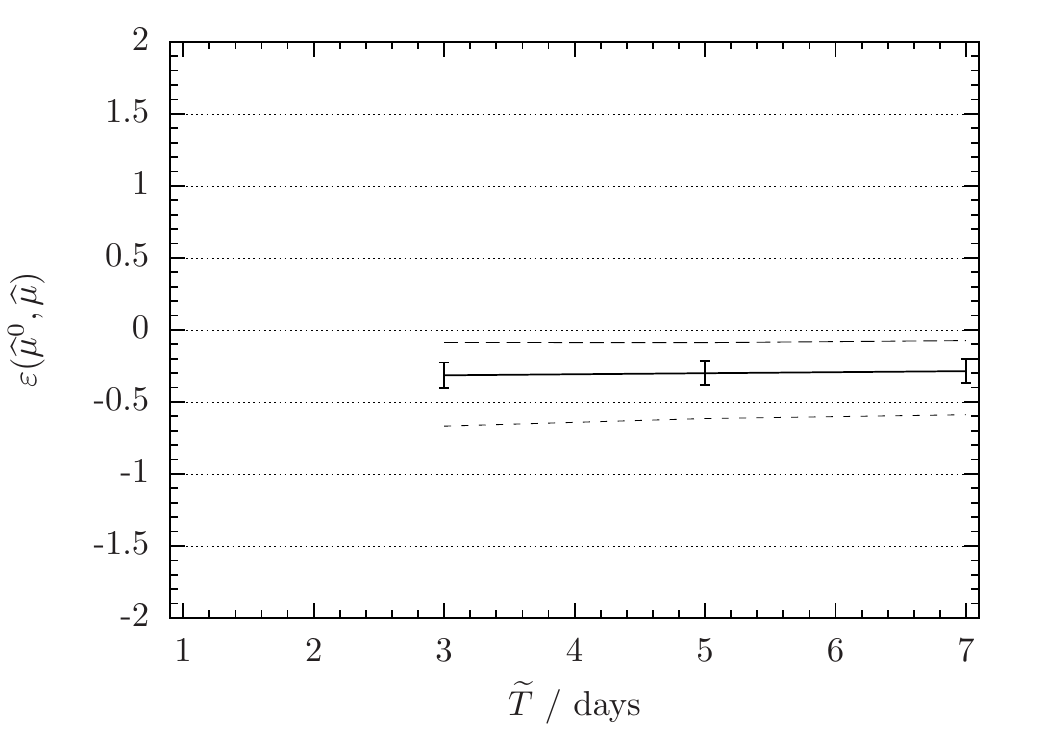}\label{fig:mu_re_twoF_rssky_cohT_cohTgt1_tot}}\\
\subfloat[]{\includegraphics[width=0.49\linewidth]{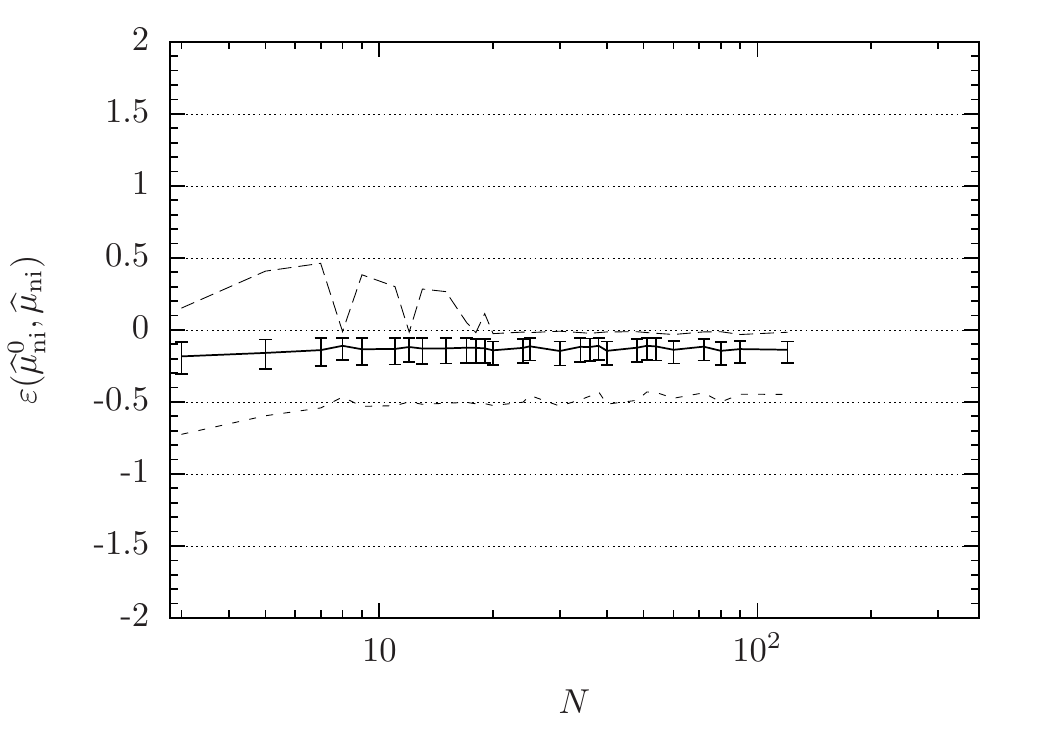}\label{fig:mu_re_twoF_rssky_N_cohTgt1_noi}}
\subfloat[]{\includegraphics[width=0.49\linewidth]{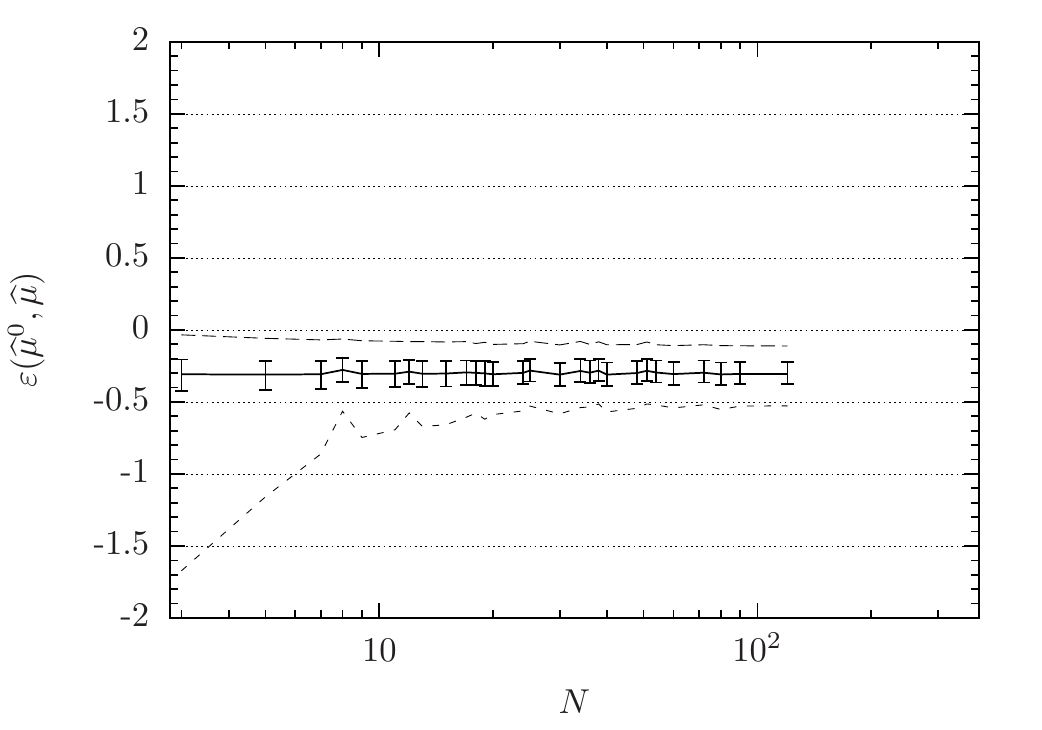}\label{fig:mu_re_twoF_rssky_N_cohTgt1_tot}}
\caption{\label{fig:mu_re_twoF_rssky_TN_cohTgt1}
Relative errors $\relerr{\semi\mu\Fstat\unoint}{\semi\mu\unoint}$ (left column) and $\relerr{\semi\mu\Fstat}{\semi\mu}$ (right column), at constant \protect\subref{fig:mu_re_twoF_rssky_semiT_cohTgt1_noi}\protect\subref{fig:mu_re_twoF_rssky_semiT_cohTgt1_tot}~total time-span $\semi T$, \protect\subref{fig:mu_re_twoF_rssky_cohT_cohTgt1_noi}\protect\subref{fig:mu_re_twoF_rssky_cohT_cohTgt1_tot}~segment time-span $\coh T$, and \protect\subref{fig:mu_re_twoF_rssky_N_cohTgt1_noi}\protect\subref{fig:mu_re_twoF_rssky_N_cohTgt1_tot}~number of segments $N$.
All other simulation parameters are averaged over, subject to the restriction $\coh T > 1$~day.
Plotted are the median (solid line), the 25th--75th percentile range (error bars), and the 2.5th (lower, short-dashed line) and 97.5th (upper, long-dashed line) percentiles.
}
\end{figure*}

\begin{figure*}
\subfloat[]{\includegraphics[width=0.49\linewidth]{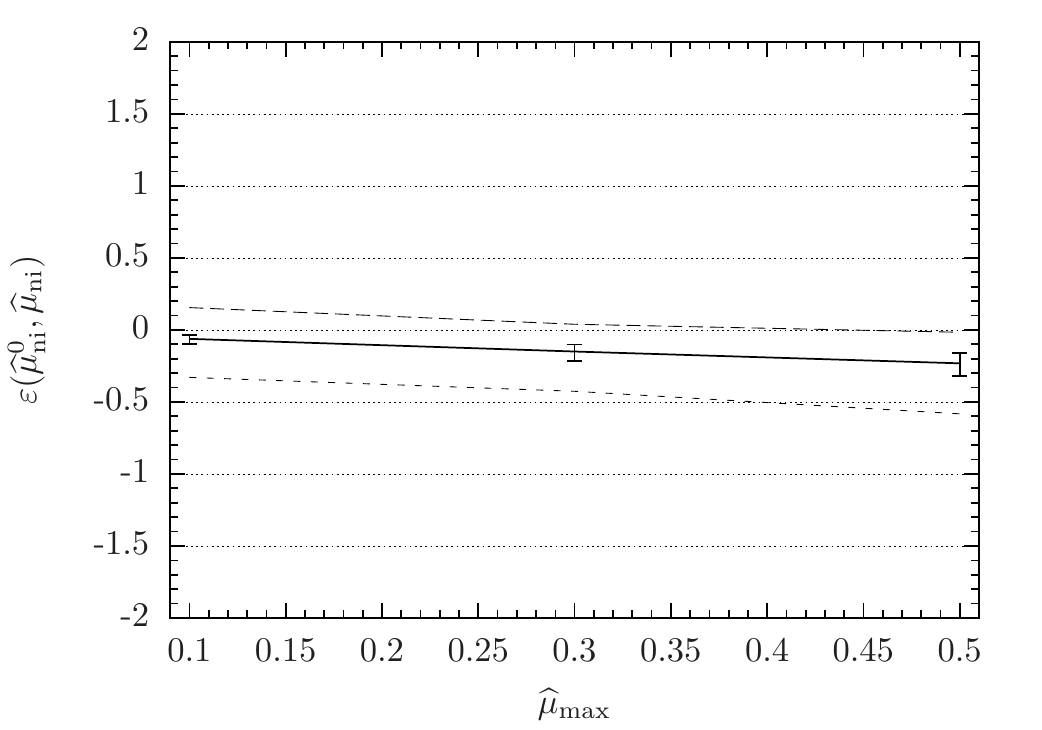}\label{fig:mu_re_twoF_rssky_semimumax_cohTgt1_noi}}
\subfloat[]{\includegraphics[width=0.49\linewidth]{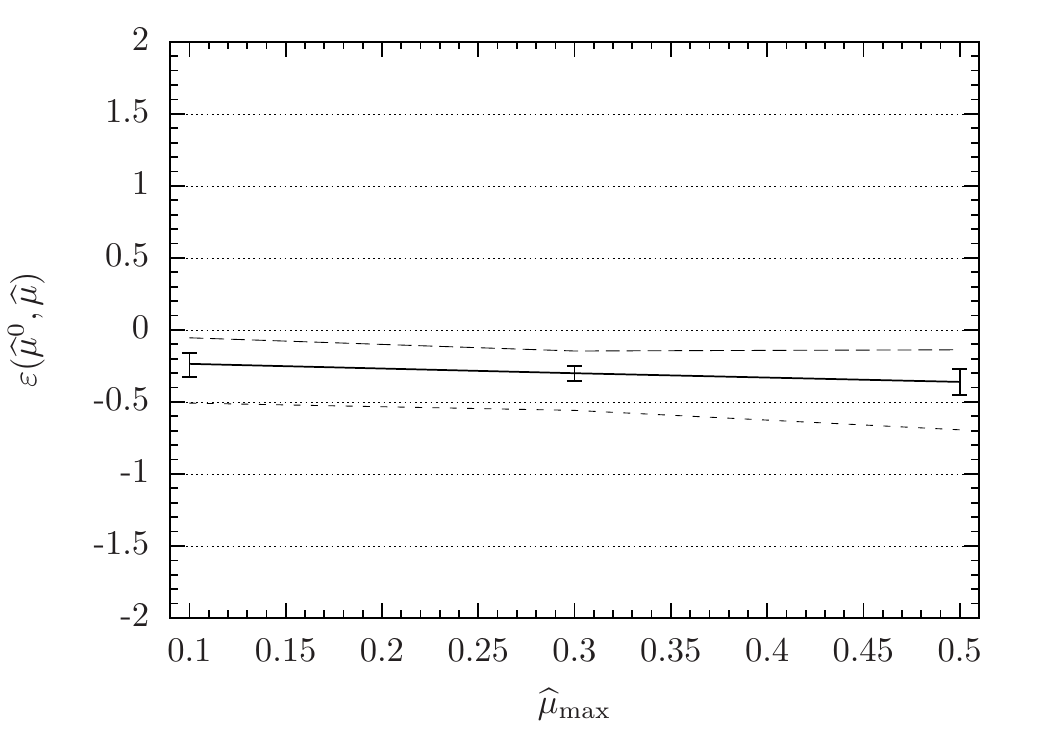}\label{fig:mu_re_twoF_rssky_semimumax_cohTgt1_tot}}\\
\subfloat[]{\includegraphics[width=0.49\linewidth]{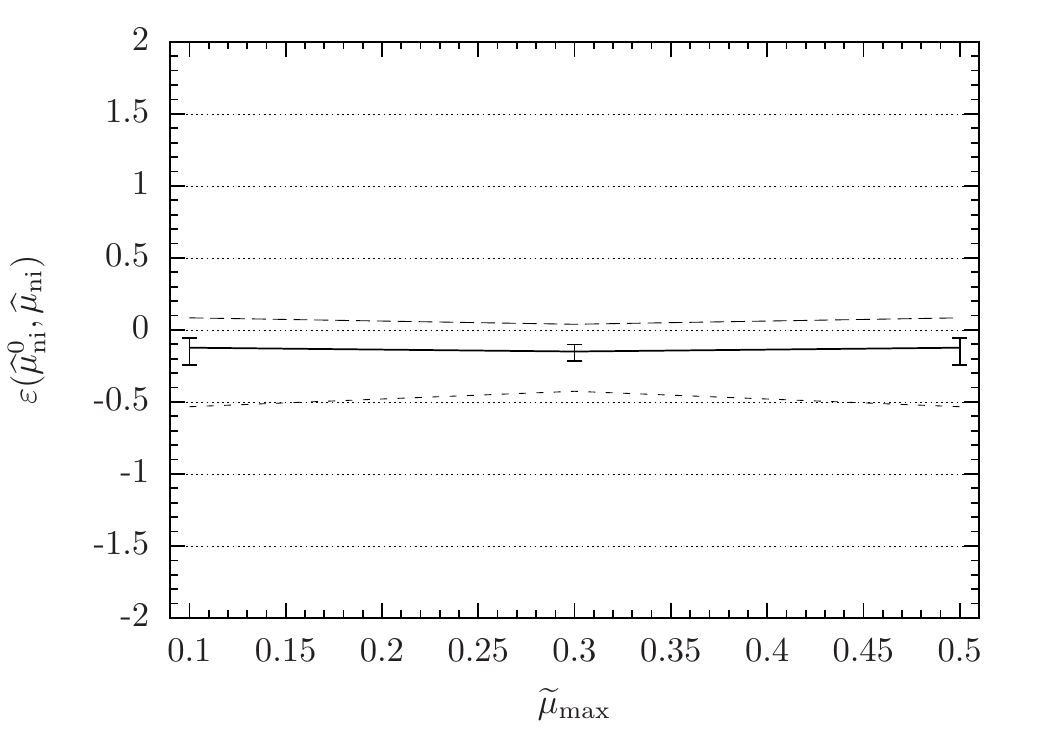}\label{fig:mu_re_twoF_rssky_cohmumax_cohTgt1_noi}}
\subfloat[]{\includegraphics[width=0.49\linewidth]{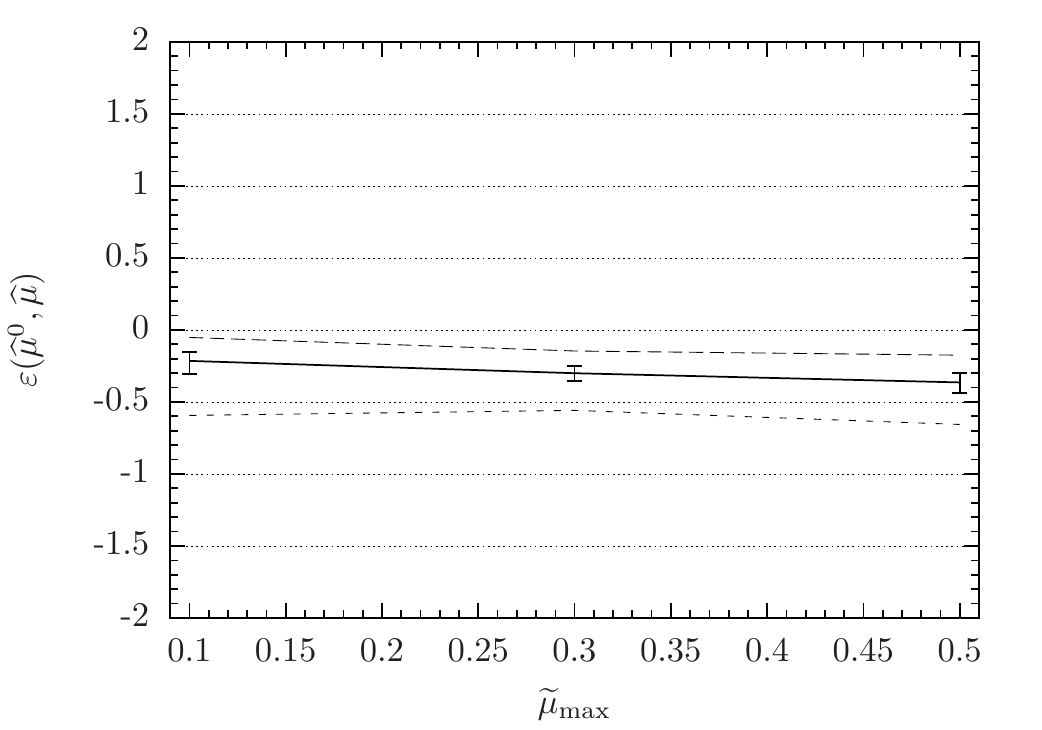}\label{fig:mu_re_twoF_rssky_cohmumax_cohTgt1_tot}}\\
\subfloat[]{\includegraphics[width=0.49\linewidth]{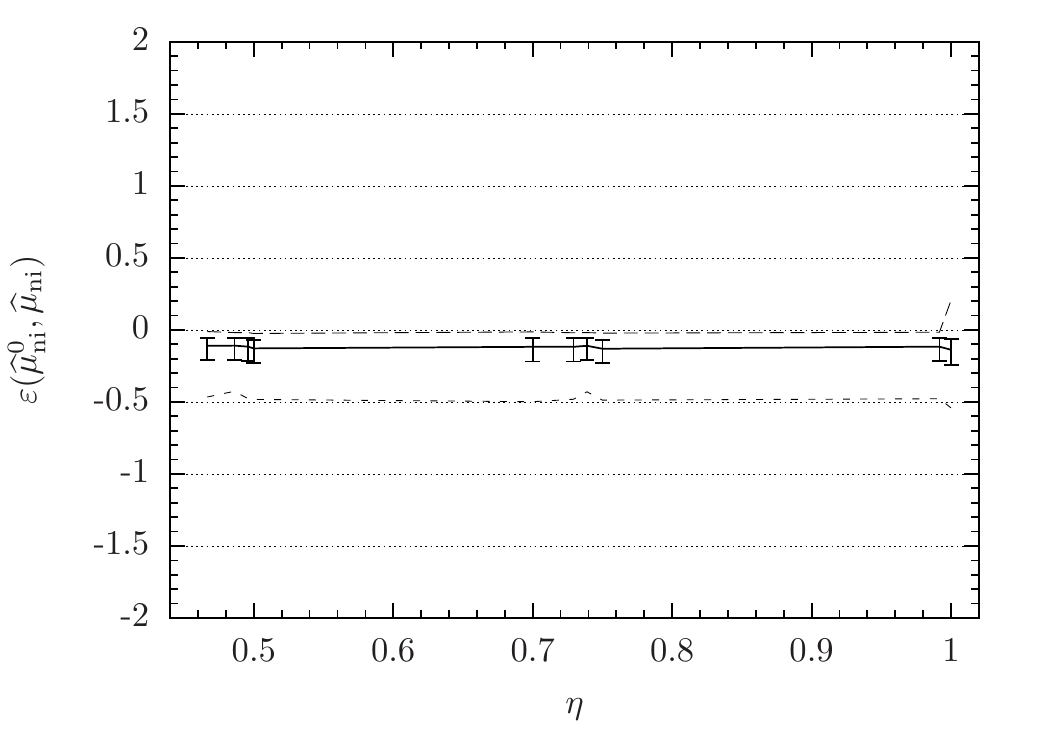}\label{fig:mu_re_twoF_rssky_eta_cohTgt1_noi}}
\subfloat[]{\includegraphics[width=0.49\linewidth]{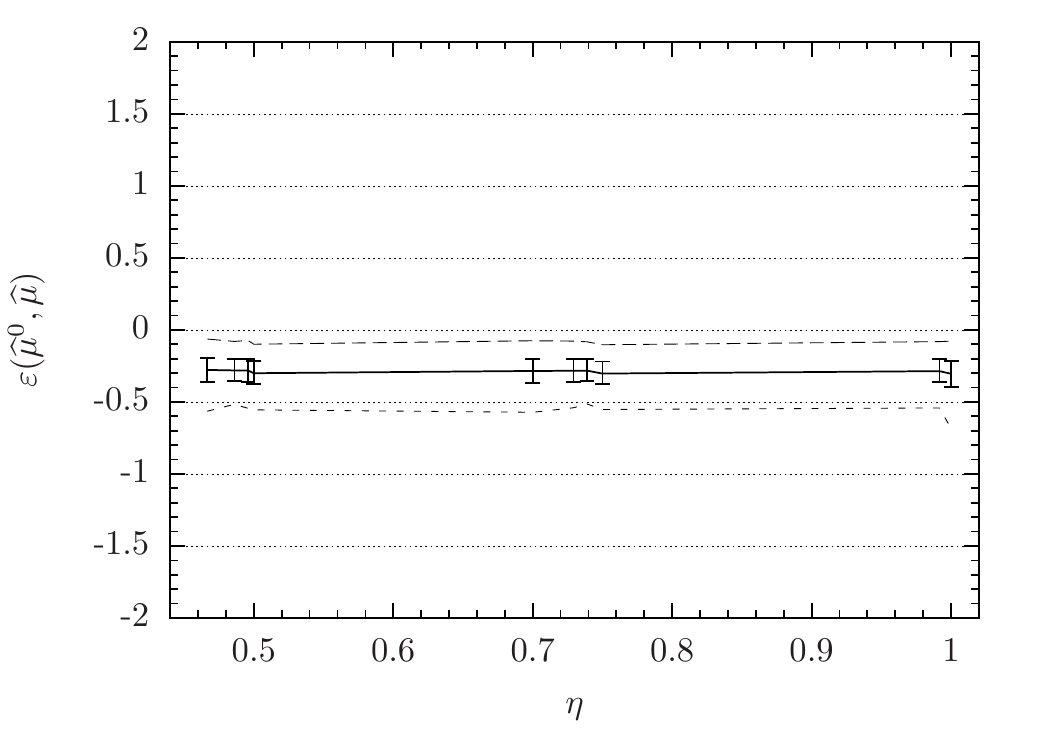}\label{fig:mu_re_twoF_rssky_eta_cohTgt1_tot}}
\caption{\label{fig:mu_re_twoF_rssky_mueta_cohTgt1}
Relative errors $\relerr{\semi\mu\Fstat\unoint}{\semi\mu\unoint}$ (left column), and $\relerr{\semi\mu\Fstat}{\semi\mu}$ (right column), at constant \protect\subref{fig:mu_re_twoF_rssky_semimumax_cohTgt1_noi}\protect\subref{fig:mu_re_twoF_rssky_semimumax_cohTgt1_tot}~maximum semicoherent template bank mismatch $\semi\mu\umax$, \protect\subref{fig:mu_re_twoF_rssky_cohmumax_cohTgt1_noi}\protect\subref{fig:mu_re_twoF_rssky_cohmumax_cohTgt1_tot}~maximum coherent template bank mismatch $\coh\mu\umax$, and \protect\subref{fig:mu_re_twoF_rssky_eta_cohTgt1_noi}\protect\subref{fig:mu_re_twoF_rssky_eta_cohTgt1_tot}~segment duty cycle $\eta = N \coh T / \semi T$.
All other simulation parameters are averaged over, subject to the restriction $\coh T > 1$~day.
Plotted are the median (solid line), the 25th--75th percentile range (error bars), and the 2.5th (lower, short-dashed line) and 97.5th (upper, long-dashed line) percentiles.
}
\end{figure*}

Following~\PaperI, two mismatches $\mu\Fstat$ and $\mu$ are compared by plotting their \emph{relative error}
\begin{equation}
\label{eq:relerr-def}
\relerr{\mu\Fstat}{\mu} \equiv \frac{ \mu\Fstat - \mu }{ 0.5( \mu\Fstat + \mu ) } \,, \quad \mu\Fstat, \mu \ge 0 \,.
\end{equation}
If $\relerr{\mu\Fstat}{\mu} < 0$, then the metric mismatch $\mu$ \emph{over}-estimates the true $\calF$-statistic mismatch $\mu\Fstat$.
This implies that the metric is a conservative estimator of the $\calF$-statistic mismatch, and therefore will restrict $\mu\Fstat \le \mu\umax$ to less than the maximum metric mismatch $\mu\umax$ prescribed by the template bank.
If $\relerr{\mu\Fstat}{\mu} > 0$, then $\mu$ \emph{under}-estimates $\mu\Fstat$; this implies that the metric is an optimistic estimator of the $\calF$-statistic mismatch, and therefore may allow $\mu\Fstat > \mu\umax$.
The behavior of the relative error can depend on how the parameter-space offsets, e.g.\ $\sig\diff\semivec\lambda$, are generated; see Figure~1 of \PaperI.
If, for example, the ellipsoid $\sig\diff\semivec\lambda \cdot \semimat g\Fstat(\vec\calA, \sigvec\lambda) \cdot \sig\diff\semivec\lambda \le \semi\mu\umax$ (where $\semimat g\Fstat(\vec\calA, \sigvec\lambda) = \sums \cohsmat g\Fstat(\vec\calA, \sigvec\lambda)$) differs in orientation from the ellipsoid $\sig\diff\semivec\lambda \cdot \semimat g \cdot \sig\diff\semivec\lambda \le \semi\mu\umax$, from which $\sig\diff\semivec\lambda$ are sampled, this will tend to $\relerr{\semi\mu\Fstat\unoint}{\semi\mu\unoint} < 0$.

Figures~\ref{fig:mu_re_twoF_rssky_TN} and~\ref{fig:mu_re_twoF_rssky_mueta} plot the relative error between $\semi\mu\Fstat\unoint$ and $\semi\mu\unoint$, $\relerr{\semi\mu\Fstat\unoint}{\semi\mu\unoint}$, and between $\semi\mu\Fstat$ and $\semi\mu$, $\relerr{\semi\mu\Fstat}{\semi\mu}$.
Overall, the metric $\semimat g$ is a good predictor of the $\calF$-statistic mismatch: with no interpolation, the average median $|\relerr{\semi\mu\Fstat\unoint}{\semi\mu\unoint}| \sim 0.20$, and the average 25th--75th percentile range $\sim 0.04$; with interpolation, the average median $|\relerr{\semi\mu\Fstat}{\semi\mu}| \sim 0.35$, and the average 25th--75th percentile range $\sim 0.03$.
The magnitudes of the relative errors with interpolation are generally larger than those without interpolation, due to the
mismatch between $\sigvec\lambda$ and $\cohsvec\lambda(\semivec\lambda)$ being generally larger than that between $\semivec\lambda$ and $\cohsvec\lambda(\semivec\lambda)$; it was observed in~\PaperI\ that the predictions of the metric worsen at higher mismatches.
The relative error magnitudes are mostly indifferent to changes in the various simulation parameters plotted in Fig.~\ref{fig:mu_re_twoF_rssky_mueta}.

In general, the metric $\semimat g$ over-estimates the $\calF$-statistic mismatch.
The dominant contribution to this effect is the Taylor expansion in Eq.\eqref{eq:coh-Fstat-mismatch-metric} of the $\calF$-statistic mismatch, used in deriving the metric $\cohsmat g\Fstat(\vec\calA, \sigvec\lambda)$ of which $\cohsmat g$ is an approximation; since $\semimat g$ is a function of $\cohsmat g$ by Eq.~\eqref{eq:semi-not-coh-avg}, it assumes the same approximation.
Since this expansion is of second order, the metric mismatch $\semi\mu$ increases quadratically (see e.g.\ Figs.~11--12 in~\cite{Prix.2007a}), while the $\calF$-statistic mismatch $\semi\mu\Fstat$ is known to increase at a slower rate (see e.g.\ Fig.~10 in~\cite{Prix.2007a}, and Fig.~7 in~\PaperI).
In addition, $\semi\mu$ is unbounded, while $\semi\mu\Fstat$ is by construction restricted to $\semi\mu\Fstat \le 1$.
This general behavior is common to all parameter-space metrics based on a second-order Taylor expansion; see e.g.\ \cite{Prix.2007a,Keppel.2012a}, and \PaperI.
In addition, as noted above, the behavior of the relative error itself is biased to reporting over-estimation when the orientation of the metric ellipsoid of $\semimat g$ differs from that of $\semimat g\Fstat(\vec\calA, \sigvec\lambda)$.

Figures~\ref{fig:mu_re_twoF_rssky_TN_cohTgt1} and~\ref{fig:mu_re_twoF_rssky_mueta_cohTgt1} plot the same quantities as Figs.~\ref{fig:mu_re_twoF_rssky_TN} and~\ref{fig:mu_re_twoF_rssky_mueta}, but subject to the restriction $\coh T > 1$~day.
This restriction improves the predictions of $\semimat g$: with no interpolation, the average median $|\relerr{\semi\mu\Fstat\unoint}{\semi\mu\unoint}| \sim 0.13$, and the average 25th--75th percentile range $\sim 0.03$; with interpolation, the average median $|\relerr{\semi\mu\Fstat}{\semi\mu}| \sim 0.30$, and the average 25th--75th percentile range $\sim 0.007$.
Many of the outlying relative errors seen in Figs.~\ref{fig:mu_re_twoF_rssky_TN} and~\ref{fig:mu_re_twoF_rssky_mueta} are also removed by this restriction: compare the 2.5th percentile lines in Figs.~\ref{fig:mu_re_twoF_rssky_semiT_cohTgt1_noi} and~\ref{fig:mu_re_twoF_rssky_semiT_cohTgt1_tot} vs.\ Figs.~\ref{fig:mu_re_twoF_rssky_semiT_noi} and~\ref{fig:mu_re_twoF_rssky_semiT_tot}; in Figs.~\ref{fig:mu_re_twoF_rssky_N_cohTgt1_noi} and~\ref{fig:mu_re_twoF_rssky_N_cohTgt1_tot} vs.\ Figs.~\ref{fig:mu_re_twoF_rssky_N_noi} and~\ref{fig:mu_re_twoF_rssky_N_tot}; and Figs.~\ref{fig:mu_re_twoF_rssky_eta_cohTgt1_noi} and~\ref{fig:mu_re_twoF_rssky_eta_cohTgt1_tot} vs.\ Figs.~\ref{fig:mu_re_twoF_rssky_eta_noi} and~\ref{fig:mu_re_twoF_rssky_eta_tot}.

As noted in \PaperI, the predictions of $\cohsmat g$ are worse at $\coh T = 1$~day for two reasons.
First, the phase metric $\cohsmat g\phase(\sigvec\lambda)$ approximates $\cohsmat g\Fstat(\vec\calA, \sigvec\lambda)$ in Eq.~\eqref{eq:coh-phase-metric-def} in part by neglecting the amplitude modulation of the signal of a gravitational-wave pulsar; since these modulations arise primarily from the rotation of the Earth, this approximation is worst at $\coh T \sim 1$~day.
Second, the numerical ill-conditionedness of the phase metric (see~\cite{Prix.2007a} and \PaperI) increases the difficulty of computing $\cohsmat g\unc$, from which both $\cohsmat g$ and $\semimat g$ are derived, at smaller $\coh T$.

One feature of interest is the 2.5th percentile lines in Figs.~\ref{fig:mu_re_twoF_rssky_semiT_cohTgt1_tot} and~\ref{fig:mu_re_twoF_rssky_N_cohTgt1_tot}.
At this line, $\relerr{\semi\mu\Fstat}{\semi\mu} \sim -2$ at $\semi T \sim 10$~days and $N \sim 3$, but reduces to $\relerr{\semi\mu\Fstat}{\semi\mu} \sim -1$ for $\semi T \gtrsim 40$~days and $N \gtrsim 6$.
The general decrease in $\relerr{\semi\mu\Fstat}{\semi\mu}$ is simply due to the larger number of segments $N$ (since $\semi T$ is approximately $\propto N$), which mitigates the effect of a large over-estimation of the $\calF$-statistic mismatch in any one segment.
(The momentary increase in $\relerr{\semi\mu\Fstat}{\semi\mu}$ along the 2.5th percentile line in Fig.~\ref{fig:mu_re_twoF_rssky_semiT_cohTgt1_tot} for $30 \lesssim \semi T / \textrm{days} \lesssim 60$ is explained in the next section.)

The jumps in the relative errors seen in Figs.~\ref{fig:mu_re_twoF_rssky_N_noi}, \ref{fig:mu_re_twoF_rssky_N_tot}, \ref{fig:mu_re_twoF_rssky_N_cohTgt1_noi}, and~\ref{fig:mu_re_twoF_rssky_N_cohTgt1_tot} -- most noticeably in the 2.5th and 97.5th percentile lines -- are artifacts of plotting the 72 search setups with respect to $N$.
For example, note that in Fig.~\ref{fig:mu_re_twoF_rssky_N_tot} the magnitude and spread of the relative error is noticeably less at $N = 8$ than at either $N = 7$ or $N = 9$.
From Table~\ref{tab:segment_list_table}, the only setup with $N = 8$ is that with $\semi T = 120$~days, while the setups at $N = 7$ and $N = 9$ have $\semi T \le 49$~days and $\semi T \le 63$~days respectively.
Given that the metric predictions improve with $\semi T$, it is expected that the relative error at $N = 8$ is reduced relative to $N = 7$ or $N = 9$.

\subsubsection{Simulation results: metric under-estimation}\label{sec:simul-results:-metr}

\begin{figure}
\includegraphics[width=\linewidth]{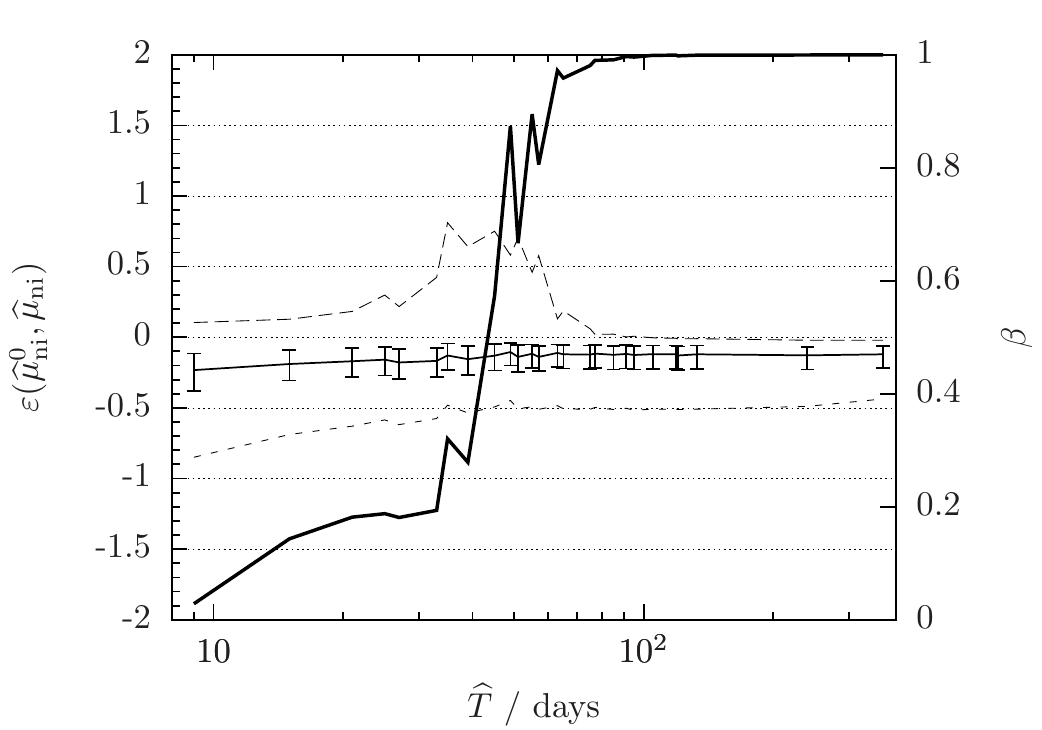}
\caption{\label{fig:mu_re_twoF_rssky_semiT_cohTgt1_noi_orient}
Relative errors $\relerr{\semi\mu\Fstat\unoint}{\semi\mu\unoint}$ (left y-axis) at constant total time-span $\semi T$, as in Fig.~\protect\ref{fig:mu_re_twoF_rssky_semiT_cohTgt1_noi}.
Overplotted in black is the orientation angle $\beta$ (right y-axis) of the plane of the reduced supersky coordinates $(\coh n\ua, \coh n\ub)$.
}
\end{figure}

\begin{figure*}
\subfloat[]{\includegraphics[width=0.49\linewidth]{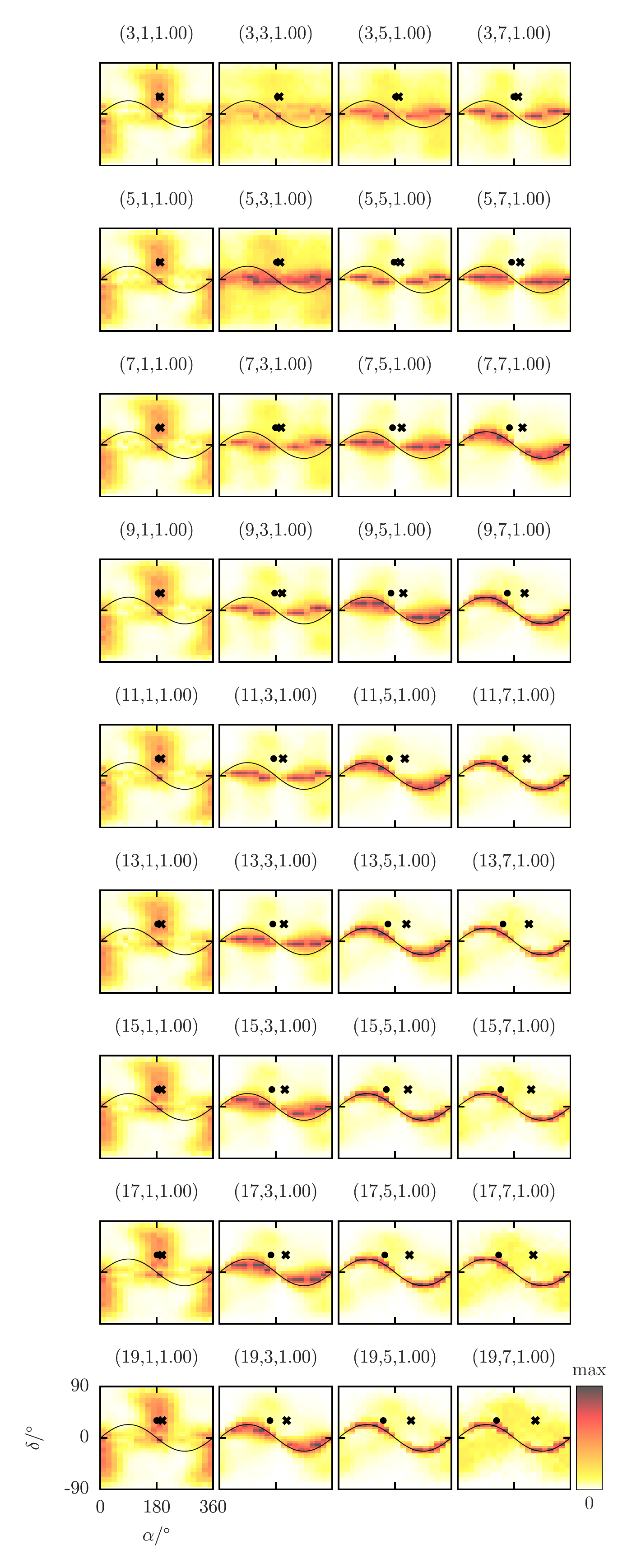}\label{fig:mu_underest_skymap_sseg_noi}}
\subfloat[]{\includegraphics[width=0.49\linewidth]{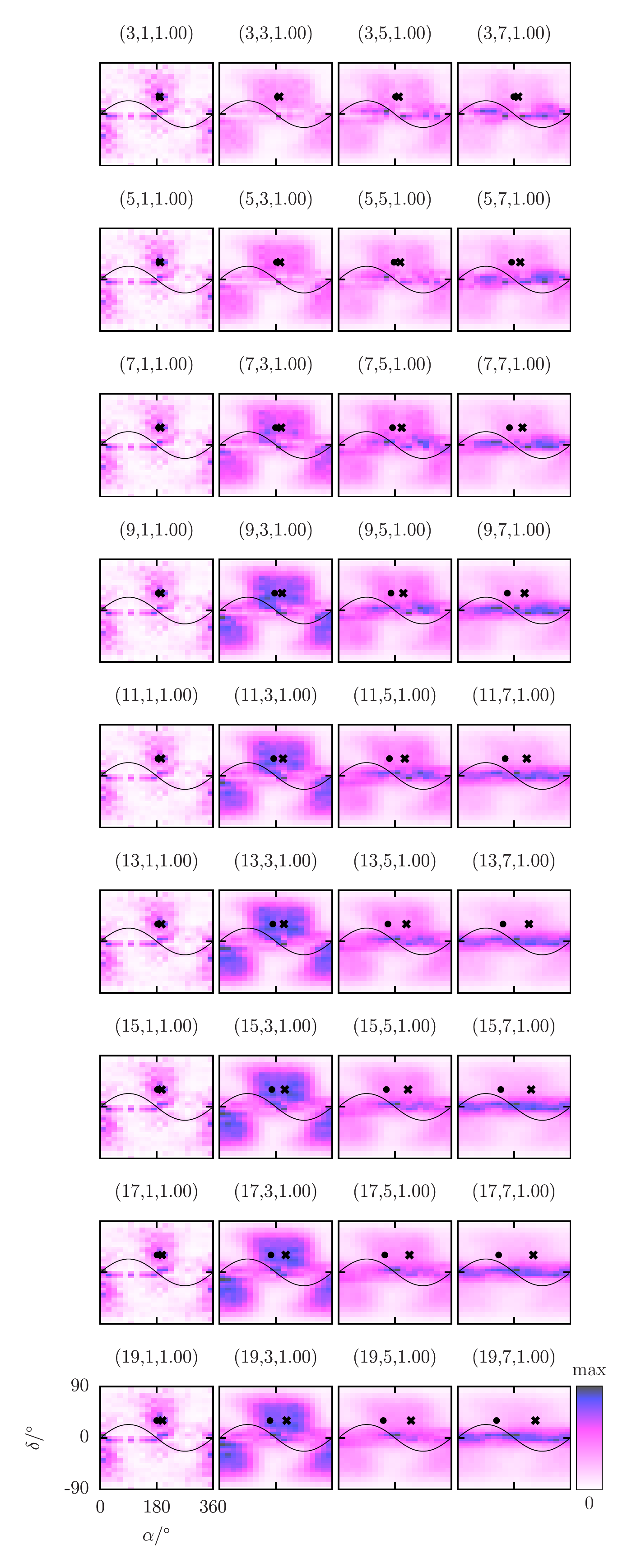}\label{fig:mu_underest_skymap_sseg_tot}}
\caption{\label{fig:mu_underest_skymap_sseg}
Intensity maps of the probabilities
\protect\subref{fig:mu_underest_skymap_sseg_noi} $p \big(\relerr{\semi\mu\Fstat\unoint}{\semi\mu\unoint} > 0 \big)$ and
\protect\subref{fig:mu_underest_skymap_sseg_tot} $p \big(\relerr{\semi\mu\Fstat}{\semi\mu} > 0 \big)$,
as functions of right ascension $\alpha$ and declination $\delta$, for the 36 setups listed in the top half of Table~\ref{tab:segment_list_table}.
The remaining simulation parameters, $f$ and $(\coh\mu\umax, \semi\mu\umax)$, are averaged over.
Overplotted are the ecliptic equator (black line), and the positions of the LIGO Livingston detector at the start times of the first (black dot) and last (black cross) segments.
Each map is titled by the triplet ($N$, $\coh T$ / days, $\eta$).
The maps are normalized such that the total probability in each map is either
\protect\subref{fig:mu_underest_skymap_sseg_noi} $10^{-S^{0}\unoint}$ or
\protect\subref{fig:mu_underest_skymap_sseg_tot} $10^{-S^{0}}$,
where $S^{0}\unoint$ and $S^{0}$ are given for each setup in Table~\ref{tab:segment_list_table}.
}
\end{figure*}

\begin{figure*}
\subfloat[]{\includegraphics[width=0.49\linewidth]{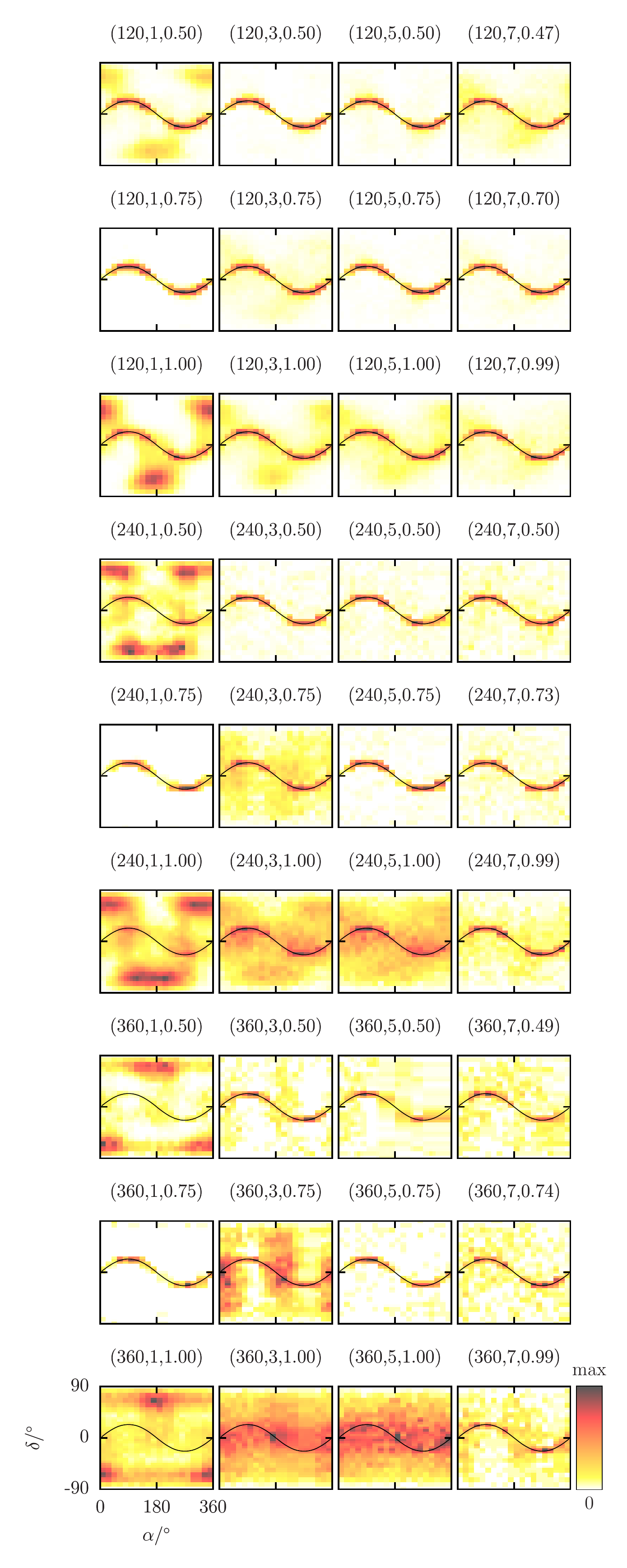}\label{fig:mu_underest_skymap_noi}}
\subfloat[]{\includegraphics[width=0.49\linewidth]{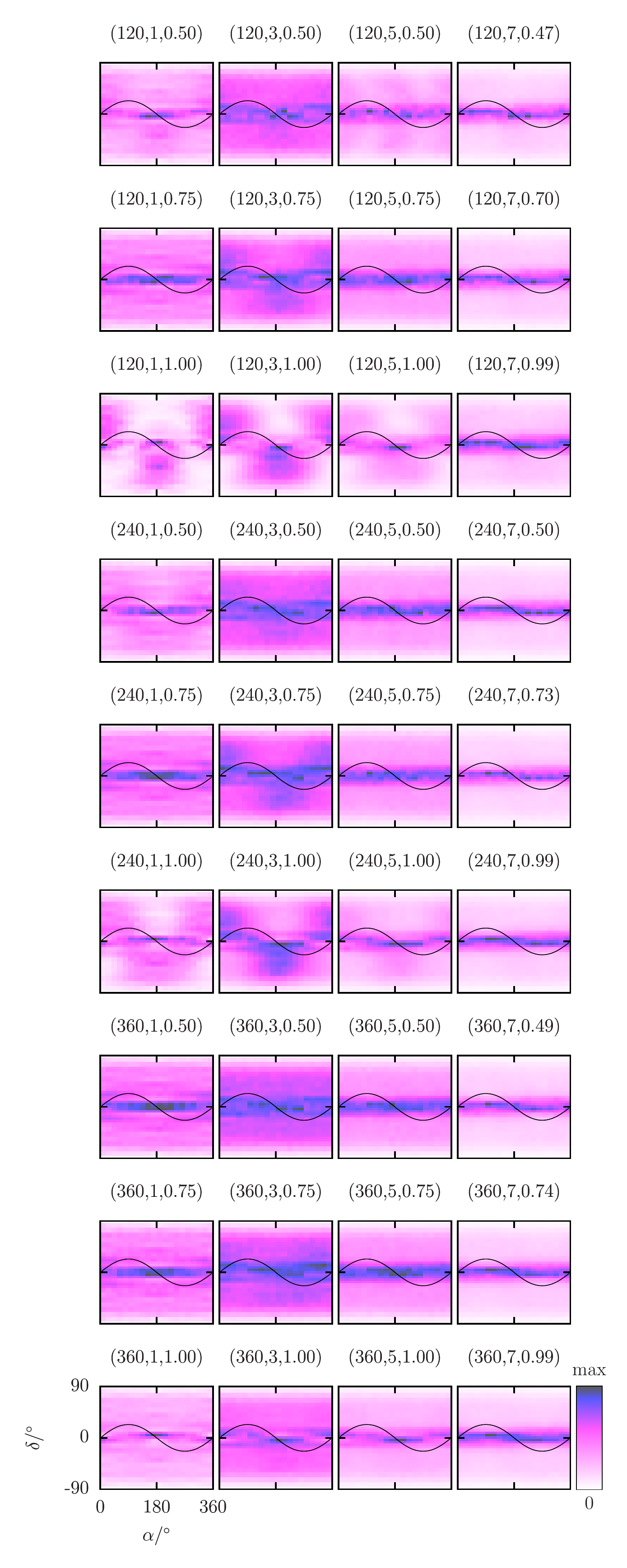}\label{fig:mu_underest_skymap_tot}}
\caption{\label{fig:mu_underest_skymap}
Intensity maps of the probabilities
\protect\subref{fig:mu_underest_skymap_sseg_noi} $p \big(\relerr{\semi\mu\Fstat\unoint}{\semi\mu\unoint} > 0 \big)$ and
\protect\subref{fig:mu_underest_skymap_sseg_tot} $p \big(\relerr{\semi\mu\Fstat}{\semi\mu} > 0 \big)$,
as functions of right ascension $\alpha$ and declination $\delta$, for the 36 setups listed in the bottom half of Table~\ref{tab:segment_list_table}.
The remaining simulation parameters, $f$ and $(\coh\mu\umax, \semi\mu\umax)$, are averaged over.
Overplotted is the ecliptic equator (black line).
Each map is titled by the triplet ($\semi T$ / days, $\coh T$ / days, $\eta$).
The maps are normalized such that the total probability in each map is either
\protect\subref{fig:mu_underest_skymap_sseg_noi} $10^{-S^{0}\unoint}$ or
\protect\subref{fig:mu_underest_skymap_sseg_tot} $10^{-S^{0}}$,
where $S^{0}\unoint$ and $S^{0}$ are given for each setup in Table~\ref{tab:segment_list_table}.
}
\end{figure*}

It is also possible for the metric $\semimat g$ to under-estimate the $\calF$-statistic mismatch.
It is important to note that, when performing a real search using an implementation of a lattice template bank, as e.g.\ described in \PaperII, metric under-estimation may not necessarily lead to loss of coverage (i.e.\ areas of parameter space where $\semi\mu > \semi\mu\umax$).
For example, metric under-estimation at parameter-space boundaries may be compensated for by additional boundary templates.
Tests of lattice template banks constructed using $\cohsmat g$ in \PaperII\ have demonstrated that near-complete coverage is attainable using the reduced supersky metric.

Table~\ref{tab:segment_list_table} lists, for each of the 72 search setups, the probability that the semicoherent mismatches $\semi\mu\Fstat$ and $\semi\mu\Fstat\unoint$ (i.e.\ with and without interpolation) will under-estimate the respective $\calF$-statistic mismatches.
The probabilities are given in logarithmic form by $S^{x} \equiv -\log_{10} p \big(\relerr{\semi\mu\Fstat}{\semi\mu} > x \big)$ and $S^{x}\unoint \equiv -\log_{10} p \big(\relerr{\semi\mu\Fstat\unoint}{\semi\mu\unoint} > x \big)$, where $x \in \{ 0, 0.5 \}$.
We observe that the probability of under-estimation for any search setup is $\lesssim 10$\% (i.e. $S^{x} \gtrsim 1$), and decreases with increasing number of segments $N$ and coherent time-span $\coh T$, and with the addition of interpolation.
This is expected: increasing $N$ reduces the potential effect of $\semimat g$ under-estimating the $\calF$-statistic mismatch in a few segments, as does the addition of interpolation which averages the coherent and semicoherent mismatches; and Figs.~\ref{fig:mu_re_twoF_rssky_semiT_noi} and~\ref{fig:mu_re_twoF_rssky_semiT_tot} demonstrate that $\semimat g$ better approximates the $\calF$-statistic at larger $\coh T$.

In Figures~\ref{fig:mu_re_twoF_rssky_semiT_noi} and~\ref{fig:mu_re_twoF_rssky_semiT_cohTgt1_noi}, we can see that, at the 97.5th percentile line, $\semi\mu\unoint$ noticeably under-estimates $\semi\mu\Fstat\unoint$ for $30 \lesssim \semi T / \textrm{days} \lesssim 60$.
This is due to a known feature of the reduced supersky metric; see Section~V~D of \PaperI.
As the time $\semi T$ spanned by the metric $\semimat g$ increases, the orientation of the plane of the reduced supersky coordinates $(\semi n\ua, \semi n\ub)$ transitions from being perpendicular to the equatorial $z$-axis to being perpendicular to the ecliptic $Z$ axis; in other words, the $(\semi n\ua, \semi n\ub)$ transition from ``equatorial-like'' coordinates to ``ecliptic-like'' coordinates.
During this transition, the reduced supersky metric is more likely to under-estimate the $\calF$-statistic mismatch; compare Fig.~\ref{fig:mu_re_twoF_rssky_semiT_noi} and~\ref{fig:mu_re_twoF_rssky_semiT_cohTgt1_noi} to Fig.~14 of \PaperI.

To illustrate the transition, Figure~\ref{fig:mu_re_twoF_rssky_semiT_cohTgt1_noi_orient} overplots Fig.~\ref{fig:mu_re_twoF_rssky_semiT_cohTgt1_noi} with the orientation angle $\beta$ (given by Eq.~(51) of \PaperI) of the plane of the reduced supersky coordinates $(\semi n\ua, \semi n\ub)$, normalized such that $\beta = 0$ signifies an ``equatorial-like'' orientation and $\beta = 1$ an ``ecliptic-like'' orientation.
The transition from ``equatorial-like'' to ``ecliptic-like'' coordinates occurs between $\semi T \sim 30$~days and 60~days, the same domain where $\semi\mu\unoint$ noticeably under-estimates $\semi\mu\Fstat\unoint$.
The momentary increase in $\relerr{\semi\mu\Fstat}{\semi\mu}$ along the 2.5th percentile line in Fig.~\ref{fig:mu_re_twoF_rssky_semiT_cohTgt1_tot} also occurs when $30 \lesssim \semi T / \textrm{days} \lesssim 60$, and therefore is also likely due to the transition.

To determine whether metric under-estimation is correlated with particular regions of parameter space, we identify pulsar sky positions, in terms of right ascension $\alpha$ and declination $\delta$, where the metric $\semimat g$ under-estimates the $\calF$-statistic mismatch.
Figures~\ref{fig:mu_underest_skymap_sseg} and~\ref{fig:mu_underest_skymap} map the probability of metric underestimation as a function of $(\alpha, \delta)$, for each setup given in Table~\ref{tab:segment_list_table}.
The maps are normalized such that the total probability in each map is either $10^{-S^{0}\unoint}$ without interpolation (Figs.~\ref{fig:mu_underest_skymap_sseg_noi} and~\ref{fig:mu_underest_skymap_noi}), or $10^{-S^{0}}$ with interpolation (Figs.~\ref{fig:mu_underest_skymap_sseg_tot} and~\ref{fig:mu_underest_skymap_tot}), where $S^{0}\unoint$ and $S^{0}$ are given for each setup in Table~\ref{tab:segment_list_table}.
Two effects, which cause metric under-estimation to be concentrated in particular regions of the sky, are identified and explained in the following paragraphs.

First, for short segment time-spans $\coh T \lesssim 3$~days, regions of the sky which (anti-)correlate with the position of the detector used in the numerical simulations -- the LIGO Livingston detector -- are more likely to result in metric under-estimation; see Figure~\ref{fig:mu_underest_skymap_sseg}.
This is likely due to the neglect of the amplitude modulation of the gravitational-wave pulsar signal in deriving the phase metric $\cohsmat g\phase(\sigvec\lambda)$, as noted above in Section~\ref{sec:simul-results:-relat}.
Without interpolation (Fig.~\ref{fig:mu_underest_skymap_sseg_noi}), where only the semicoherent metric $\semimat g$ is involved, the effect appears at $\coh T \sim 1$~day and generally fades for longer $\coh T$ and more $N$; with interpolation (Fig.~\ref{fig:mu_underest_skymap_sseg_tot}), where both $\semimat g$ and the coherent metrics $\cohsmat g$ are involved, the effect is most noticeable at $\coh T \sim 3$~days.
It is probable that the effect of neglecting the amplitude modulation becomes irrelevant more quickly for the semicoherent metric as $N$ increases and more coherent metrics are averaged together.

Second, at longer $\coh T$, the probability of metric under-estimation becomes concentrated at either the ecliptic equator (without interpolation), or at the equatorial equator $\delta = 0$ (without interpolation); see Figures~\ref{fig:mu_underest_skymap_sseg} and~\ref{fig:mu_underest_skymap}.
This is likely due to the transition of the reduced supersky coordinates from ``equatorial-like'' coordinates to ``ecliptic-like'' coordinates, which begins at $\semi T \gtrsim 30$~days for the semicoherent metric (see Section~\ref{sec:simul-results:-relat} and Fig.~\ref{fig:mu_re_twoF_rssky_semiT_cohTgt1_noi_orient}), and at $\coh T \gtrsim 10$~days for the coherent metric (see Section~V~D of \PaperI).
In Figs.~\ref{fig:mu_underest_skymap_sseg_noi} and~\ref{fig:mu_underest_skymap_noi} (i.e.\ without interpolation), only the semicoherent metric $\semimat g$ is in use, and the maps are generally consistent with a transition from ``equatorial-like'' to ``ecliptic-like'' coordinates on the timescale $\semi T \gtrsim 30$~days.
In Figs.~\ref{fig:mu_underest_skymap_sseg_tot} and~\ref{fig:mu_underest_skymap_tot} (i.e.\ with interpolation), $\semimat g$ contributes only once to the total metric mismatch, whereas the coherent metrics $\cohsmat g$ contribute $N \gg 1$ times; see the vector summation in Fig.~\ref{fig:mismatch_diagram}.
The maps therefore tend to reflect the behavior of the coherent metric, which for $\coh T \le 7$~days remains ``equatorial-like''.

\subsubsection{Simulation results: mismatch histograms}\label{sec:simul-results:-mism}

\begin{figure*}
\subfloat[]{\includegraphics[width=\linewidth]{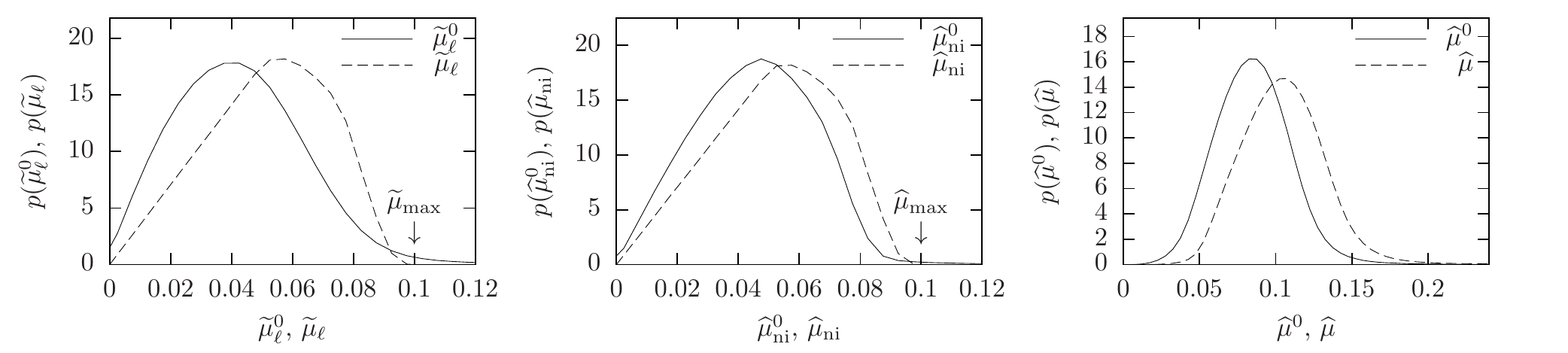}\label{fig:mu_hgrm_twoF_rssky_coh0p1_semi0p1}}\\
\subfloat[]{\includegraphics[width=\linewidth]{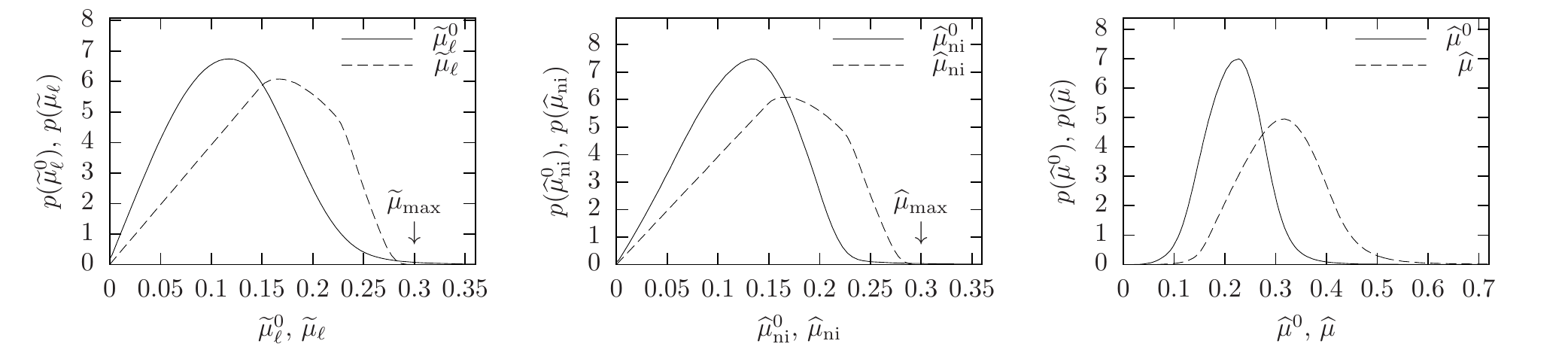}\label{fig:mu_hgrm_twoF_rssky_coh0p3_semi0p3}}\\
\subfloat[]{\includegraphics[width=\linewidth]{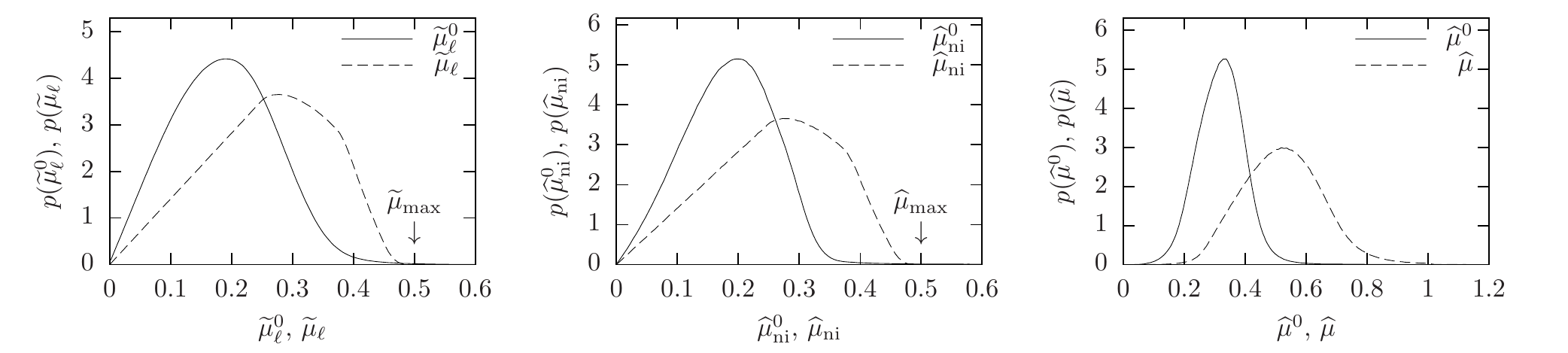}\label{fig:mu_hgrm_twoF_rssky_coh0p5_semi0p5}}\\
\subfloat[]{\includegraphics[width=\linewidth]{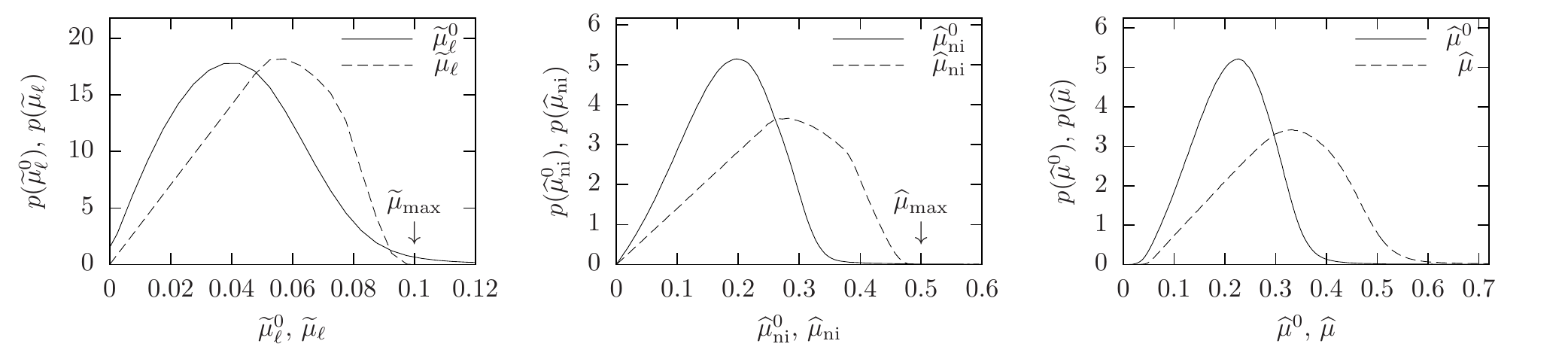}\label{fig:mu_hgrm_twoF_rssky_coh0p1_semi0p5}}\\
\subfloat[]{\includegraphics[width=\linewidth]{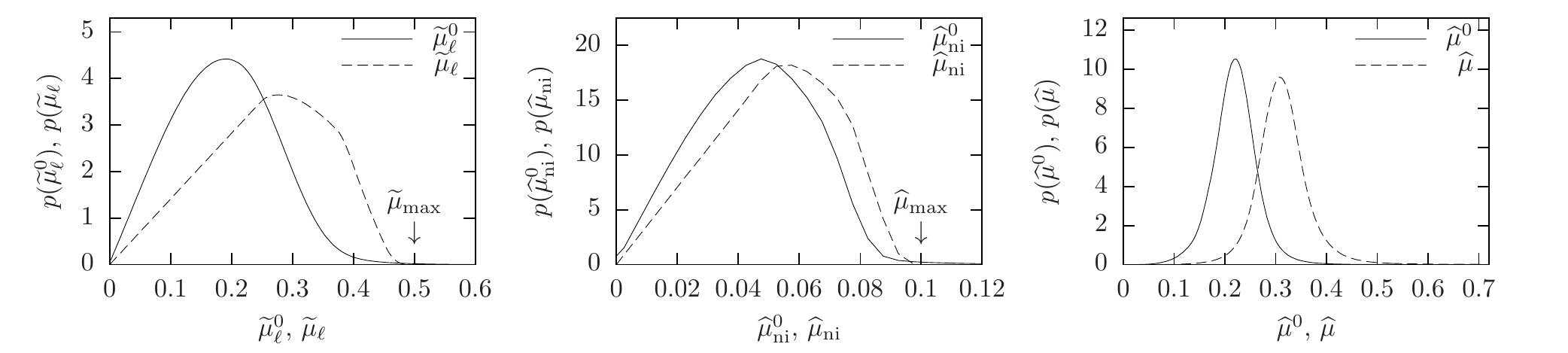}\label{fig:mu_hgrm_twoF_rssky_coh0p5_semi0p1}}
\caption{\label{fig:mu_hgrm_twoF_rssky}
Histograms in mismatches in: the coherent mismatches in each segment $\cohs\mu\Fstat$, $\cohs\mu$ (left column); the semicoherent mismatches with no interpolation $\semi\mu\Fstat\unoint$, $\semi\mu\unoint$ (middle column); and the semicoherent mismatches with interpolation $\semi\mu\Fstat$, $\semi\mu$ (right column).
Histograms are at fixed maximum template-bank mismatches $(\coh\mu\umax, \semi\mu\umax)$ equal to
\protect\subref{fig:mu_hgrm_twoF_rssky_coh0p1_semi0p1} $(0.1,0.1)$,
\protect\subref{fig:mu_hgrm_twoF_rssky_coh0p3_semi0p3} $(0.3,0.3)$,
\protect\subref{fig:mu_hgrm_twoF_rssky_coh0p5_semi0p5} $(0.5,0.5)$,
\protect\subref{fig:mu_hgrm_twoF_rssky_coh0p1_semi0p5} $(0.1,0.5)$,
\protect\subref{fig:mu_hgrm_twoF_rssky_coh0p5_semi0p1} $(0.5,0.1)$.
All other simulation parameters are averaged over.
}
\end{figure*}

Figure~\ref{fig:mu_hgrm_twoF_rssky} plots histograms of the coherent mismatches in each segment, $\cohs\mu\Fstat$, $\cohs\mu$, and the semicoherent mismatches with and without interpolation, $\semi\mu\Fstat$, $\semi\mu$ and $\semi\mu\Fstat\unoint$, $\semi\mu\unoint$ respectively.
The coherent mismatches reproduce the results obtained in Fig.~4 of~\PaperII.
The histograms again demonstrate the conservative over-estimation of the metric approximation, in that the metric mismatch histograms (dashed lines in Fig.~\ref{fig:mu_hgrm_twoF_rssky}) are peaked at higher mismatches than the $\calF$-statistic mismatch histograms (solid lines).

The shape of the histograms in $\cohs\mu$ and $\semi\mu\unoint$ in Figure~\ref{fig:mu_hgrm_twoF_rssky} conform to that expected for four-dimensional template banks constructed from an $\Ans[4]$ lattice; see Fig.~1 of~\PaperII.
The shape of the histograms in $\semi\mu$, to which both the coherent and semicoherent template bank mismatches contribute, is closer to a Gaussian in shape; this is likely due to the central limit theorem applying when the $N$ coherent mismatches $\cohs\mu\Fstat\big( \semivec\lambda; \cohsvec\lambda(\semivec\lambda) \big)$ are averaged in Eq.~\eqref{eq:semi-Fstat-mismatch-pure-interp}, assuming $N$ is sufficiently large.
Note that, however, the histogram in $\semi\mu$ for $(\coh\mu\umax, \semi\mu\umax) = (0.1,0.5)$ (Fig.~\ref{fig:mu_hgrm_twoF_rssky_coh0p1_semi0p5}) instead more closely resembles that of $\semi\mu\unoint$, as might be expected since $\coh\mu\umax \ll \semi\mu\umax$, i.e.\ the coherent template bank mismatches $\cohs\mu$ contribute little to the total mismatch $\semi\mu$.

\section{Comparisons to previous work}\label{sec:comp-prev-work}

\begin{figure*}
\subfloat[]{\includegraphics[width=0.49\linewidth]{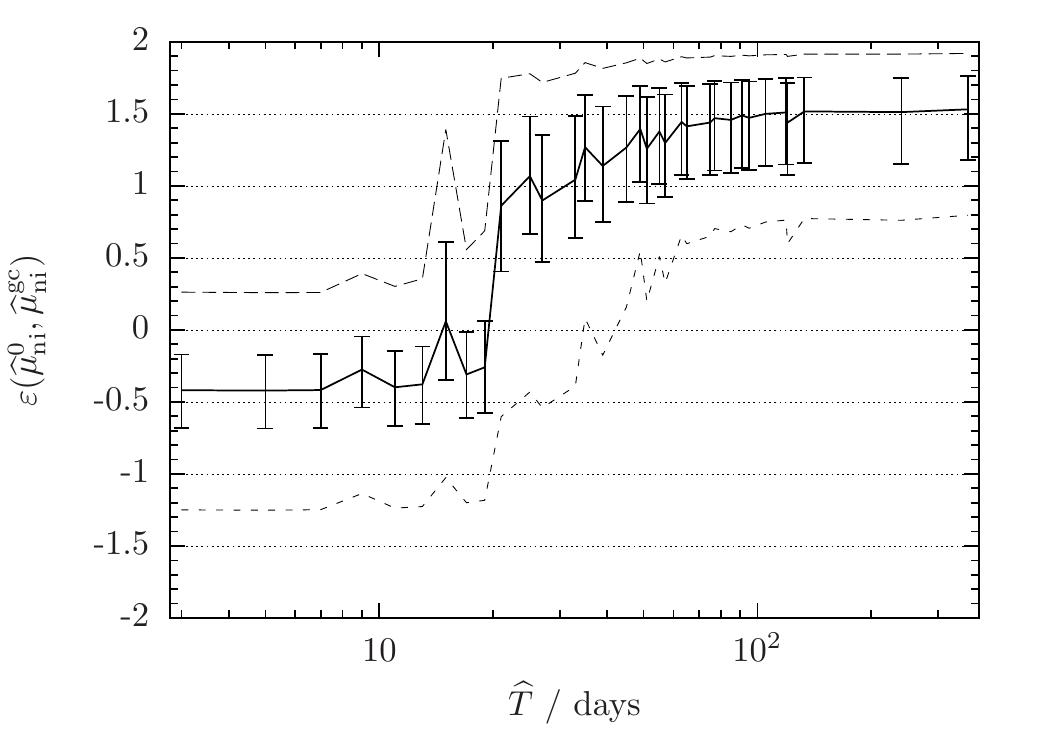}\label{fig:mu_re_twoF_gctco_semiT_noi}}
\subfloat[]{\includegraphics[width=0.49\linewidth]{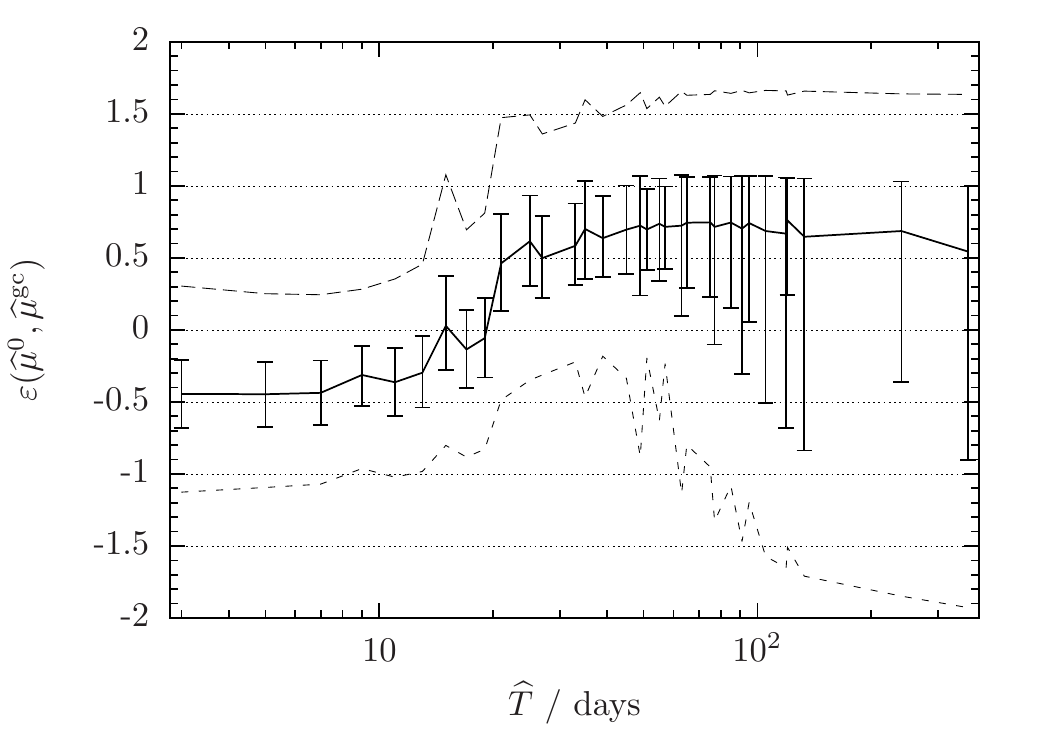}\label{fig:mu_re_twoF_gctco_semiT_tot}}\\
\subfloat[]{\includegraphics[width=0.49\linewidth]{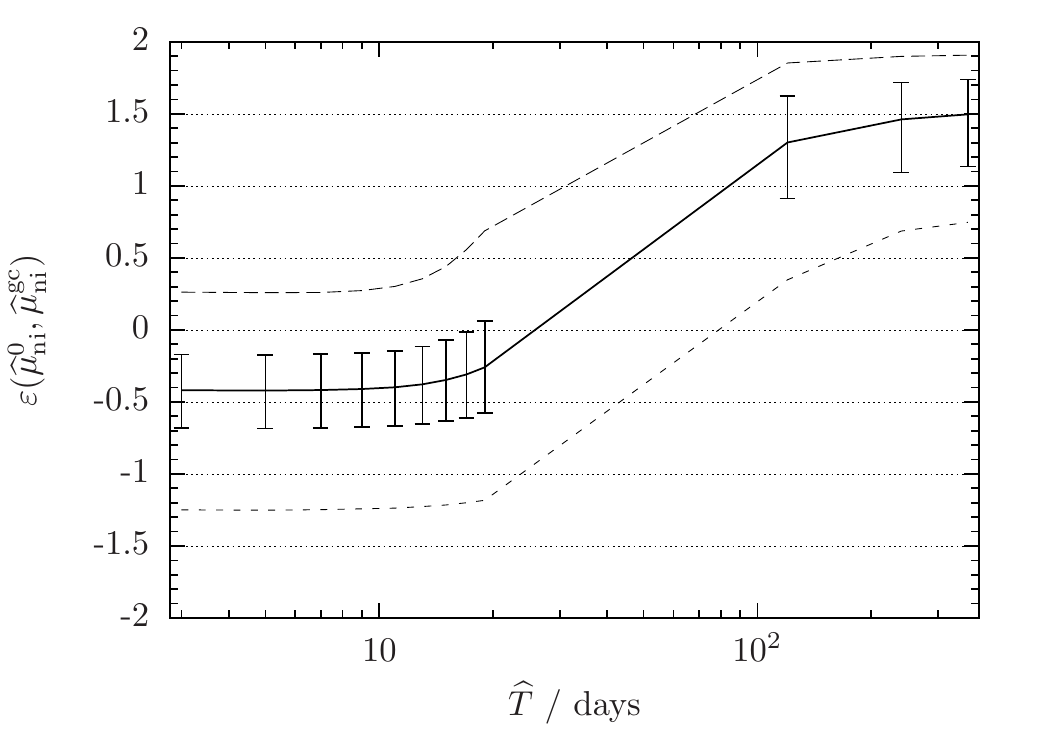}\label{fig:mu_re_twoF_gctco_semiT_cohTeq1_dutyeq1_noi}}
\subfloat[]{\includegraphics[width=0.49\linewidth]{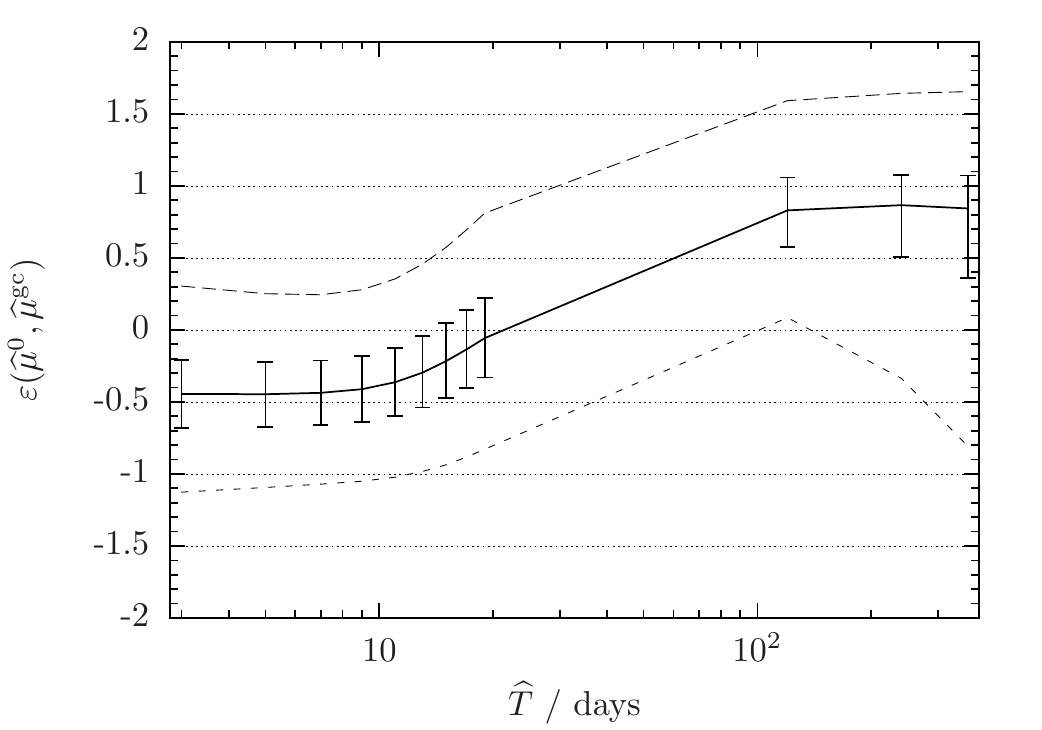}\label{fig:mu_re_twoF_gctco_semiT_cohTeq1_dutyeq1_tot}}\\
\subfloat[]{\includegraphics[width=0.49\linewidth]{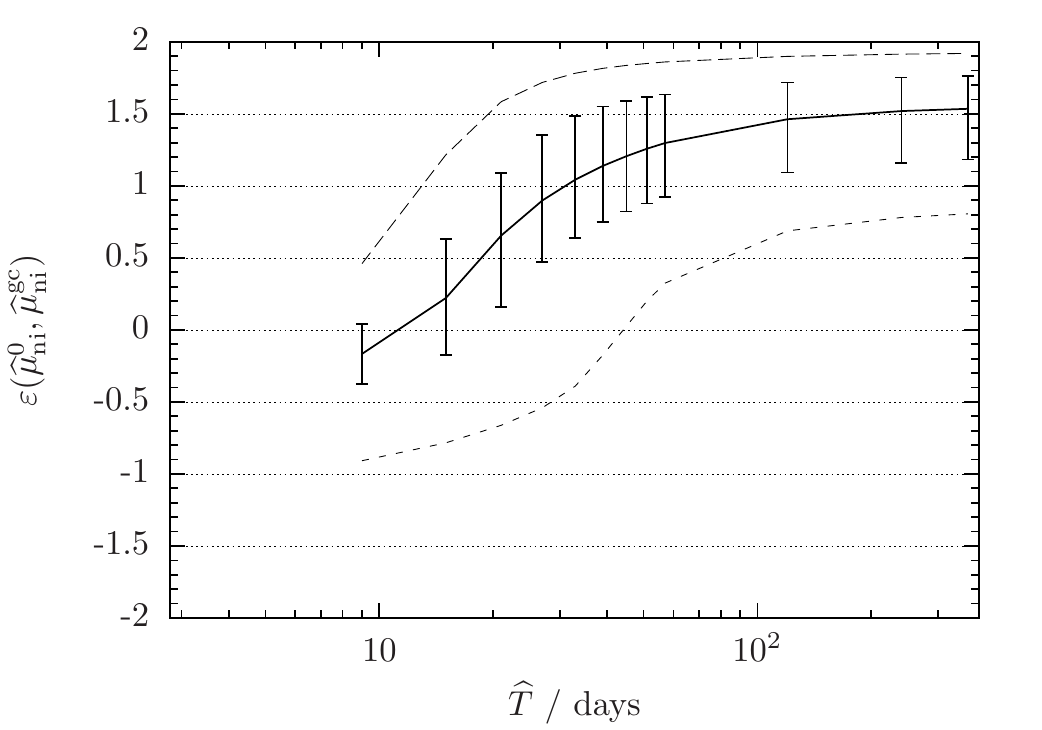}\label{fig:mu_re_twoF_gctco_semiT_cohTeq3_dutyeq1_noi}}
\subfloat[]{\includegraphics[width=0.49\linewidth]{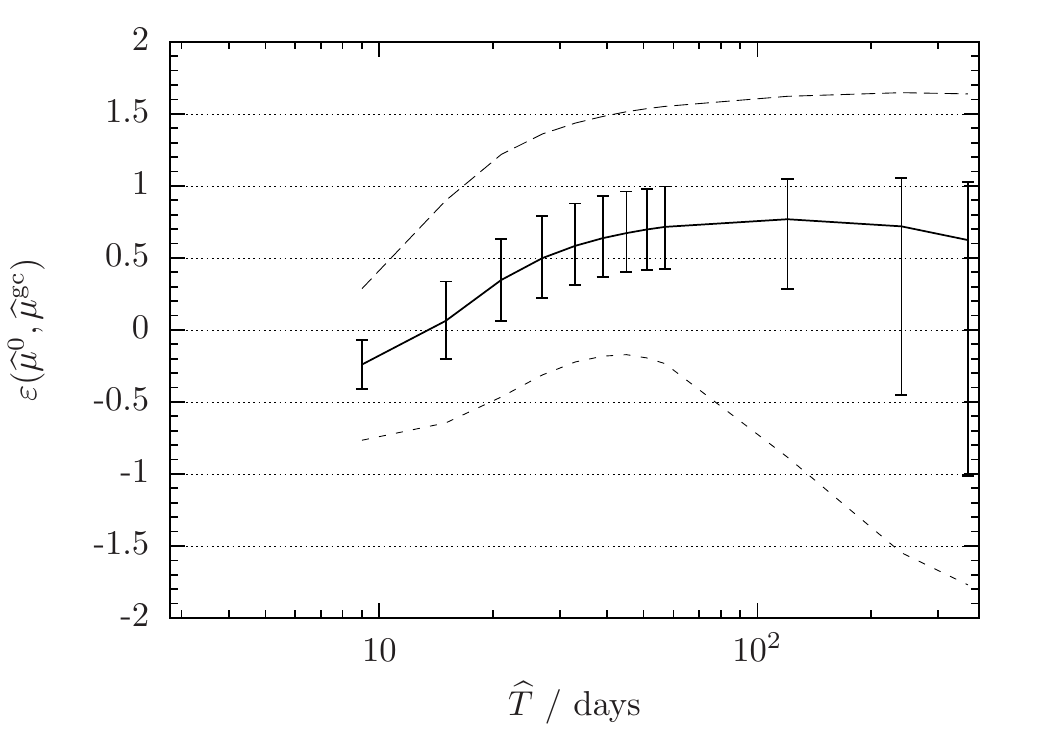}\label{fig:mu_re_twoF_gctco_semiT_cohTeq3_dutyeq1_tot}}
\caption{\label{fig:mu_re_twoF_gctco_semiT}
Relative errors $\relerr{\semi\mu\Fstat\unoint}{\semi\mu\gct\unoint}$ (left column) and $\relerr{\semi\mu\Fstat}{\semi\mu\gct}$ (right column), at constant total time-span $\semi T$.
All other simulation parameters are averaged over, subject to the following restrictions on the search setups listed in Table~\ref{tab:segment_list_table}: \protect\subref{fig:mu_re_twoF_gctco_semiT_noi}\protect\subref{fig:mu_re_twoF_gctco_semiT_tot}~no restrictions; \protect\subref{fig:mu_re_twoF_gctco_semiT_cohTeq1_dutyeq1_noi}\protect\subref{fig:mu_re_twoF_gctco_semiT_cohTeq1_dutyeq1_tot}~$\coh T = 1$~day, $\eta = 1$; and \protect\subref{fig:mu_re_twoF_gctco_semiT_cohTeq3_dutyeq1_noi}\protect\subref{fig:mu_re_twoF_gctco_semiT_cohTeq3_dutyeq1_tot}~$\coh T = 3$~days, $\eta = 1$.
Plotted are the median (solid line), the 25th--75th percentile range (error bars), and the 2.5th (lower, short-dashed line) and 97.5th (upper, long-dashed line) percentiles.
}
\end{figure*}

Section~\ref{sec:comp-glob-corr} compares the semicoherent metric $\semimat g$, presented in the previous section, to the global correlation metric of~\cite{Pletsch.2010a}.
Section~\ref{sec:numb-semic-templ} compares predictions for the number of semicoherent templates given by $\semimat g$ to predictions given by~\cite{Brady.Creighton.2000a,Pletsch.2010a}.

\begin{figure}
\includegraphics[width=\linewidth]{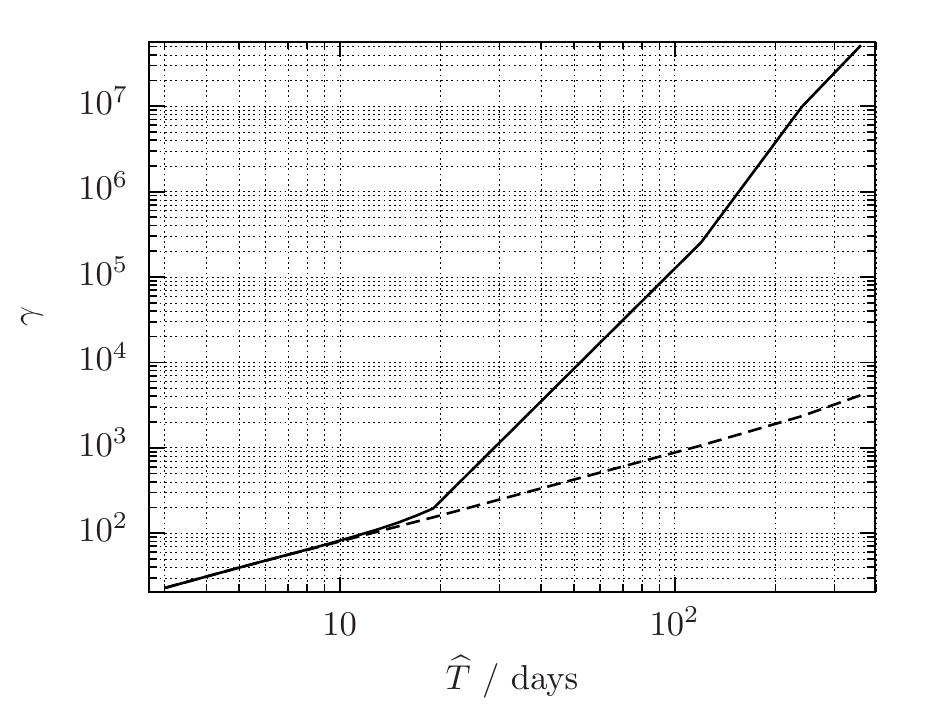}
\caption{\label{fig:refinement_factor}
Refinement of the supersky metric given by Eq.~\eqref{eq:refinement-factor-def} (solid line), and of the global correlation metric given by~\cite{Pletsch.2010a} (dashed line), as a function of total time-span $\semi T$.
The search setups listed in Table~\ref{tab:segment_list_table} are restricted to $\coh T = 1$~day and $\eta = 1$.
}
\end{figure}

\begin{figure}
\includegraphics[width=\linewidth]{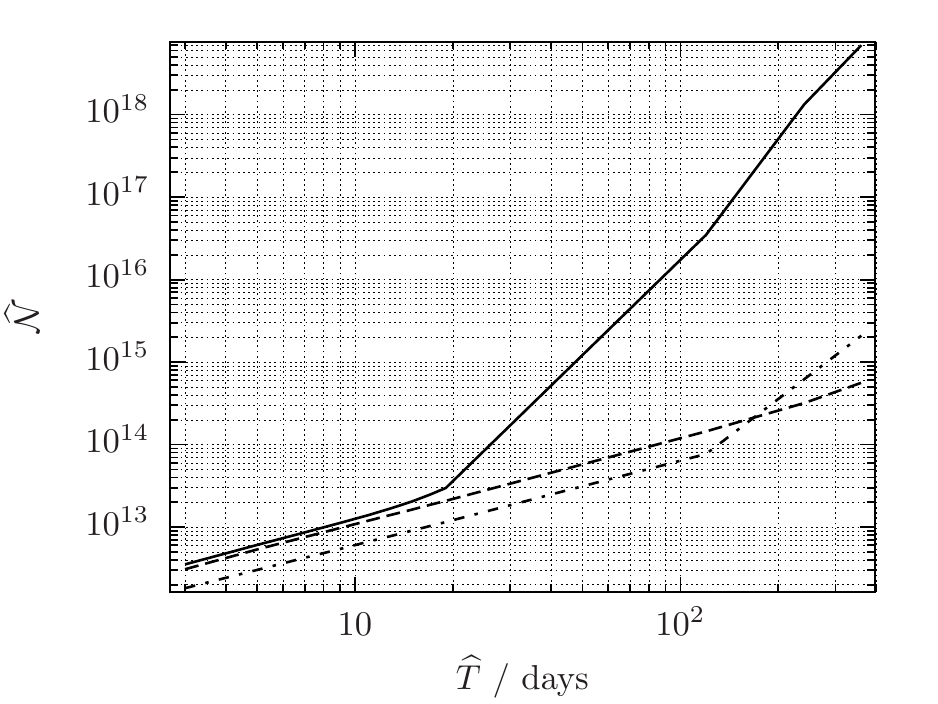}
\caption{\label{fig:number_of_templates}
Number of semicoherent templates predicted by the supersky metric (solid line), by the global correlation metric of~\cite{Pletsch.2010a} (dashed line), and by~\cite{Brady.Creighton.2000a} (dash-dotted line), as a function of total time-span $\semi T$.
The search setups listed in Table~\ref{tab:segment_list_table} are restricted to $\coh T = 1$~day and $\eta = 1$.
}
\end{figure}

\subsection{Comparison to the global correlation metric}\label{sec:comp-glob-corr}

The numerical simulations described in Section~\ref{sec:valid-using-numer} were repeated with two changes: the coherent global correlation metrics $\cohsmat g\gct$ given by Eqs.~(50) of~\cite{Pletsch.2010a} were substituted for the coherent supersky metrics $\cohsmat g$, and the semicoherent global correlation metric $\semimat g\gct$ given by Eqs.~(59) of~\cite{Pletsch.2010a} was substituted for the semicoherent supersky metric $\semimat g$.
In other words, the simulations were repeated assuming that the global correlation metrics are used to construct the coherent and semicoherent template banks in steps~\eqref{item:num-sim-nearest-semi} and~\eqref{item:num-sim-nearest-coh} of the procedure described in Section~\ref{sec:valid-using-numer}.

Figure~\ref{fig:mu_re_twoF_gctco_semiT} plots the relative error between $\semi\mu\Fstat\unoint$ and $\semi\mu\gct\unoint$, $\relerr{\semi\mu\Fstat\unoint}{\semi\mu\gct\unoint}$, and between $\semi\mu\Fstat$ and $\semi\mu\gct$, $\relerr{\semi\mu\Fstat}{\semi\mu\gct}$.
Figs.~\ref{fig:mu_re_twoF_gctco_semiT_noi} and~\ref{fig:mu_re_twoF_gctco_semiT_tot} are directly comparable to Figs.~\ref{fig:mu_re_twoF_rssky_semiT_noi} and~\ref{fig:mu_re_twoF_rssky_semiT_tot} for the supersky metrics respectively.
The global correlation metric over-estimates the $\calF$-statistic mismatch, with the median $|\relerr{\semi\mu\Fstat\unoint}{\semi\mu\gct\unoint}| \sim |\relerr{\semi\mu\Fstat}{\semi\mu\gct}| \lesssim 0.4$, provided that $\semi T \lesssim 20$~days; for $\semi T \gtrsim 20$~days, the global correlation metric severely under-estimates the $\calF$-statistic mismatch, with the median $\relerr{\semi\mu\Fstat\unoint}{\semi\mu\gct\unoint} \gtrsim 1$ and $\relerr{\semi\mu\Fstat}{\semi\mu\gct} \gtrsim 0.5$.
Figs.~\ref{fig:mu_re_twoF_gctco_semiT_cohTeq1_dutyeq1_noi}--\ref{fig:mu_re_twoF_gctco_semiT_cohTeq3_dutyeq1_tot} restrict the search setups listed in Table~\ref{tab:segment_list_table} to $\coh T \in \{1,3\}$~days and $\eta = 1$ (i.e.\ contiguous segments).
The behavior of the median relative error is similar under these restrictions: in Fig.~\ref{fig:mu_re_twoF_gctco_semiT_cohTeq1_dutyeq1_tot}, $\relerr{\semi\mu\Fstat}{\semi\mu\gct} \gtrsim 0$ for $\semi T \gtrsim 20$~days and $\relerr{\semi\mu\Fstat}{\semi\mu\gct} \gtrsim 0.5$ for $\semi T \gtrsim 120$~days; in Fig.~\ref{fig:mu_re_twoF_gctco_semiT_cohTeq3_dutyeq1_tot}, $\relerr{\semi\mu\Fstat}{\semi\mu\gct} \gtrsim 0$ for $\semi T \gtrsim 14$~days and $\relerr{\semi\mu\Fstat}{\semi\mu\gct} \gtrsim 0.5$ for $\semi T \gtrsim 30$~days.

The relationship between the semicoherent and coherent metrics is often quantified using the \emph{refinement},
\begin{equation}
\label{eq:refinement-factor-def}
\gamma \equiv \frac{ \sqrt{\det \semimat g} }{ \median\useg \{ \sqrt{\det \cohsmat g} \} } \,.
\end{equation}
This quantifies the ratio of the number of templates in the semicoherent template bank $\semi\calN \propto \sqrt{\det \semimat g}$ to the median~\footnote{
In contrast to e.g.~\cite{Pletsch.2010a}, we use the median rather than the mean in the denominator of Eq.~\eqref{eq:refinement-factor-def}, for the following reason.
Throughout this paper we calculate detector positions $\protect\vec r(t)$ [Eq.~\eqref{eq:phase-def}] using standard Jet Propulsion Laboratory ephemerides for the motion of the Earth and Sun.
For segments with $\protect\coh T \sim 1$~day and specific segment mid-times, the ephemeris-derived orbital motion of the Earth can potentially interact with step~\eqref{item:coh-metric-lin-transf} of the procedure described in Section~\ref{sec:coher-supersky-param} to predict anomalously large numbers of coherent templates.
Using the median instead of the mean, therefore, gives a more representative value for the refinement in these circumstances.
} number of templates in the coherent template banks $\cohs\calN \propto \sqrt{\det \cohsmat g}$, assuming $\semi\mu\umax = \coh\mu\umax$.
Figure~\ref{fig:refinement_factor} plots the refinement for both supersky and global correlation metrics, the former given by Eq.~\eqref{eq:refinement-factor-def}, the latter given by Eqs.~(50), (59), and (73) of~\cite{Pletsch.2010a}.
The search setups listed in Table~\ref{tab:segment_list_table} are restricted to $\coh T = 1$~day and $\eta = 1$.
For $\semi T \gtrsim 20$~days, the rate of increase of the supersky metric refinement increases markedly, while that of the global correlation metric refinement remains constant.
The lack of additional refinement by the global correlation metric for $\semi T \gtrsim 20$~days is consistent with the deterioration in its ability to predict the $\calF$-statistic mismatch above the same threshold, as observed in Fig.~\ref{fig:mu_re_twoF_gctco_semiT}.

The semicoherent global correlation metric is derived under a number of assumptions, which were studied in unpublished work by~\cite{Manca.2013a}.
One assumption is that the sky position parameters do not contribute to the refinement, i.e.\ that the number of distinct sky positions that must be searched remains the same in both the semicoherent and coherent template banks.
We conclude from Figs.~\ref{fig:mu_re_twoF_gctco_semiT} and~\ref{fig:refinement_factor}, however, that this assumption remains valid only for $\semi T \lesssim 20$~days.
That the sky position parameters do contribute to the refinement has also been found in work on the Hough method, a similar semicoherent search method for gravitational-wave pulsars~\cite{Krishnan.etal.2004a,Astone.etal.2014a}.

\subsection{Number of semicoherent templates}\label{sec:numb-semic-templ}

Figure~\ref{fig:number_of_templates} plots the number of semicoherent templates $\semi\calN$ required to search the whole sky, frequencies up to $f\umax$, and spindowns $|\dot f| \le f / \tau\umin$ (motivated by a minimum ``spindown age'' of the gravitational-wave pulsar), with a prescribed maximum mismatch $\mu\umax$.
This parameter space is used in~\cite{Brady.etal.1998a} and~\cite{Brady.Creighton.2000a} for predictions of the number of coherent and semicoherent templates respectively.
Setting $f\umax = 200$~Hz, $\tau\umin = 10^3$~yr, and $\mu\umax = \coh\mu\umax = 0.3$, the prediction for the number of coherent templates in~\cite{Brady.etal.1998a} was compared in~\PaperII\ to that given by $\cohsmat g$.
Here, using the same values of $f\umax$, $\tau\umin$, and $\mu\umax = \semi\mu\umax$, the prediction for $\semi\calN$ given by $\semimat g$, which follows from Eq.~(29) of~\PaperII, is compared to the prediction of~\cite{Brady.Creighton.2000a}, given by Eqs.~(2.15)--(2.19) and~(2.22)--(2.27) of that paper, and to the prediction of the global correlation metric, given by Eqs.~(56), (59), and~(65) of~\cite{Pletsch.2010a}.
The search setups listed in Table~\ref{tab:segment_list_table} are restricted to $\coh T = 1$~day and $\eta = 1$.

In~\cite{Brady.Creighton.2000a}, the number of semicoherent sky templates is computed assuming that \emph{only} the rotational motion of the Earth is important; a similar assumption is made in deriving the global correlation metric~\cite{Pletsch.2010a}.
\PaperI\ demonstrated, however, that \emph{both} the rotational and orbital motion of the Earth are important to the coherent metric, and the same is true of the semicoherent metric.
The neglect of the orbital motion by both~\cite{Brady.Creighton.2000a} and~\cite{Pletsch.2010a} is consistent with their predictions for $\semi\calN$ not following the increase predicted by $\semimat g$ for $\semi T \gtrsim 20$~days, once the orbital motion begins to dominate, as seen in Fig.~\ref{fig:number_of_templates}.

In Section~\ref{sec:simul-results:-relat} we saw that the semicoherent metric $\semimat g$ generally over-estimates the $\calF$-statistic mismatch $\semi\mu\Fstat$.
The metric will therefore generate a template bank with more templates than are strictly required to satisfy the condition $\semi\mu\Fstat \le \semi\mu\umax$.
This effect cannot, however, account for the increase in the number of templates seen in Figure~\ref{fig:number_of_templates}, for two reasons.
First, the number of templates scales with the maximum metric mismatch as $\calN \propto \semi\mu\umax^{-n/2}$, where $n = 4$ for a four-dimensional parameter space.
Let us assume \emph{very} conservatively that $\semimat g$ \emph{always} over-estimates $\semi\mu\Fstat$ by a factor of 2, i.e. $\relerr{\semi\mu\Fstat}{\semi\mu} = -0.667$ (cf.\ Figs.~\ref{fig:mu_re_twoF_rssky_TN} and~\ref{fig:mu_re_twoF_rssky_mueta}).
We could therefore increase $\semi\mu\umax$ by a factor of 2 to achieve the desired maximum $\semi\mu\Fstat$, which would reduce the number of templates by a factor $(\semi\mu\umax/2)^{-4/2} / \semi\mu\umax^{-4/2} = 4$.
This factor is still \emph{much} smaller than the $10^{\lesssim 4}$ increase in the number of templates predicted by $\semimat g$, seen in Fig.~\ref{fig:number_of_templates}; this increase cannot, therefore, be accounted for by metric over-estimation.
Second, as discussed in Section~\ref{sec:simul-results:-relat}, the dominant contribution to the over-estimation of $\semi\mu\Fstat$ by $\semimat g$ is the Taylor expansion in Eq.~\eqref{eq:coh-Fstat-mismatch-metric} used in deriving the phase metric.
Both the metrics of~\cite{Brady.Creighton.2000a} and of~\cite{Pletsch.2010a}, against which $\semimat g$ is compared in Fig.~\ref{fig:number_of_templates}, are also derived from a phase metric via Taylor expansion.
The metrics of~\cite{Brady.Creighton.2000a,Pletsch.2010a} will therefore, in addition to their other assumptions, also tend to over-estimate the $\calF$-statistic mismatch.

\section{Summary}\label{sec:summary}

This paper presents a useful approximation to the semicoherent parameter-space metric: the semicoherent supersky metric $\semimat g$.
Together with the coherent supersky metric introduced in \PaperI, it constitutes a complete description of the parameter-space metrics for all-sky semicoherent searches for isolated gravitational-wave pulsars.
The number of semicoherent templates required to maintain a prescribed maximum mismatch is found to be several orders of magnitude larger than previous predictions.

The consequences of the required increase in the number of semicoherent templates for the sensitivity of gravitational-wave pulsar searches is left for future work.

\acknowledgments

I thank Reinhard Prix for many valuable discussions, and David Keitel for helpful comments on the manuscript.
I also thank the anonymous referee for suggesting valuable improvements to the paper.
Numerical simulations were performed on the ATLAS computer cluster of the Max-Planck-Institut f\"ur Gravitationsphysik.
This paper has document number LIGO-P1500126.

\bibliography{paper}

\end{document}